\documentclass[aps,showpacs,showkeys]{revtex4}
\usepackage{graphicx}
\usepackage{dcolumn}
\usepackage{bm}
\usepackage{graphicx}
\usepackage{amssymb}
\usepackage{amsmath}
\usepackage{hyperref}
\usepackage{latexsym,color}
\usepackage{amsmath,amssymb,amsthm,mathrsfs,amsfonts,dsfont}
\usepackage{amssymb}
\begin{document}
\newcommand{\ti}[1]{\mbox{\tiny{#1}}}
\newcommand{\im}{\mathop{\mathrm{Im}}}
\def\be{\begin{equation}}
\def\ee{\end{equation}}
\def\bea{\begin{eqnarray}}
\def\eea{\end{eqnarray}}
\def\beas{\begin{eqnarray*}}
\def\eeas{\end{eqnarray*}}
\newcommand{\tb}[1]{\textbf{\texttt{#1}}}
\newcommand{\rtb}[1]{\textcolor[rgb]{1.00,0.00,0.00}{\tb{#1}}}
\newcommand{\btb}[1]{\textcolor[rgb]{0.00,0.00,1.00}{\tb{#1}}}
\newcommand{\il}{~}
\newcommand{\dd}{\mathcal{D}}
\def\0o{\overset{o}}
\newcommand{\lie}{\mathcal{L}}
\newcommand{\la}{\mathcal{A}}
\newcommand{\Tem}{T^{\rm{em}}}
\newcommand{\g}[1]{\Gamma^{\phantom\ #1}}
\newcommand{\para}{{}^{\ti{$\parallel$}}}
\newcommand{\ort}{{{}^{\ti{$\perp$}}}}
\newcommand{\rin}[1]{\mathring{#1}}
\newcommand{\oB}{\0o{\mathbf{B}}}

\title{On the   locally
rotationally symmetric    Einstein-{Maxwell} perfect fluid
}

\author{Daniela Pugliese$^{1,\,2}$}
 \email{d.pugliese@qmul.ac.uk}
\author{Juan A. Valiente Kroon $^1$}%
 \email{j.a.valiente-kroon@qmul.ac.uk}
\affiliation{
$^1$ School of Mathematical Sciences, Queen Mary University of London\\
Mile End Road, London E1 4NS, UK%
\\
$^2$Institute of Physics, Faculty of Philosophy \& Science,
  Silesian University in Opava,
 Bezru\v{c}ovo n\'{a}m\v{e}st\'{i} 13, CZ-74601 Opava, Czech Republic}

\date{\today}
\begin{abstract}
We examine {the stability of an }   Einstein-Maxwell perfect
fluid configuration  with a privileged direction of symmetry by means of a $1+1+2$-tetrad formalism.  We use
 this formalism  to cast, in a quasi linear symmetric hyperbolic form the
equations describing the evolution of the system. This hyperbolic
reduction is used  to discuss the stability of solutions of
the linear perturbation.  {
By restricting the analysis to isotropic
fluid configurations, we made use of a constant electrical
conductivity coefficient for the fluid (plasma), and
the nonlinear stability for the case of an infinitely conducting plasma is also considered.  As a result of this analysis we provide a complete
classification and characterization  of various stable and unstable
 configurations. We found in particular that in many cases the stability conditions
is strongly determined by the constitutive equations by means of the square of the velocity of sound and the electric
conductivity, and a threshold for the emergence of the instability appears in both contracting and expanding systems.}
\end{abstract}
\maketitle
\section{Introduction}
The stability problem of plasma configurations is an important issue
in a variety  of astrophysical scenarios involving for example stellar objects and accretion
disks, and  various phenomena of the high energy Astrophysics,  related  to the accretion disks with the instability processes as the accretion   or the Jet emission.  In this article we consider the  situation where gravity plays a decisive role in determining both the equilibrium states of the  configurations then the dynamical  phases  associated  to the  instability, requiring    a full  general relativistic analysis \cite
{EtiLiuSha10,Fon03,Font2008,
Alc08,GiaRez07,
Lichnerowicz67,
Anton2006,
Radice:2013xpa,
Witek:2013ora}.
Consequently these systems are described by the coupled Einstein-Maxwell-Euler equations. Notable examples  are the  general relativistic (GR) magnetohydrodynamic (MHD)  systems. Often very complicated, the perturbation  analysis must be conducted with suitable assumptions on the  symmetries for the system (for example in the toroidal accretion disks) and the  dynamics,  numerical  approaches are often required.
A major challenge  in dealing with  these  systems is to find an appropriate formulation of the problem: from one side to set up a    formulation  adapted to the  configuration symmetries  and, on the other side a suitable  formulation of an initial value
problem is  necessary for the construction of the  numerical solutions,
in order to ensure the  local
and global existence problems and the analysis of the stability of certain reference
solutions\cite{PugKroon12} for a general discussion see for example\cite{ShiSek05,
BauSha03,
PalGarLehLie10,Anile89,Lichnerowicz67,Disconzi(2014),Radice:2013xpa,Witek:2013ora,Anton2006}.

In this work we set  the Einstein-Maxwell-Euler  equations in a  quasi-symmetric hyperbolic form,
 we explore the stability properties of a
    perfect fluid
configuration with a preferred direction of symmetry coupled with the electromagnetic field.  The formalism is adapted to the  description of a general  locally rotationally symmetric system, a remarkable example is the simple case of a spherically symmetric configurations.
For a more specific discussion on the stability of spherically symmetric plasma  see for example   \cite{Ho60,Mult-12,Tan-B86,Las-Lun07,VCxL97,Gu-Sha1999}.
The special case of an infinitely conducting plasma
describes an adiabatic flow so that the entropy per particle is
conserved along the flow lines.

The plasma configuration instability, especially in the  geometrically thin toroidal structures  orbiting around an attractor  (for example in the Shakura-Sunyaev accretion disks)  is often  described by  the  magneto-rotational instability (MRI). The dissipative (visco-resistive) effects are essential in these models as they  allow the transport of angular momentum in the configurations in accretion on the central object. 
In fact, in the geometrically thin configurations  it is assumed that the time scales of the  dynamical process  (balance in pressure and gravitational and centrifugal forces) are less then  the  thermal one (for dissipative heating and radiation) that is less then the viscous ones  (dissipative stresses and consequent angular momentum transport).
The magnetic field,  the dissipative effects and the radial gradient of the plasma relativistic angular velocity are therefore essential for the  MRI instability. However, some aspects of the theoretical framework of the  MRI and accretion process  are still to be clarified. An  intriguing  issue  for example is  the so-called visco-resistive puzzle: eventually high values of and resistivity  and  viscosity  have to be assumed.
In this work we  investigate the stability problem for the  systems with the magnetic field but not dissipative effects, showing  the emergence of the instability and a threshold for the occurrence of the unstable modes which is  essentially regulated  by the conductivity  parameter $\sigma_J$ and the speed of sound $v_s$, even for  configurations  with more specific symmetries,  considered  here  in classes and subclasses of solutions, stable and unstable for linear perturbation. Considering  both   contracting and expanding systems (according to the sign of  the kinematic expansion scalar $ \Theta $), we show  a threshold for the instability of the system determining two ranges for the   density of matter and especially for  the shear scalar  $\Sigma$ along the privileged direction of the system.
 More specifically we consider a barotropic equation of state:
 when the fluid entropy is a constant
of both space and time,  an equation of state to link the pressure $p$
to the matter density $\rho$ can be given in the form $p = p(\rho)$
---see e.g. \cite{PugKroon12}. In the present work we restrict our
attention to isotropic fluids. We consider a one species particle
fluid (\emph{simple fluid}) and, since no particle
annihilation or creation processes is expected, we use  the equation of
conservation of particle number. Moreover, we assume a polytropic
equation of state with a constant velocity of sound and we specify the
form of the conduction current using the  \emph{Ohm's law}, so that a linear relation between the conduction current
and the electric field holds.   By restricting our attention to the isotropic
fluids configurations we can make use of a constant electrical
conductivity coefficient for the fluid (plasma). These assumptions
simplify considerably the  analysis of the stability problem of
linear perturbations.

A central aspect of the stability analysis to be pursued in this
article is the construction
of a \emph{quasi linear symmetric hyperbolic evolution system} for the
variables of the configuration. This, in turn, ensures the  well-posed Cauchy problem for the
system ---in other words, the  local existence and uniqueness result
for the Einstein-Maxwell-Euler equations. By prescribing suitable
initial data on an initial hypersurface, a unique solution exists in a neighbourhood of that
hypersurface. This solution depends  continuously on the initial
data. Accordingly, we first write the  evolution equations for the
independent components of $n$ variables collected in  $n$-dimensional
vector $\textbf{{v}}$ used to obtain a suitable symmetric hyperbolic
evolution system of the form
\begin{equation}
\label{Eq:symm_hyp_forma}
\mathbf{A}^t\partial_t \textbf{{v}}-\mathbf{A}^j\partial_j \textbf{{v}}=\mathbf{B} \mathbf{v},
\end{equation}
where $\mathbf{A}^t$ and $\mathbf{A}^j$ and $\textbf{B}$ are smooth
matrix valued functions of the coordinates $(t, x)$ and the variables
$\textbf{{v}}$ with the index $j$ associated to some spatial
coordinates $x$.  The system is   symmetric hyperbolic if the
matrices $\mathbf{A}^t$ and $\mathbf{A}^j$ are symmetric and if
$\mathbf{A}^t$ is a positive-definite matrix. The evolution equations
is complemented by {constraint and  the constitutive
  equations}. As the purpose of this work this is the study of the
properties of the evolution system and the analysis of its linear
stability,  the problem of the propagation of the constraints will
only be briefly discussed  referring  further details to the
literature ---see in particular \cite{Reu98,Reu99}.

Our analysis is based on an adapted \emph{$1+1+2$-tetrad formalism} for
the \emph{locally
rotationally symmetric spacetimes} (LRS)\cite{Clarkson07}, the simplest example being the spherical symmetric   configurations ---that is, a covariant
decomposition of  Einstein-Maxwell perfect fluid field equations
which is particularly suitable  for   LRS systems. This
formalism is an extension of the usual $1+3$-formalism in which the
existence of a privileged timelike  vector field $u^a$ assumed ---in
applications involving the description of fluids, it is natural to let
$u^a$ to follow the congruence generated by the fluid
\cite{EllEls98}. In the $1+1+2$-formalism  the presence of a further
(spatial) vector field $n^a$ is assumed. This gives rise to a
a second split of the $1+3$-reduced equations on the plane parallel
and orthogonal to $n^a$. This type of decomposition is particularly
useful in the presence of symmetries as one can naturally fix the spatial vector on the privileged symmetry direction (from now on, for simplicity ``radial direction'') at
each point of  locally rotationally symmetric
spacetimes \cite{Clarkson07}. In the case of spherically symmetric
spacetimes  for example (here indicated with LSS),  it is natural to choose $n^a$
to point in the {radial direction} of the spherical symmetry. After the decomposition, all tensors are covariantly split into scalars, vectors, and
transverse-traceless 2-tensors, with respect to $n^a$. In the
  case considered here, as a result of this split, it can be shown
that it is only necessary to consider radially projected tensors. Further discussion on the decomposition of
Einstein-Maxwell-perfect fluid equations can be found in
\cite{Clarkson07,Clar-Mark-Bet-Dusns04,Bets-Clark04}. In
\cite{Las-Lun07} the same formalism has been used to analyse self
gravitating spherically symmetric charged perfect fluid
configurations in hydrostatic equilibrium. Details on the $1+3$ and
$1+1+2$-decompositions of the  Einstein-Maxwell  equations in LRS
spacetimes can be found in  \cite{Bur-cqg-08,Clark-Bar03,Els-Ellis96}.
We  consider  perturbations of the metric tensor,  of the matter and EM fields, all the quantities  share  the same   preferred direction of symmetry, as in  locally rotationally symmetric
(LRS) classes II space-times as described in \cite{Clarkson07} \footnote{We refer to  \cite{Las-Lun07,Clarkson07,Clar-Mark-Bet-Dusns04,Bets-Clark04,Bur-cqg-08,Burston:2007wt,Burston:2007ws} for a  general discussion concerning the  configurations  with a unique  symmetry direction, described by the  Einstein-Maxwell-Euler system. We mention also the well known solution of toroidal magnetic field widely adopted  in the  axes-symmetric  accretion configurations \cite{Komiss}, and \cite{Zanotti:2014haa} for a discussion on the case of a poloidal magnetic field where a  metric representation  is  adapted to the direction of the field,  as proved by Bekenstein \& Oron \cite{10,11}. Moreover for a deeper and more general discussion about the MHD configurations  in spherical symmetry, see for example the work \cite{Ho60,Mult-12,Tan-B86,Las-Lun07,VCxL97,Gu-Sha1999}}.

Taking  into account  the  gravito-electromagnetic (GEM) effects,  we provide a complete classification of the solutions in terms of the scalars of  the Weyl's conformal tensor. A general discussion and classification, of the solutions of the  Einstein-Maxwell equations,  through the scalars of the  Weyl tensor 
can be found in\cite{KramerSte}.

Here the consider the electromagnetic fields considered through real vector functions, while in many of the LSR spacetimes in \cite{Clar-Mark-Bet-Dusns04,
Bets-Clark04,
Bur-cqg-08,
Burston:2007wt,
Burston:2007ws}
was naturally used  the complex variable  $\psi=E+i B$ and $\psi^*=E-i B$ in order to
decouple the equations with the  appropriate symmetries and  obtain linear equations in the fields.
As pointed out in
\cite{Clar-Mark-Bet-Dusns04}
there are two ways to proceed
depending of how one considers the  coupled fields and test fields: specifically  on how one views the  Maxwell's and Einstein's
equations, and the   gravitational effects on the EM field, in
other words the analysis of test  fields  on a
background. The gravitational background  must, in any case, have the same symmetries of LRS systems: a suitable example is therefore
  the spherically symmetric  schwarzschild solution.
 Alternatively, one can consider other scenarios, we
refer to a general discussion of the analysis in \cite{Clar-Mark-Bet-Dusns04,
Bets-Clark04,
Bur-cqg-08,
Burston:2007wt,
Burston:2007ws}.
Here we simply observe that such a  system can be
 self-gravitating or not; one can consider the  perturbations of the  background or otherwise to fix the spacetime.

In this article we address the more general case, considering also the perturbations of the  gravitational part, in terms of scalars  of the Weyl tensor.
We assumed  the symmetries to be preserved by the perturbation  and limiting  the analysis to  constant velocity of sound and conductivity parameter. To simplify  the discussion of the results,   the calculation procedures, and the complete systems before  the  assumption of one privileged symmetry of direction, are specified  in  some   Appendix Sections.

The procedure of  hyperbolic reduction used in the present article
follows the presentation given in  \cite{LubbeKroon2011kz}. This
particular analysis is independent of geometric
gauge considerations\footnote{As pointed out in \cite{Clarkson07} in  a locally
rotationally symmetric  spacetime any  background quantities are scalars, implying   that  the  vector and tensor quantities are automatically gauge invariant, under linear perturbations
as a consequence of the Stewart-Walker lemma \cite{proc}.
}. In the present work we also provide a suitable
propagation equation for the fluid radial acceleration.
To this end, we introduce  an auxiliary  field  corresponding
to the derivative of the matter density projected along the radial
direction. suitable field and evolution
equations can be obtained for this quantity. The resulting evolution system
is then used to analyse the stability  problem for small nonlinear
perturbations of a background solution. More precisely, we perform a
first order perturbation to $\mathbf{v}$ of the form
$\mathbf{v}\mapsto \epsilon \mathring{\mathbf{v}}+
\breve{\mathbf{v}}$,  where the parameter $\epsilon$ sets  the order
of the perturbation while $\breve{\mathbf{v}}$ describes  the (linear)
perturbation of  the background  solution. Now, assuming the
background variables $\mathring{\mathbf{v}}$ to satisfy the evolutions
we end up with an evolution system for the perturbations of the form
\[
\mathring{\mathbf{A}}^t\partial_t \breve{\textbf{{v}}}-\mathring{\mathbf{A}}^j\partial_j \breve{\textbf{{v}}}=\mathring{\mathbf{B}} \breve{\mathbf{v}}.
\]
The core of the stability analysis consists of the study  of the term $\mathring{\mathbf{B}} $  using some
relaxed stability eigenvalue conditions. The procedure to analyse
stability used here is adapted from\cite{Reu99} ---see also
discussion in \cite{AlhMenVal10} and cited references. Under suitable
circumstances it can be regarded as a first step toward de analysis of
\emph{non-linear} stability. In our case, the elements of matrix
$\mathring{\mathbf{B}} $ are, in general, functions of the  space and
time coordinates. For a general discussion on the time dependent case
and the case of non constant matrix coefficient (depending on both
time and space) we refer to  \cite{Reu99}. The case of a linearised
system where the coefficients are constant matrices is discussed in \cite{Acta98}. Finally, the case of systems with
vanishing eigenvalues has been discussed in
\cite{1997funct.an..3003K,JMP1,JMP2}. In our case, a fully analysis of
the stability properties of the system turns  out extremely cumbersome
because of the  form of  $\mathring{\mathbf{B}}$  associated to the
present problem. We will proceed with the analysis of the values of
the eigenvalues using  indirect methods based on the inspection of the
characteristic polynomial. In order to keep the problem manageable, we
analyse a number of simplified systems obtained by making some
assumptions about the configuration. More precisely, we consider
background configurations with a vanishing radial acceleration case
for the reference solution, and we explore particular models with
fixed values of the kinematic scalars. This analysis constitutes the
main result of the article. We provide a detailed classification by
considering systems with particular kinematic configurations defined
by fixing the radial acceleration of the four velocity, and the
$1+1+2$-projected expansion, shear, twisting and the vorticity of the
system:{  the stability conditions
can be strongly determined by the constitutive equations by means of the square of the velocity of sound and the electric
conductivity.}

Since a significant part of this work was dedicated to the formalization of the problem in symmetric hyperbolic form, a first part  of the article  was necessarily devoted to the presentation of the formalism and the explanation of the adopted notation. To simplify the discussion of the problem and the illustration of the results the article has been developed  into three parts: in the first (\textbf{I}) part, from Sec.\il(\ref{Sec:prel}) to Sec.\il(\ref{Sec:varandeq}),  we introduced the $1 + 1 + 2$ formalism   decomposing the equations  and set system in   symmetric hyperbolic form, which is the first outcome of this work.
The second (\textbf{II}) part,  Sec.\il(\ref{Sec:Perturbation}) and
Sec.\il(\ref{Sec:discussion}), develops the perturbations and the   system stability is analysed. Part \textbf{II} constitutes the main part of this article with a major  discussion  of the main results on the  system stability  in the fundamental classes of the solutions. A more specific discussion on the other subclasses can be found  in the final part of this article.
The third (\textbf{III}) part  is constituted by the Appendix Sections,  deepening details of the \textbf{I} and \textbf{II} parts, and explaining some important aspects of the decomposition procedure. We  show the perturbed equations  in symmetric hyperbolic form for the general case of non-zero radial acceleration  discussing in detail the conditions for  the  unstable configurations belonging to the various  classes and subclasses of solutions.

\medskip
In details, the present article is structured   as follows: the $1+3$-formalism
 is briefly reviewed in
Section \ref{Sec:1+3}. The $1+1+2$-decomposition that will be used in
our present analysis is discussed in Section \ref{Sec:1+1+2}. In
Section \ref{Sec:1+1+2equations} we write the  $1+1+2$-equations for the LRS
system. Section \ref{Sec:on-the-thermod} provides
a discussion on the thermodynamical quantities of the system. A summary of the evolution
equations is given in Section \ref{Sec:varandeq}. The re-parametrised
set of evolution equations considered for the stability analysis is
given in Section \ref{Sec:Re-parametrized}. Section
\ref{Sec:Perturbation} discusses  the perturbation to the first order
of the variables. Section \ref{Sec:Remarks} provides some general
remarks on the set of perturbed equations and system
stability. Section \ref{Sec:discussion} contains the main results
concerning the nonlinear stability of the symmetric hyperbolic
system. Some concluding remarks are given in Section
\ref{Sec:conclusions}. Finally, in Appendix \ref{Sec:fluidfield}  we
provide an alternative  symmetric hyperbolic system for the fluid variables. The
$1+1+2$-decomposition of these equations is in Appendix
\ref{App:2+1+1decom}. Some general notes on the evolution equations
and hyperbolicity considerations can be found in Appendix \ref{Sec:summary}.
\section{Preliminaries}\label{Sec:prel}
We consider  the stability problem for a  configuration  with a spatial direction of symmetry (privileged direction of the system  admitting a
one-dimensional isotropy group), described by the Einstein-Maxwell-Euler  equations for a perfect fluid. By applying the $1+1+2$ decomposition i
we  can take full advantage of the symmetries of the
LRS system, and moreover
this procedure allows  to construct in an easy, and relatively immediate manner,  a quasi-linear hyperbolic system.  This ensures   also the consistency of possible  numerical approaches to the problems, without requiring the introduction of any auxiliary variable to handle both the propagation equations   along the timelike direction then  the constraint part of the system. Morover, the covariant and gauge-invariant
perturbation formalism    turn to be  especially  suitable for dealing
with spacetimes with some preferred spatial direction, not
necessary spherical symmetry in the background, and to the application
 possibly to the case of gravitational wave propagation by
introducing a radial unit
vector, decomposing all covariant quantities with
respect
to this \cite{Clarkson07,Clar-Mark-Bet-Dusns04,
Marklund:2004qz,
BauSha03}.  All  the equations and the quantities related  to the  fields and the curvature tensor will be decomposed according to the $1 + 1 + 2$ procedure. For this purpose
it will be necessary   to  first fix  the notation induced by the  $1 + 3$-decomposition as introduced  Sec.\il(\ref{Sec:1+3}).  The procedure of  $1+1+2$-decomposition, adapted to the symmetries of the system,  will be discussed   with some details in Sec.\il(\ref{Sec:1+1+2}).
\subsection{The $1 + 3$-formalism}\label{Sec:1+3}
The implementation of the $1 + 3$-formalism used in the present
article follows, as much as possible, the notation and conventions of
\cite{EllEls98}. We consider 4-dimensional metrics $g_{ab}$ with
signature $(-,+,+,+)$. The Latin indices $a, b, c...$ will denote
spacetime tensorial indices taking the values $(0,1,2,3)$ while
$i,j,k...$ will correspond to spatial frame indices ranging over $(1,
2, 3)$. The Levi-Civita covariant derivative of $g_{ab}$ will be
denoted by $\nabla_a$. Whenever convenient, we use the semicolon
notation. As usual, one has that $\nabla_a g_{bc}=g_{bc;a}=0$.

\medskip

In what follows, the timelike vector field (\emph{flow vector}) $u^a$
will describe the normalised future directed 4-velocity of the
fluid. It satisfies $u^a u_a=-1$. Indices are raised and lowered with
$g_{ab}$. The tensor $h^{ab}\equiv g^{ab} + u^{a} u^{b}$ is the
projector onto the three dimensional subspace orthogonal to $u^a$,
thus, one has that
\[
h^a{}_{b}=\delta^a{}_{b}+ u^a u_b,\quad h^b{}_ah^a{}_c=h^b{}_c,\quad h^a{}_b h^b_a=3, \quad h^a{}_bu_a=0.
\]
Following the standard approach of $1+3$-formalisms, we split the
first covariant derivative of $u^b$ as
\be\label{E:deco3+1u}
\nabla_{a}u_{b}= \sigma_{ab}+ \omega_{ab}+ {\frac{1}{3}}\,\Theta
h_{ab}- \dot{u}_{b}u_{a},
\ee
where $\sigma_{ab}\equiv D_{\langle b}u_{a\rangle}$ with
$\sigma_{ab}u^b=0$, $\omega_{ab}\equiv D_{[a}u_{b]}$ with
$\omega_{ab}u^b=0$, $\Theta\equiv D^{a}u_{a}$ and
$\dot{u}_{a}=u^{b}\nabla_{b}u_{a}$ are, respectively, the \emph{shear}
and the \emph{vorticity} tensors, the \emph{volume expansion scalar},
and the \emph{4--acceleration} vector. Moreover, we introduce the
vorticity vector $\omega^a\equiv\epsilon^{abc}\omega_{bc}/2$ where
$\epsilon_{abc}=u^d\epsilon_{dabc}$, $\epsilon_{abe}\epsilon^{abe}=6$
and $\epsilon_{dabc}$ stands for the totally antisymmetric tensor with
$\epsilon_{0123}=\sqrt{-\det{g_{ab}}}$. Then,
$\sigma_{ab}u^{a}=0=\omega_{ab}u^{a}=\dot{u}_{a}u^{a}$ by
construction. In the above expressions, the operator $D_a$ corresponds
to the 3-dimensional covariant derivative obtained from projecting the
spacetime covariant derivative in the distribution orthogonal to
$u_b$. As an example of this procedure, for a generic 2-rank tensor
$T_{bc}$, one has that $D_aT_{bc}=h^s{}_a h^t{}_b h^p{}_c T_{st;p}$, and
$T_{st;p}=\nabla_pT_{st}$ while $\dot{T}_{st}\equiv
u^a\nabla_{a}T_{st}$. For clarity, whenever necessary, projected
indices of a tensor will be highlighted by $\langle
\rangle$- brackets. For example, we write $T_{\langle a b \rangle} = h_a{}^c
h_b{}^d T_{cd}$.
\subsection{The $1+1+2$-formalism}\label{Sec:1+1+2}
The implementation of a $1+1+2$-formalism to be used in the present
article follows the notation and conventions of \cite{Clarkson07}. In
what follows, let $n_a$ denote a spacelike normalised vector $n_a$
chosen along a preferred direction of the spacetime and which
define a  further  projector tensor $N_a^b$
\be
n_a n^a=1,\quad u_a n^a=0,\quad N_a^b\equiv h_a^b-n_a n^b=\delta_a^b+u_au^b-n_a n^b,\quad N_a^bu^a=N_a^bn^a=0,
\ee
on the space orthogonal to both $n_a$ and $u_a$ ---a 2-dimensional \emph{sheet}. In addition to this projector, we also introduce the following rank 2 totally antisymmetric  tensor
\bea
\epsilon_{ab}\equiv\epsilon_{abc}n^c=\epsilon_{dabc}u^dn^c,\quad\epsilon_{(ab)}=\epsilon_{ab}n^b=0,
\quad
\epsilon_{abc}&=&n_a\epsilon_{bc}
-n_{b}\epsilon_{ac}+n_c\epsilon_{ab},
\\
\epsilon_{ab}\epsilon^{cd}=N_a^cN_b^d - N_a^dN_b^c,
\quad
\epsilon_a^{\phantom\ c}\epsilon_{bc}=N_{ab},\quad \epsilon^{ab}\epsilon_{ab} = 2.
\eea
Given  3-vector $\psi^a$, one can use the projector $N_a{}^b$ to split it as
\(
\psi^a=\Psi n^a+\Psi^{\bar{a}},
\)
with $\psi^{\bar{a}}\equiv\psi_{b}N^{ab}$, lying in the sheet
orthogonal to $n_a$. In order to avoid any confusion, we make use of
an overbar on index to denote the projection with $N_a{}^b$. By means of the projector $N_a^b$ one can decompose the kinematical
quantities $\dot{u}_a$ and $\omega_a$ in the form
\(
\dot{u}^a=\mathcal{A}n^a+\mathcal{A}^a, \qquad
\omega^a=\Omega n^a+\Omega^a,
\)
with $\mathcal{A}\equiv n^au^{b}\nabla_{b}u_{a}$ and $\Omega$ describing the components of the acceleration and the vorticity in the direction of $n^a$. Now, any spatially projected, symmetric,
trace-free (PSTF) tensor $\psi_{ab}$ can be split into a scalar ${{\Psi}}=\psi^bn_b$, a 2-tensor ${\Psi}_{ab}$ and a vector \textbf{${\Psi}_{a}$}
\be\label{E:tracefreet}
\psi_{ab}=\psi_{\langle ab \rangle }=\Psi \left(n_an_b-\frac{1}{2}N_{ab}\right)+2\Psi_{(a}n_{b)}+\Psi_{ab},
\ee
where
%
\bea\label{E:defidemi}
\Psi&\equiv&\psi_{ab}n^an^b=-N^{ab}\psi_{ab},\quad
\Psi_a\equiv N_a^bn^c\psi_{bc}=\psi_{\bar{a}},\quad
\Psi_{ab}\equiv\psi_{\{ab\}}\equiv\left( N_{(a}^{\phantom\ c}N_{b)}^{\phantom\ d}-\frac{1}{2}N_{ab}N^{cd}\right)\phi_{cd}.
\eea
%
In the above expressions, the curly brackets indicate the
transverse-traceless part of the corresponding tensor. Note that in
the present context,the term ``transverse'' only refers to the fact
that the tensor is orthogonal to $n_a$. In the following, where
useful, we will denote with $(\ort)$ the component of any tensor
orthogonal to $n_a$ and with $(\para)$ the component parallel to
it. It is convenient to further define the following two derivatives:
\bea
&&\hat{\phi}^{\phantom\ \phantom\ \phantom\ \phantom\ c \cdots d}_{a \cdots b}\equiv n^e D_e \phi^{\phantom\ \phantom\ \phantom\ \phantom\ c \cdots d}_{a \cdots b},
\quad \delta_e\phi^{\phantom\ \phantom\ \phantom\ \phantom\ c \cdots d}_{a \cdots b}\equiv N_e{}^j N_a{}^f\cdots N_b{}^g N_h{}^c\cdots N_i{} ^dD_j \phi^{\phantom\ \phantom\ \phantom\ \phantom\ h\cdots i}_{f\cdots g}.
\eea
Accordingly, one can write
\bea\label{Eq:trek}
D_a n_b=n_a a_b+\frac{1}{2}\Phi N_{ab}+\xi \epsilon_{ab}+\zeta_{ab},
\quad
 a_a\equiv n^cD_cn_a=\hat{n}_a,
\qquad
\Phi\equiv \delta_a n^a,
\quad\xi\equiv \frac{1}{2}\epsilon^{ab}\delta_a n_b,
\quad
\zeta_{ab}\equiv\delta_{\{a}n_{b\}}.
\eea
The scalar $\Phi$ denotes is the \emph{expansion} of the 2-sheet
generated by $n_a$, $\zeta_{ab}$ the shear of $n^a$ and $\xi$ is the
\emph{rotation} of $n^a$ (i.e. the \emph{twisting} of the 2 - sheet),
finally $a^a$ is the acceleration. Finally, also define the derivative
\be
\dot{n}_a=\mathcal{A}u_a+\alpha_a, \quad \alpha_a\equiv\dot{n}_{\bar{a}}.
\ee
It can then be verified that
\bea
&& \dot{N}_{ab}= 2u_{(a}\dot{u}_{b)} - 2n_{(a}\dot{n}_{b)} =
2 u_{(a}\mathcal{A}_{b)} -2 n_{(a}\alpha_{b)},
\quad\hat{N}_{ab} = -2n_{(a}a_{b)}, \quad
\delta_cN_{ab} =0,
\eea
and, furthermore, that
\begin{eqnarray}
&& \dot{\epsilon}_{ab} = -2u_{[a}
\epsilon_{b]c}\mathcal{A}^c + 2n_{[a}\epsilon_{b]c}\alpha^c,
\quad\hat{\epsilon}_{ab} = 2n_{[a}\epsilon_{b]c}a^c,
\qquad
\delta_c\epsilon_{ab} = 0.
\end{eqnarray}
\begin{description}
\item{On the decomposition of the kinematic quantities}
Applying the decomposition \eqref{E:tracefreet} to the symmetric trace
free tensor $\sigma_{ab}$ appearing in the $1+3$-decomposition given
by equation \eqref{E:deco3+1u} one obtains
\be
\sigma_{ab}=\Sigma \left(n_an_b-\frac{1}{2}N_{ab}\right)+2\Sigma_{(a}n_{b)}+\Sigma_{ab},
\ee
where we have introduced the scalar $\Sigma$ (the totally projected
part of the \emph{shear of the 3-sheet}), the vector $\Sigma_a$ and
the 2-tensor $\Sigma_{ab}$ and defined as in equations
\eqref{E:defidemi}.

The full covariant derivative
of $n^a$  and $u^a$ can now be written as
\begin{eqnarray}
&& \nabla_a n_b = - \mathcal{A} u_au_b - u_a\alpha_b + (
\Sigma + \tfrac{1}{3}\Theta ) n_a u_b + (\Sigma_a  -
\epsilon_{ac}\Omega^c ) u_b+ n_aa_b + \frac12\phi
N_{ab} + \xi\epsilon_{ab} + \zeta_{ab}
\\\nonumber
&& \nabla_au_b = -u_a\left( \mathcal{A} n_b + \mathcal{A}_b\right) + n_an_b(
\tfrac{1}{3}\Theta + \Sigma) + n_a( \Sigma_b + \epsilon_{bc}\Omega^c)
 + \left( \Sigma_a-\epsilon_{ac}\Omega^c \right) n_b + N_{ab}(
\tfrac{1}{3}\Theta - \tfrac{1}{2}\Sigma) + \Omega\epsilon_{ab}+\Sigma_{ab}.
\end{eqnarray}
\item[{The decomposition of the Weyl tensor}]
Let $C_{abcd}$ denote the Weyl curvature tensor of the metric
$g_{ab}$. As it is well known, it can be $1+3$-decomposed in terms of
symmetric, traceless 2-rank tensors defined by
\bea
&&
\mathcal{E}_{ab}\equiv C_{abcd}u^c u^d=\mathcal{E}(n_a n_b
- \tfrac{1}{2}N_{ab})+2\mathcal{E}_{(a}n_{b)}+\mathcal{E}_{ab},
\\
&&
\mathcal{B}_{ab}\equiv\frac{1}{2}\epsilon_{ade}C^{de}_{\phantom\ \phantom\ bc}u^c=\mathcal{B}(n_a n_b -\tfrac{1}{2}N_{ab})+2{\mathcal{B}}_{(a}n_{b)}+\mathcal{B}_{ab},
\eea
---the so-called electric $\mathcal{E}_{ab}$ and magnetic
$\mathcal{B}_{ab}$ parts with respect to $u^a$. Using equation
\eqref{E:tracefreet}, these rank 2 tensors are $1+1+2$-decomposed
where the scalars $\mathcal{E}$ and $\mathcal{B}$ denote,
respectively, the totally projected electric and magnetic parts of the
Weyl tensor.

\item[{Decomposition of the matter fields}]
The energy momentum tensor $T_{ab}$ describing the matter-field
content of the system under consideration is of the form
\be\label{E:Tm}
T_{ab}=T^{\rm{f}}_{a b}+T^{\rm{em}}_{a b},\quad   T^{\rm{f}}_{a b}=\rho u_{a}
u_{b}+ p h_{a b},
\ee
where $T^{\rm{f}}_{a b}$ denotes the energy-momentum tensor of the
perfect fluid given by the well-known expression
 with $\rho$ and $p$ denoting, respectively, the \emph{total energy
   density} and \emph{pressure} as measured by an observer moving with
 the fluid.  Accordingly, the timelike vector field (\emph{flow
   vector}) $u^a$ denotes the normalised future directed 4-velocity of
 the fluid. The electromagnetic energy momentum tensor $T^{\rm{em}}_{a
   b}$ is given by
\be\label{E:ff}
T^{\rm{em}}_{a
b}=\left (F_{a c}F^{\phantom\ c}_{b}-\frac{1}{4} F_{c d} F^{c
d} g_{ab}\right),\quad F_{ab}=-( 2 E_{[a}u_{b]}
-\epsilon_{abc}B^{c}).
 \ee
where $F_{ab}$ is the \emph{electromagnetic
field (Faraday) tensor}, the Faraday tensor  has been split in its \emph{electric
part}, $E_{a}\equiv F_{ab}u^{b}$, and its \emph{magnetic part},
$B^{a}\equiv\frac{1}{2}\epsilon^{abc}F_{cd}$, with respect to
the flow.
Alternatively, one can
rewrite equation \eqref{E:ff} in the form
\be\label{E:ff2}
T^{\rm{em}}_{a b}= \tfrac{1}{2} u_au_b (E^2+B^2)+\tfrac{1}{6}h_{ab}(E^2+B^2)+P_{ab}+2 \mathcal{G}_{(a}u_{b)},
\ee
where we have written $E^2\equiv E_aE^a$ and $B^2 \equiv B^aB_a$, and
$P_{ab}$ denotes the symmetric, trace-free tensor given by
\[
P_{ab}=P_{(ab)}\equiv\tfrac{1}{3}h_{ab}(E^2+B^2)-(E_aE_b+B_aB_b).
\]
Furthermore, $\mathcal{G}_{a}\equiv\epsilon_{auv}E^uB^v$,
denotes the \emph{Poynting vector}. Thus we obtain
\be
T_{ab}^{\rm{em}}n^an^b=\frac{1}{2}(E^2+B^2)-(\para E^2+\para B^2),\quad E^a=\para E n^a+\ort E^a,
\quad
B^a=\para B n^a+\ort B^a
\ee
 the electric
$E^a$ and magnetic $B^a$ are  split  relative to the vector $n^a$ .
Now, recall that he Maxwell equations can be written as
\be\label{E:maxz}
\nabla_b F^{ab}=J^a, \qquad \nabla_{[a} F_{bc]}=0 \quad\mbox{where}\quad J_a=\varrho_C u_a +{}^{\ti{(3)}}j_a,\quad {}^{\ti{(3)}}j^a=\para j n^a+\ort j^a,
\ee
where $J^a$ is the 4-current.
%
%
where $\varrho_C$ is the \emph{charge density}, and the 3-current is ${}^{\ti{(3)}}j_a$.
\end{description}
\section{$1+1+2$-equations for LRS
system}\label{Sec:1+1+2equations}
Evolution equations adapted to the $1+1+2$-decomposition of spacetime
discussed in the previous Sections can be readily obtained from the
$1+3$-evolution equations using the split of equation
\eqref{E:tracefreet} for a symmetric, trace-free tensor and projecting
the original propagation equations along the longitudinal and
orthogonal directions given by the vector $n^a$. For a generic
symmetric, trace-free tensor one obtains three equations for the
projected components $\Psi$, $\Psi_a$ and $\Psi_{ab}$. In the case of
a LRS and LSS systems it is only necessary to consider the
evolution equation of the radially projected component ---i.e. the
scalar $\Phi$. This approach is allowed in any spacetimes with a
preferred spatial direction at each point, the so-called locally
rotationally symmetric spacetimes (LRS) ---see e.g. \cite{Clarkson07}.

In the case {LSS
  spacetimes}, $n^a$ is a vector pointing along the axis of symmetry
and can therefore be thought of as being a radial vector.  As there is
no preferred directions in the 2-surface sheet all the non-radial
components of the various tensors are assumed to vanish. Consequently,
the kinematical quantities can be decomposed as
\[
\dot{u}^a=\mathcal{A}n^a,\qquad
\omega^a=\Omega n^a, \qquad
\sigma_{ab}=\Sigma \big(n_an_b-\tfrac{1}{2}N_{ab}\big).
\]
so that
\bea
&& n_{ab}\equiv \nabla_a n_b = - \mathcal{A} u_au_b  + \big(
\Sigma + \tfrac{1}{3}\Theta \big) n_a u_b + \tfrac{1}{2}\Phi
N_{ab} + \xi\epsilon_{ab} ,
\\
&&  u_{ab}\equiv \nabla_au_b = - \mathcal{A}u_a n_b+ n_an_b\big(
 \Sigma+\tfrac{1}{3}\Theta\big) + N_{ab}\big(
\tfrac{1}{3}\Theta - \frac12\Sigma\big) + \Omega\epsilon_{ab},
\eea
while the decomposition of the Weyl tensor reduces to
\(
\mathcal{E}_{ab} = \para\mathcal{E}\big(n_a n_b
- \tfrac{1}{2}N_{ab}\big),\qquad
\mathcal{B}_{ab} = \para\mathcal{{B}}\big(n_a n_b -\tfrac{1}{2}N_{ab}\big).
\)
In an analogous manner, it can be seen that the only non-vanishing
components for the matter fields are given by the scalars $(\rho, p,
\para E,\para B, \varrho_C,\para j)$.

\begin{description}
\item[{Evolution equation for the Maxwell fields}]
Starting from the usual $1+3$-decomposition of the Maxwell equations
---see for example \cite{EllEls98}--- and applying the general
procedure described at the beginning of the previous Section one
readily obtains the following $1+1+2$-evolution equations
\bea\label{E:ma}
&& \para\dot{{E}} =2\xi\para{B} -\big(\tfrac{2}{3}\Theta-\Sigma\big)\para{E}+\ort{E}^a(\alpha_a
+\Sigma_a)  +\epsilon_{ab}\delta^a\ort {B}^b+\epsilon_{ab}\big(\mathcal{A}^a\ort{B}^b +\Omega^a \ort{E}^b\big)-
\para {j},
\\\label{E:mb}
&&  \para\dot{{B}}=-2\xi\para{E}-\big(\tfrac{2}{3}\Theta-\Sigma\big)\para{B} -\epsilon_{ab}\delta^a\ort{E}^b+\ort{B}^a(\alpha_a
+\Sigma_a) -\epsilon_{ab}(\mathcal{A}^a \ort{E}^b-\Omega^a \ort{B}^b).
\eea
The above equations are supplemented by a pair of constraint equations
for for the scalars $\para E$ and $\para B$ ---see below. If one now assumes
LRS, equations \eqref{E:ma} and \eqref{E:mb} reduce to
\bea\label{E:ma-re}
&&  \para\dot{{E}} =2\xi\para{B} -\left(\frac{2}{3}\Theta-\Sigma\right)\para{E}-\para{j},
\quad \para\dot{{B}}=-2\xi\para{E}-\left(\frac{2}{3}\Theta-\Sigma\right)\para{B},
\eea
while the constraint equations are given by
\begin{equation}
\label{E:contE}
\para\hat{E}+\para E\Phi-2\Omega\para B-\varrho_C=0,
\qquad
\para\hat{B}+\para B\Phi+2\Omega\para E=0.
\end{equation}
As in this  LRS one has that $E_a= E n_a$, $B_a= B n_a$
and $J_a=j n_a+\varrho_C u_a$, the superscript $\para$ can be dropped
from the above equations without giving rise to any ambiguity.
\item[{Perfect fluid equations}]
From the conservation of the energy-momentum tensor
%
$\nabla^a T_{ab}=0$,
it readily follows that
\be\label{E:Tor}
u_bu_a\nabla^a (p+\rho)+(p+\rho)\big(u_b(\nabla^au_a)+u_a \nabla^a u_b\big)+\nabla_b p+\left(\nabla^a F_{ac}\right)F_{b}^{\phantom\ c}=0.
\ee
In what follows, we will consider the projections of equation \eqref{E:Tor}
along the directions parallel and orthogonal to the flow lines of the
fluid. Contracting equation \eqref{E:Tor} with $u^b$ we obtain the
\emph{conservation equation}
\be\label{Eq:conservazione}
u_a\nabla^a\rho+(p+\rho)\nabla^au_a- u^bF_{b}^{\phantom\ c}(\nabla^aF_{ac})=0.
\ee
Now, in the case of an ideal conducting fluid, where
$E_a=F_{ab}u^{b}=0$, the last term of equation
\eqref{Eq:conservazione} is identically zero and the electromagnetic
field does not have a direct effect on the conservation equation along
the flow lines. Contracting equation \eqref{E:Tor} with $h^{bc}$ one
obtains the so-called \emph{Euler equation}
\be\label{E:qwety}
(p+\rho)u^a\nabla_au^c+ h^{bc}\nabla_b p+(\nabla^aF_{ad})F^{\phantom\ d}_b h^{bc}=0.
\ee
This last equation can also be written as
\be\label{E:qwety2}
(p+\rho)u_a\nabla^au_b+(u_bu^d\nabla_d+ \nabla_b) p+(\nabla^aF_{ad})\left(F^{\phantom\ d}_b+ F^{ed}u_e u_b\right)=0.
\ee
The last term of equation \eqref{E:qwety2} is identically zero as a
consequence of the Maxwell equation so that at the end of the day one
obtains the simpler expression
\(
(p+\rho)u^a\nabla_au^c+ h^{bc}\nabla_b p+(\nabla^aF_{ad})F^{cd}=0.
\)
Now, using equation \eqref{E:maxz} in the above one obtains that
\bea\nonumber
&& n^c(n^b\nabla_bp)+(\rho+p)(\para \mathcal{A} n^c+\ort\mathcal{A}^c)+N^b_c\nabla_b p -\para E(n^c \varrho_C+\para j u^c)\\
&& \hspace{2cm}-\ort E^c \varrho_C-\ort j_d \ort E^d u^c-\para B \epsilon^{cd}\ort j_d-(\epsilon^{cdu}\ort j_d-\para j \epsilon^{cu})\ort B_u=0,\label{E:peq}.
\eea
Consequently, projecting equation \eqref{E:peq} along $n_a$ we obtain
a constraint equation for the pressure $p$. Namely, one has that
\be\label{E:mapor}
\hat{p}+(p+\rho)\mathcal{A}-\varrho_C\para E-\epsilon^{de}\ort j_d \ort B_e=0, \quad
%
\dot{\rho}+(\rho+p)\Theta-\para E\para j-\ort j^c\ort E_c=0.
\ee
where \eqref{Eq:conservazione} has been considered for  the conservation equation.
In the particular case of a LSS, equation (\ref{E:mapor}) read
\be\label{E:HP_rho}
\dot{\rho}+(\rho+p)\Theta-\para E \para j=0,\quad
\hat{p}+(p+\rho)\mathcal{A}-\varrho_C\para E=0.
\ee

\item[{Evolution equation for the electric part of the Weyl tensor}]
Using the expressions obtained in the previous Section it can be
verified that the evolution equation for the electric part of the Weyl
tensor takes the form
\bea\nonumber
&& \para\dot{\mathcal{E}}=-\frac{1}{2}(\rho+p)\Sigma+\big(\tfrac{3}{2}\Sigma-\Theta\big)\para \mathcal{E}+3\para\mathcal{B}\xi+\epsilon^{cd}\delta_c \mathcal{B}_d+\epsilon^{cd}\ort\mathcal{B}_{db}\zeta_{c}^{\phantom\ b}
\\\label{E:apo}
&& \hspace{2cm} +\Sigma^a\mathcal{E}_a-\Sigma^e_c\ort \mathcal{E}_e^c+2\mathcal{A}_c\epsilon^{cd}\ort\mathcal{B}_d+\Omega_c\epsilon^{cd}\mathcal{E}_d
+2\mathcal{E}^c\alpha_c+\rm{\dot{F}}^{\ti{em}},
\eea
where
\bea\label{orol}
{\rm{\dot{F}}^{\ti{em}}}=n^a n^b\big(-\tfrac{1}{3}\sigma_{ab}(E^2+B^2)-\tfrac{1}{2}\dot{P}_{ab}-\tfrac{1}{6}\Theta P_{ab}
-\tfrac{1}{2}D_{\langle a}\mathcal{G}_{b \rangle}-\la_{\langle a}\mathcal{G}_{b \rangle}-\tfrac{1}{2}\sigma_{\langle a}^cP_{b \rangle c}+\tfrac{1}{2}\epsilon_{cd \langle a}\omega^cP_{b\rangle }^d
\big).
\eea
Now, following Clarkson's procedure for the $1+1+2$-decomposition of a
LSS, equation \eqref{E:apo} becomes
\[
\para\dot{{\mathcal{E}}}=\big(\tfrac{3}{2}\Sigma
-\Theta\big)\para{{\mathcal{E}}}+3\xi\para{{\mathcal{B}}}-\tfrac{1}{2}
(\rho+p)\Sigma-\tfrac{1}{3}(\para E^2+\para B^2)\big(\frac{\Sigma}{2}-\frac{\Theta}{3}\big)+\frac{1}{3}u^a\nabla_a(\para E^2+\para B^2),
\]
while the constraint equation is given by
\bea
\para\widehat{\mathcal{E}}+\frac{3}{2}\para\mathcal{E}\Phi-\frac{1}{3}\hat{\rho}-3\Omega\para
\mathcal{B}-{\rm{F}}^{\ti{em}} =0,
\quad
{\rm{F}}^{\ti{em}}=\frac{1}{2}\bigg[ \big(\para E^2 +\para B^2\big) \phi+\para E\widehat{\para E}+\para B\widehat{\para B} \bigg].
\eea
Using, respectively, the evolution equations \eqref{E:ma-re} and the equation for the electric parts of the Maxwell field magnetic
field we obtain
\begin{equation}\label{Eq:HP_Weyl-E}
u^{a} D_{a}\mathcal{E}+\tfrac{2}{3} E j +\Theta\left[ \tfrac{1}{3}(B^2+E^2)   + \mathcal{E}\right] - 3 \mathcal{B}  \xi +\left[ \tfrac{1}{2}(p+\rho) -  \tfrac{1}{2}(B^2+E^2) -  \tfrac{3}{2} \mathcal{E} \right]
 \Sigma  =0.
\end{equation}
Moreover, the constraint equation  is given by
\be
 n^{a}  D_{a}\mathcal{E}  -  n^{a}D_{a}p - \tfrac{1}{3} n^{a}D_{a}\rho=
 \mathcal{A}  (p + \rho) - \tfrac{1}{2} \big[(B^2 +E^2 + 3 \mathcal{E})\Phi - 6 \mathcal{B}  \Omega\big].
\ee
Notice that this last equation involves also the derivatives of $\rho$
and the pressure $p$ projected along the radial direction.
\item[{Evolution equation for the magnetic part of the Weyl tensor}]
The evolution equation for the magnetic part of the Weyl tensor is given by
\bea\nonumber
\para\dot{\mathcal{B}}=\big(\tfrac{3}{2}\Sigma-\Theta\big)\para \mathcal{B}-3\para\mathcal{E}\xi+ 2\bar{B}^c\alpha_c-\epsilon^{cd}\delta_c \mathcal{E}_d +\Sigma^a\mathcal{B}_a-\Sigma_{ac}\ort B^{ac}-2\epsilon_{cd}\mathcal{A}^c\mathcal{E}^d+\epsilon^{cd}B_d \Omega_c-\tfrac{1}{2}\epsilon^{cd}\ort\mathcal{E}_d^b\zeta_{cb}+\rm{\dot{B}}^{\ti{em}},\label{E:apoB}
\eea
where
\begin{equation}\label{be}
{\rm{\dot{B}}}^{\ti{em}}=n^an^b\big(\tfrac{1}{2}\mathrm{curl}\,P_{ab}+\tfrac{3}{2}\omega_{\langle a}\mathcal{G}_{b \rangle}+\tfrac{1}{2}\epsilon_{cd \langle a}
\sigma^c_{b\rangle }\mathcal{G}^d\big).
\end{equation}
Thus, in a LSS one obtains the propagation equation and the associated constraint equation
\be\label{Eq:HP_Weyl-B}
\para\dot{{\mathcal{B}}}=\big(\tfrac{3}{2}\Sigma
-\Theta\big)\para{{\mathcal{B}}}-3\xi{}\para {{\mathcal{E}}}-\xi \left(\para E^2+\para B^2\right),
\quad
\para\widehat{\mathcal{B}}+\tfrac{3}{2}\para\mathcal{{B}}\Phi+(\rho+p)\Omega+3\Omega\para \mathcal{E}-\Omega (\para E^2+\para B^2)=0.
\ee
\item[{Evolution equations for the kinematic quantities}]
In order to discuss the evolution equations it is convenient to
consider the following third rank tensors
\(
R^{(n)}_{abc}\equiv2\nabla_{[a}\nabla_{b]}n_c-R_{abcd}n^d=0, \qquad R^{(u)}_{abc}\equiv2\nabla_{[a}\nabla_{b]}u_c-R_{abcd}u^d=0.
\)
The required evolution and associated constraint equations are then
obtained from the projection of these \emph{zero quantities} along the
tensors $(u^a, n^a, \epsilon_{ab},  \epsilon_{abc})$, we are able to find out the
evolution and constraint equations for the kinematic quantities.

\medskip
From the constraint $R^{(u)}_{abc}u^a\epsilon^{bc}=0$ we readily
deduce the following evolution equation for the vorticity projected
along the radial direction
\(
\dot{\Omega}-\mathcal{A}\xi-\tfrac{1}{2}\epsilon^{bc}D_b\mathcal{A}_c
=\big(\Sigma-\tfrac{2}{3}\Theta\big)\Omega +(\ort \Sigma_b+\alpha_b)\ort \Omega^b.
\)
In the LSS case it readily reduces to
\be\label{Eq:HP_omega}
\dot{\Omega}
=\mathcal{A}\xi+\big(\Sigma -\tfrac{2}{3}\Theta \big)\Omega,\quad n^{a} D_{a}\Omega+(\Phi- \mathcal{A}) \Omega=0.
\ee
we obtain (in a
LSS) the constraint equation through  the contraction $R_{abc}^{(u)}\epsilon^{abc}=0$.
Computing $ u^aN^{bc}R^{(n)}_{abc}=0$ we find an evolution equation for $\Phi$, namely,
\begin{eqnarray*}
&& \dot{\Phi}-\big(\tfrac{1}{3}\Theta-\tfrac{1}{2}\Sigma\big)\big(2 \mathcal{A}-\Phi \big)-2\xi\Omega= \delta_c\alpha^c+\mathcal{A}^d\big(\epsilon_{dc}\ort\Omega^c-
\ort\Sigma_d-a_d+\alpha_d\big)\\
&& \hspace{3cm}-\zeta^{cd}\ort\Sigma_{cd}+a^c\big(\ort\Sigma_c-\epsilon_{cu}\ort \Omega^u\big)+\epsilon_{uv}\ort E^u\ort B^v
\end{eqnarray*}
which, for a LSS, reduces to
\begin{equation}\label{Eq:HP_phi}
\dot{\Phi}-\big(\tfrac{1}{3}\Theta-\tfrac{1}{2}\Sigma\big)\big(2\mathcal{A}-\Phi \big)-2\xi\Omega=0,\quad  n^{a} D_{a}\Phi+\mathcal{E}-  \tfrac{2}{9} \Theta^2 - 2 \xi^2 + \tfrac{2}{3} \rho -  \tfrac{1}{3} \Theta \Sigma + \Sigma^2 + \tfrac{1}{2} \Phi^2=0
\end{equation}
constraint equation for $\Phi$ in a LSS is found by the contraction $R_{abc}^{(n)}n^aN^{bc}$.
Computing the contraction $u^a\epsilon^{bc}R^{(n)}_{abc}=0$ one obtains
\[
\dot{\xi}=\tfrac{1}{2}\mathcal{B}-\big(\frac{1}{3}\Theta-\tfrac{1}{2}\Sigma\big)\xi+
\tfrac{1}{2}\epsilon_{ab}\delta^a\alpha^b+\big(
\mathcal{A}-\tfrac{1}{2}\Phi \Omega\big)\Omega-\frac{1}{2}\epsilon_{ab}\zeta^a_u
\ort\Sigma^{bu}+\tfrac{1}{2}\big[\epsilon_{ab}(\ort \Sigma^b+\alpha^b)+\ort\Omega_a\big[\big(a^a+ \mathcal{A}^a\big),
\]
which, for a LSS reduces to
\be\label{Eq:HP_xi}
\dot{\xi}=\tfrac{1}{2}\mathcal{B}-\big(\tfrac{1}{3}\Theta-\tfrac{1}{2}\Sigma\big)\xi
+\big(\mathcal{A}-\tfrac{1}{2}\Phi\big)\Omega,\quad n^{a}D_{a}\xi+ \xi \Phi - \big(\tfrac{1}{3} \Theta  +  \Sigma \big)\Omega=0,
\ee
the associated constraint is given, for a LSS, by the constraint $R_{abc}^{(n)}\epsilon^{bca}=0$.
Through similar calculations, one obtains from the contractions  $R_{abc}^{(u)}u^ag^{ab}=0$ and  $R_{abc}^{(n)}u^an^bu^c=0$ two alternative equations for the expansion scalar:
\begin{subequations}
\begin{eqnarray}
&& u^{a} D_{a}\Theta -  n^{a} D_{a}\mathcal{A} + \frac{1}{2}(B^2+E^2)  + \tfrac{1}{2}\big(3 p + \rho\big)+ \tfrac{1}{3} \Theta^2 + \tfrac{3}{2} \Sigma^2 - \mathcal{A}^2-  \mathcal{A} \Phi - 2 \Omega^2 =0,\label{Eq:HP_theta} \\
&&  \tfrac{1}{3} u^{a} D_{a}\Theta +  u^{a} D_{a}\Sigma-n^{a} D_{a}\mathcal{A}  =\mathcal{A}^2 -  \tfrac{1}{2}(B^2+E^2) -  \mathcal{E}  - \tfrac{1}{2}\big(p+\tfrac{1}{3} \rho\big) - \big( \tfrac{1}{3} \Theta  + \Sigma\big)^2.\label{Eq:HP_sigma}
\end{eqnarray}
\end{subequations}

Finally, in the sequel it will be useful to consider the following
equations obtained, respectively, from the contractions
$R_{abc}^{(n)}\epsilon^{ab}u^c=0$, $R_{abc}^{(u)}u^aN^{bc}=0$,
$R_{abc}^{(u)}n^aN^{bc}=0$:
\begin{subequations}
\begin{eqnarray}
&& \mathcal{B} - 3 \xi \Sigma + \big(2 \mathcal{A}  -  \Phi \big)\Omega=0, \label{Eq:vincolo}\\
&& \tfrac{2}{3}u^{a} D_{a}\Theta -  u^{a}D_{a}\Sigma + p + \tfrac{1}{3} \rho + 2\big( \tfrac{1}{3} \Theta - \tfrac{1}{2} \Sigma \big)^2- \mathcal{E}-  \mathcal{A} \Phi - 2 \Omega^2=0, \\
&& \tfrac{2}{3} n^{a}D_{a}\
\Theta -  n^{a}D_{a}\Sigma- \tfrac{3}{2} \Sigma \Phi - 2 \xi \Omega=0.
\end{eqnarray}
\end{subequations}
\end{description}
\section{Remarks on the thermodynamical quantities}\label{Sec:on-the-thermod}
In the present article we consider a one species particle fluid (\emph{simple
fluid}), and  denote, respectively, by $n,\,s,\, T$ the \emph{particle number density},
the \emph{entropy per particle} and the \emph{absolute temperature},
as measured by comoving observers. We also introduce the \emph{volume
per particle} $v$ and the \emph{energy per particle}  $e$ via the relations
$
v\equiv {1}/{n}$ and  $ e \equiv {\rho}/{n}$.
In terms of these variables the first law of Thermodynamics,
$\mbox{d}e=-p\mbox{d}v+T\mbox{d}s$, takes the form
 \be\label{E:1law}
\mbox{d}\rho=\frac{p+\rho}{n} \mbox{d}n +n T \mbox{d}s.
\ee
Assuming an equation of state of the form $\rho=f(n,s)\geq0$, one obtains from
 equation \eqref{E:1law} that
\be\label{E:the}
p(n,s)=n\left(\frac{\partial \rho}{\partial n}\right)_{s}-\rho(n,s),
\qquad T(n,s)=\frac{1}{\rho}\left(\frac{\partial \rho}{\partial s}\right)_n.
\ee
Assuming that $\partial p/\partial \rho>0$ we define the \emph{speed
of sound} $\nu_s=\nu_s(n,s)$ by
\be\label{E:vs}
\nu_s^2\equiv\left(\frac{\partial p}{\partial \rho}\right)_s=\frac{n}{\rho+p}\frac{\partial p}{\partial n}>0.
\ee
Since we are not considering particle annihilation or creation processes we
consider the equation of conservation of particle number:
\be\label{E:then}
u^a\nabla_a n+n \nabla_a u^a=0.
\ee
Combining this equation with equations \eqref{Eq:conservazione} and \eqref{E:1law} we obtain
\be\label{E:fors}
u^a\nabla_a s=\frac{1}{nT}u^b F_{b}^{\phantom\ c}\nabla^a F_{ac}.
\ee

Where $n$ is subject to equation \eqref{E:then} and $T$ is given by equation \eqref{E:the},  thus,
%
using the $1+1+2$ decomposition
%
in a LSS one obtain the couple of equations
\begin{equation}\label{E:fosev2}
\dot{s}=+\frac{1}{nT}\para E\para j
\qquad
\dot{n}+n \para \Theta=0.
\end{equation}
In the case of an infinitely conducting plasma, where the last term of
equation \eqref{E:1law} vanishes, equation \eqref{E:fors} describes an
adiabatic flow ---that is, $u^a\nabla_as=0$, so that the entropy per
particle is conserved along the flow lines. A particular case of
interest is when $s$ is a constant of both space and time.  In this
case the equation of state can be given in the form $p=p(\rho)$. In
the following for convenience, it is assumed that
$\nabla_ap=v_s\nabla_a\rho$ where we have defined
$v_s\equiv\nu_s^2$. Further discussion on the thermodynamical
quantities and assumptions can be found in Appendix \ref{Sec:summary}.
\section{Variables and equations}\label{Sec:varandeq}
In this subsection we summarise our analysis so far. The evolution
equations for the \emph{independent} components of the vector-valued
unknown
\be\label{E:v} \vec{v}=\left(\rho,s, n,\Omega,\xi, \Phi, E, B,\mathcal{B},  \mathcal{E},  \Theta, \Sigma, \mathcal{A}\right),
\ee
describing the evolution of a LSS  in   equations \eqref{E:HP_rho}, \eqref{E:fosev2}, \eqref{Eq:HP_omega}, \eqref{Eq:HP_xi}, \eqref{Eq:HP_phi}, \eqref{E:ma-re},\eqref{Eq:HP_Weyl-B}, \eqref{Eq:HP_Weyl-E}, \eqref{Eq:HP_theta} and \eqref{Eq:HP_sigma} are given by
\begin{subequations}
\begin{eqnarray}\label{permu}
&& \dot{\rho}=-(\rho+p)\Theta+ E \para j,
\\
\label{E:osev2}
&& \dot{s}=\frac{1}{nT} E\para j,
\\
\label{E:nosev}
&& \dot{n}=-n  \Theta,
\\
\label{qomega}
&& \dot{\Omega}=\mathcal{A}\xi
-\tfrac{2}{3}   \Theta \Omega+\Sigma \Omega,
\\
\label{perxi}
&& \dot{\xi}=\tfrac{1}{2}\mathcal{B}-\big(\tfrac{1}{3}\Theta-\tfrac{1}{2}\Sigma\big)\xi
+\big(\mathcal{A}-\tfrac{1}{2}\Phi\big)\Omega,
\\
\label{perfi}
&& \dot{\Phi}=\big(\tfrac{1}{3}\Theta+\tfrac{1}{2}\Sigma\big)\big(2 \mathcal{A}+\Phi \big)-2\xi\Omega=0,
\\
\label{E:mawire}
&& \dot{{E}} = 2\xi{B} -\big(\tfrac{2}{3}\Theta-\Sigma\big){E}-\para{j},
\\
\label{E:mawirb}
&& \dot{{B}}=-2\xi{E}-\big(\tfrac{2}{3}\Theta-\Sigma\big){B},
\\\label{bdotv}
&& \dot{{\mathcal{B}}}=\big(\tfrac{3}{2}\Sigma
-\Theta\big){{\mathcal{B}}}-3\xi{} {{\mathcal{E}}}-\xi \big( E^2+ B^2\big),
\\
\label{E:We_R1}
&& \dot{{\mathcal{E}}}=\big(\frac{3}{2}\Sigma
-\Theta\big){{\mathcal{E}}}+3\xi{{\mathcal{B}}}-\tfrac{1}{2}
(\rho+p)\Sigma+\left(\frac{\Sigma}{2}-\frac{\Theta}{3}\right)(E^2+ B^2)-\tfrac{2}{3} E \para j,
\\\label{Ec:thetaaA}
&& u^{a} D_{a}\Theta-  n^{a} D_{a}\mathcal{A}= - \tfrac{1}{2}(B^2+E^2)   -\tfrac{1}{2}\big(3p + \rho\big) - \tfrac{1}{3} \Theta^2  - \tfrac{3}{2} \Sigma^2 +\mathcal{A}^2 +  \mathcal{A} \Phi + 2 \Omega^2,
\\\label{Ec:thetaaSigma}
&& u^{a} D_{a}\Sigma+\tfrac{1}{3} u^{a} D_{a}\Theta  -  n^{a} D_{a}\mathcal{A}=\mathcal{A}^2 -  \tfrac{1}{2}(B^2+E^2)  -  \mathcal{E}  -\tfrac{1}{2}\big(p + \tfrac{1}{3} \rho\big)-  \big(\tfrac{\Theta}{3}  + \Sigma\big)^2.
\end{eqnarray}
\end{subequations}

In the subsequent stability analysis, we will assume a vanishing
radial acceleration. It is worth pointing out that the
derivative along the radial direction of the acceleration
$\mathcal{A}$ appears in the evolution equations for $\Theta$ and
$\Sigma$.  In Section \ref{Sec:Re-parametrized} and in Appendices
\ref{Sec:fluidfield},\ref{App:2+1+1decom} and \ref{Sec:summary} we
provide a suitable propagation equation for this variable so as to
complete a symmetric hyperbolic evolution system.

As discussed in more detail in Section \ref{Sec:on-the-thermod}, the
system will be supplemented by an equation of state of the form
$p=p(\rho)$.  Furthermore, it will be necessary to specify the form of
the conduction current, $j^a$. Consistent with \emph{Ohm's law}, we
assume a linear relation between the conduction current $j^a$ and the
electric field. More precisely, we set
$
j^a=\sigma^{ab}E_b
$,
where $\sigma^{ab}$ denotes the \emph{conductivity} of the fluid
(plasma). We will restrict our attention to isotropic fluids for which
$\sigma^{ab}=\sigma_J g^{ab}$, so that
\be\label{E:lg}
J^a=\varrho_C u^a +\sigma_J E^a,
\ee
with $\sigma_J$ the \emph{electrical conductivity coefficient}.  In
terms of a $1+1+2$-decomposition one has that in the particular case
of a LSS
\be\label{E:jpara}
\para j^a=\para E \sigma_J  n^a.
\ee
The system \eqref{permu}-\eqref{E:We_R1} is  complemented by the
constraint equations for the components of the vector-valued function
$\vec{v}$.  Some of these constraint equations were obtained in the
previous Sections. Further discussion can be found in the Appendices
to this article. As the primary aim of this work is the study of the
time evolution and the analysis of its stability, the discussion of
the constraint equations will remain in a secondary level.  In what
follows, it is assumed that the constraint equations are satisfied at
all time. This assumption can be removed by fairly general arguments
---see \cite{Reu98,Reu99}.
\subsection{Re-parametrised set of evolution equations}\label{Sec:Re-parametrized}
In what follows we  assume that  $\partial_av_s=0$ and we introduce
the variables $Q$ and $T$ via the relations
\be
Q\equiv T+3 \Sigma,\qquad T\equiv \tfrac{2}{3}\Theta- \Sigma\quad\mbox{so that}\quad\Sigma = \tfrac{1}{3}(Q-T),\qquad  \Theta =\tfrac{3}{2}(T+\Sigma ).
\ee
In terms of these one obtains a vector-valued unknown with components
given by
\be\label{Eq:vu-vectorvdef}
\mathbf{v}\equiv\big(\rho,{E},{B},{\mathcal{E}},{\mathcal{B}},{Q},{T},{\xi },{\Phi },{\Omega },\la\big).
\ee
and equations \eqref{permu}, \eqref{E:mawire}, \eqref{E:mawirb},
\eqref{E:We_R1}, \eqref{bdotv}, \eqref{perfi}, \eqref{perxi},
and\eqref{qomega} take the form
\begin{subequations}
\begin{eqnarray}\label{Eq:last-rhodot}
&&\dot{\rho}- E j+\frac{3}{2} (p+\rho ) \big(\tfrac{1}{3} (Q-T)+T\big)=0,
\\
&&\label{Eq:dotE}
\dot{E}-2  B \xi
   + E T+j=0,
\\\label{Eq:dotB}
   &&
\dot{B}+ B T+2  E \xi =0,
\\
&&
\dot{\mathcal{E}}-3  \mathcal{B} \xi +\tfrac{1}{2}
   (B^2 +  E^2) T+\tfrac{3}{2}  \mathcal{E} T+\frac{2}{3}  E j+\tfrac{1}{6} (p+\rho)( Q-T)=0,
\\
&&\label{Eq:bcaldot}
\dot{\mathcal{B}}+\tfrac{3}{2}  \mathcal{B} T+ 3  \mathcal{E} \xi +(B^2 + E^2) \xi =0,
\\
&&\label{Eq:dotphi}
\dot{\Phi}-\la T-2 \xi  \Omega +\tfrac{1}{2} T \Phi=0,
\\
\label{Eq:xi_Ev}
&&
 \dot{\xi} - \la \Omega +\tfrac{1}{2}\left(\xi  T+\Phi
\Omega- \mathcal{B}\right) =0,
\\
&&\label{eq:Omega}
\dot{\Omega}-\la \xi +T \Omega=0.
\end{eqnarray}
\end{subequations}
Moreover,  from equation \eqref{Eq:vincolo} one finds that
\(
\mathcal{B} -3 \la \xi + (2 \la  - \Phi) \Omega = 0,
\)
while from equations \eqref{Ec:thetaaSigma} and \eqref{Ec:thetaaA}
one obtains that
\begin{subequations}
\begin{eqnarray}\label{Eq:T_Ev}
&& \dot{T}-\la \Phi - \mathcal{E}+p+\tfrac{1}{3}\rho+\frac{1}{2}T^2-2 \Omega ^2=0,
\\
\label{Eq:Q_Ev}
&& \dot{Q}-2\hat{\la}-2 \la^2+ B^2+E^2+2  \mathcal{E}+
p+\frac{1}{3}\rho +\frac{1}{2}Q^2=0, \\
&& \frac{4 \dot{\la}}{v_s}  - 2 \hat{Q} + \frac{4 E \hat{j}}{p + \rho}-  \frac{4 E \dot{\varrho}_C}{v_s (p +
\rho)}- \frac{2}{v_s(p+\rho)}\bigg(\la \big[
Q (v_s-1) + 2 T v_s\big] (p + \rho) -2\la  E j (1+ 2 v_s)
\nonumber \\
&& \hspace{2cm} +
\varrho_C\big[E Q  (2+ v_s) + 2(E T v_s + 2 B \xi)\big] +  v_s
(p+\rho)\big[\Phi (Q -T)  + 4 \xi \Omega\big]\nonumber \\
&& \hspace{2cm} - 2 j
\big[\varrho_C (1+ v_s)+v_s(2 B
\Omega-  E  \Phi)\big]\bigg)=0. \label{Eq:A_Ev}
\end{eqnarray}
\end{subequations}
As already  discussed  in Sections \ref{Sec:on-the-thermod} and
\ref{Sec:varandeq}) the fields  $p$ and $\varrho_C$ can be regarded as
functions of $\rho$, and while $j_c$ is regarded as a function of the
electric field  $E$. Equations \eqref{Eq:last-rhodot}-\eqref{Eq:Q_Ev}
do not contain explicit dependence on  the derivative of the matter
density along the radial direction.  In equation \eqref{Eq:A_Ev} the
equation of state and the constraint equation have
been used to obtain a simpler expression. The equation for the radial
acceleration can be recovered after some further calculations ---see
the Appendix Section. In order to simplify the stability analysis of
the subsequent Sections and to be able to extract some information
from the rather complicated perturbed equations we will focus on the
null radial acceleration case.
Summarizing, our analysis has lead to a   symmetric hyperbolic system
consisting of  eleven evolution equations for  eleven  scalar variables
of the form
 \begin{equation}
\label{Eq:symm_hyp_form}
\mathbf{A}^t\partial_t \textbf{{v}}-\mathbf{A}^j\partial_j \textbf{{v}}=\mathbf{B} \mathbf{v},
\end{equation}
\section{Perturbations}\label{Sec:Perturbation}
We will perform a perturbation to first order of the
variables $\mathbf{v}$ so that $\mathbf{v}\mapsto
\mathring{\mathbf{v}}+ \epsilon \breve{\mathbf{v}}$ with the parameter
$\epsilon$ setting the perturbation order that controls the size of
perturbation and $\breve{\mathbf{v}}$ describing the
perturbation of the background solution $\mathring{\mathbf{v}}$. Thus,
assuming that the background variables $\mathring{\mathbf{v}}$ satisfy
the evolution equations, and defining
$(\mbox{d}\mathbf{v}^{(a)}/\mbox{d}\epsilon)_{\epsilon=0}\equiv
\breve{\mathbf{v}}^{(a)}$, we can obtain from equations
\eqref{Eq:last-rhodot}-\eqref{eq:Omega} the following equations for the
linear perturbations:
\begin{eqnarray}
\nonumber
&& \dot{\breve{\rho }}-\bigg(\breve{j} \mathring{E}+\breve{E}
   \mathring{j}-\frac{1}{2} \big[(\breve{p}+\breve{\rho})(\mathring{Q}+2
       \mathring{T})+(\breve{Q}+2 \breve{T})
     \left(\mathring{p}+\mathring{\rho }\right)\big]\bigg)=0, \\\nonumber
 && \dot{\breve{E}}+\breve{j}-2 \breve{\xi } \mathring{B}+\breve{T}
 \mathring{E}+\breve{E} \mathring{T}-2 \breve{B} \mathring{\xi}=0, \\\nonumber
&& \dot{\breve{B}}+\breve{T} \mathring{B}+2 \breve{\xi } \mathring{E}+\breve{B} \mathring{T}+2 \breve{E} \mathring{\xi}=0,\\\nonumber
&& \dot{\breve{\mathcal{E}}}-
3\breve{\xi}
\mathring{\mathcal{B}}+\tfrac{1}{6} \breve{T} \left[3 (\mathring{B}^2+\mathring{E}^2+3  \mathring{\mathcal{E}})-
(\mathring{p}+\mathring{\rho})\right]
+\tfrac{2}{3} \breve{j}
\mathring{E}+ \tfrac{1}{3}\breve{E} (2\mathring{j}+ 3 \mathring{E}
\mathring{T})+\tfrac{1}{6}\breve{Q} (
\mathring{p}+\mathring{\rho})+\tfrac{1}{6}(\breve{p} +\breve{\rho })
(\mathring{Q}- \mathring{T})\\
\nonumber
&& \hspace{3cm}+\tfrac{3}{2} \breve{\mathcal{E}} \mathring{T} -3
\breve{{\mathcal{B}}} \mathring{\xi }+ \breve{B} \mathring{B}
\mathring{T}=0,\\\nonumber
&& \dot{\breve{\mathcal{B}}}+\frac{3}{2} \breve{T}
\mathring{\mathcal{B}}+\breve{\xi }( \mathring{B}^2+3
\mathring{\mathcal{E}})+\tfrac{3}{2} \breve{\mathcal{B}}
\mathring{T}+3 \breve{\mathcal{E}} \mathring{\xi }+2 \breve{B}
\mathring{B} \mathring{\xi }=0,\\\nonumber
&& \dot{\breve{\Phi }}- \breve{T}(\mathring{\la}-
\tfrac{1}{2}\mathring{\Phi})- \breve{\la}
\mathring{T}+\tfrac{1}{2}\breve{\Phi } \mathring{T}-2 \breve{\Omega }
\mathring{\xi }-2 \breve{\xi } \mathring{\Omega }=0,\\\nonumber
&& \dot{\breve{\xi }}-\tfrac{1}{2}\breve{\mathcal{B}} - \breve{\Omega}
( \mathring{\la}-\tfrac{1}{2}\mathring{\Phi})+\tfrac{1}{2}\breve{\xi}
\mathring{T}+\tfrac{1}{2}\breve{T} \mathring{\xi}- \breve{\la}
\mathring{\Omega }+\tfrac{1}{2}\breve{\Phi } \mathring{\Omega}=0,\\
&& \dot{\breve{\Omega }}-\breve{\xi } \mathring{\la}+\breve{\Omega } \mathring{T}-\breve{\la} \mathring{\xi }+\breve{T} \mathring{\Omega }=0.
\end{eqnarray}
In addition one obtains from \eqref{Eq:vincolo} the
linearised constraint
\(
\breve{\mathcal{B}}  +\breve{ \la}(2 \mathring{\Omega}-3 \mathring{\xi}) -3\mathring{\la} \breve{\xi}   - \breve{\Phi} \mathring{\Omega} +  \breve{\Omega} (2 \mathring{\la}  - \mathring{\Phi})= 0,
\)
while perturbing to first order equations
\eqref{Eq:T_Ev}-\eqref{Eq:Q_Ev}) describing the evolution of the variables $T$ and $Q$ we
obtain
\begin{eqnarray*}
&&
\dot{\breve{T}}-\breve{\mathcal{E}}+\breve{p}+\tfrac{1}{3}\breve{\rho
}-\breve{\Phi } \mathring{\la}+\breve{T} \mathring{T}-\breve{\la}
\mathring{\Phi }-4 \breve{\Omega } \mathring{\Omega }=0,\quad \dot{\breve{Q}}-2
\breve{{\la}}+
2 \breve{\mathcal{E}}+\breve{p}+\tfrac{1}{3}\breve{\rho }-4 \breve{\la} \mathring{\la}+2 \breve{B} \mathring{B}+2 \breve{E} \mathring{E}+\breve{Q} \mathring{Q}=0.
\end{eqnarray*}
Similar calculations render a complicated linearised equation for the
perturbation of acceleration $\la$. Thus, in order to undertake a
stability analysis we make some assumptions about the configuration so
as to obtain a simplified system. First, we will analyse the set
\eqref{Eq:last-rhodot}-\eqref{Eq:Q_Ev})
 with the assumption $\la=0$. Thus, we do not consider any more
 equation \eqref{Eq:A_Ev}. Accordingly, we consider the simplified
 vector-valued unknown
\(
\mathbf{v}\equiv ({\rho },{E},{B},{\mathcal{E}},{\mathcal{B}},{Q},{T},{\xi },{\Phi },{\Omega }).
\)
\subsection{Remarks on the set of equations}\label{Sec:Remarks}
We consider the  set of linearised evolution equations in the form

\begin{equation}
\label{Eq:Lin_Per_Sys}
\mathring{\mathbf{A}^t}\partial_t \breve{\textbf{{v}}}-\mathring{\mathbf{A}^j}\partial_j \breve{\textbf{{v}}}=\mathring{\mathbf{B}} \breve{\mathbf{v}}.
\end{equation}
General theory of linear of partial differential equations shows that
systems of this form can either converge (in an oscillating manner or exponentially)
to constant values, have an asymptotic exponential (oscillating) decay
or be unstable ---see e.g. \cite{Acta98}. In the rest of this article, we look at  solutions in
which the perturbations are stabilised on constant values or decay. In order to study
the asymptotical behaviour of the characteristic solutions of perturbed system
it is  necessary the study of the eigenvalues of the matrix
$\mathring{\mathbf{B}}$ as $t\rightarrow\infty$. Notice that in the
the matrix-valued function $\mathring{\mathbf{B}} $ are, in general,
functions of the coordinates. Under these circumstances, the core of
the stability analysis consists on analysing whether the matrix
$\mathring{\mathbf{B}}$ satisfies some appropriate \emph{relaxed
stability eigenvalue conditions}. More precisely, we study then the
eigenvalue problem of $\mathring{\mathbf{B}}$ by looking at the sign
of the real parts of the corresponding eigenvalues $\lambda_i$ (or
\emph{modes of the perturbed system}). A similar approach has been
used in \cite{Reu99} ---see also discussion in
\cite{AlhMenVal10}. General theory concerning the case where the
coefficients linearised system are constant is discussed in
\cite{Acta98}. The case of systems with vanishing eigenvalues has been
discussed in \cite{1997funct.an..3003K,JMP1,JMP2} while for a general
discussion on the time dependent case and the case of non constant
matrix coefficient see \cite{Reu99}.

\medskip
Due to the form of the matrix $\mathring{\mathbf{B}}$ associated to
the present problem,  a fully analysis of the stability properties of
this system is  extremely cumbersome. We proceed by analysing the sign
of the eigenvalues using the fact that a sufficient condition for the
\emph{instability} of the system is the existence of al last one
positive-real part  eigenvalues.  For a more extended discussion concerning the
requirements of the
so-called of the relaxed stability eigenvalue condition see
\cite{1997funct.an..3003K,JMP2}. This simple requirement will provide
a immediate way to show the conditions where  instability occurs
---namely, if there is at least one $i$ such that
$Re(\lambda_i)>0$. Crucially, in the discussion of  the
stability one does not need explicitly to compute the eigenvalues of
$\mathring{\mathbf{B}}$ ---the sign of the real part of
  eigenvalues can be determined for examination of the structure of
  the matrix $\mathring{\mathbf{B}}$. Given the \emph{characteristic
    polynomial}
  ${{\mathfrak{P}}}(\mathring{\mathbf{B}})(x)=\sum_{i=0}^{n}c_i x^i$
of the matrix $\mathring{\mathbf{B}}$, we will make use of the
following well-known results on the roots sign: the trace of a matrix
is equal to the sum of its eigenvalues, hence if the trace of the
matrix is positive then the system is \emph{unstable}. Now, recall
that the determinant of a matrix is the product of the
eigenvalues. Accordingly,  a \emph{necessary condition} for
\emph{stability} is that $\det\, \mathring{\mathbf{B}}\neq0$ and
$(-1)^ n \det\, \mathring{\mathbf{B}}> 0 $ where $n$ is the number of
distinct eigenvalues. Other stability criteria  based on inspecting the
characteristic polynomial are the \emph{Routh-Hurwitz criterion} and the
\emph{Li\'enard-Chipart theorem} ---see
\cite{AlhMenVal10,AnalPropPol2002,Barnett1971}. Wherever possible, we
will also make use of  the Descartes criterion  to determine the
\emph{maximum} number of positive and negative real roots of the
polynomial ${{\mathfrak{P}}}(\mathring{\mathbf{B}})(x)$   for
$c_i\in\mathbb{R}$. In particularly simple cases one can exploit a
generalisation of the Descartes rule  considering the Routh-Hurwitz
criterion to determine the number of roots with positive and negative
real part of  ${{\mathfrak{P}}}(\mathring{\mathbf{B}})(x)$ by
constructing the associated \emph{Routh matrix}.

\medskip
In the case under consideration the matrix $\mathring{\mathbf{B}}$ has
eleven columns and the coefficients of the characteristic polynomial are, unfortunately,
large expressions without an obvious structure. Therefore, we approach
the analysis of the stability  problem making the following
assumption:
\(
\mathring{j}=\mathring{E} \sigma_J,\qquad \breve{j}=\breve{E}
\sigma_J,\qquad
\mathring{p}=\mathring{\rho } v_s,\qquad \breve{p}=\breve{\rho } v_s.
\)
Thus, we take the square  of the sound velocity $v_s=\nu_s^2$  to be
constant and assume \emph{Ohm's law} $j=E \sigma_J$ with
$\mathring{\sigma}_j\equiv \breve{\sigma_j}$ and restrict our
attention to the case of null radial acceleration ---i. e.  $\la=0$.

Before proceeding to the analysis of particular cases, it is
convenient to  note here some general properties  of the system of
symmetric hyperbolic equations under consideration. The  matrix
$\mathring{\mathbf{B}}$ can be explicitly written as the $10\times
10$-matrix
\begin{equation}
\left(
\begin{smallmatrix}
 -\tfrac{1}{2} (1+v_s) \left(\mathring{Q}+2 \mathring{T}\right) & 2 \sigma_J \mathring{E} & 0 & 0 & 0 & -\tfrac{1}{2} (1+v_s) \mathring{\rho } & -(1+v_s) \mathring{\rho } & 0 & 0 & 0 \\
 0 & -\sigma_J-\mathring{T} & 2 \mathring{\xi } & 0 & 0 & 0 & -\mathring{E} & 2 \mathring{B} & 0 & 0 \\
 0 & -2 \mathring{\xi } & -\mathring{T} & 0 & 0 & 0 & -\mathring{B} & -2 \mathring{E} & 0 & 0 \\
 -\tfrac{1}{6} (1+v_s) \left(\mathring{Q}-\mathring{T}\right) & -\tfrac{1}{3} \mathring{E} \left(4 \sigma_J+3 \mathring{T}\right) & -\mathring{B} \mathring{T} & -\tfrac{3 \mathring{T}}{2} & 3 \mathring{\xi } & -\tfrac{1}{6} (1+v_s) \mathring{\rho } & \tfrac{1}{6} \left[(1+v_s) \mathring{\rho }-3 \left(\mathring{B}^2+3 \mathring{\mathcal{E}}+\mathring{E}^2\right)\right] & 3 \mathring{\mathcal{B}} & 0 & 0 \\
 0 & -2 \mathring{E} \mathring{\xi } & -2 \mathring{B} \mathring{\xi } & -3 \mathring{\xi } & -\tfrac{3 \mathring{T}}{2} & 0 & -\tfrac{3 \mathring{\mathcal{B}}}{2} & -\mathring{B}^2-3 \mathring{\mathcal{E}}-\mathring{E}^2 & 0 & 0 \\
 -\tfrac{1}{3}-v_s & -2 \mathring{E} & -2 \mathring{B} & -2 & 0 & -\mathring{Q} & 0 & 0 & 0 & 0 \\
 -\tfrac{1}{3}-v_s & 0 & 0 & 1 & 0 & 0 & -\mathring{T} & 0 & 0 & 4 \mathring{\Omega } \\
 0 & 0 & 0 & 0 & \tfrac{1}{2} & 0 & -\tfrac{\mathring{\xi }}{2} & -\tfrac{\mathring{T}}{2} & -\tfrac{\mathring{\Omega }}{2} & -\tfrac{\mathring{\Phi }}{2} \\
 0 & 0 & 0 & 0 & 0 & 0 & -\tfrac{\mathring{\Phi }}{2} & 2 \mathring{\Omega } & -\tfrac{\mathring{T}}{2} & 2 \mathring{\xi } \\
 0 & 0 & 0 & 0 & 0 & 0 & -\mathring{\Omega } & 0 & 0 & -\mathring{T}
\end{smallmatrix}
\right).
\label{Matrix:General}
\end{equation}
\section{Discussion on the nonlinear stability of the  symmetric hyperbolic system}\label{Sec:discussion}
In this Section we discuss necessary conditions for the nonlinear
stability of the configuration under consideration. We focus on the
reference solution described by matrix $\mathring{\mathbf{B}}$ in \eqref{Matrix:General}. We present the analysis
of the system stability for some special configurations defined by
proper assumptions on the kinematic variables given by the
$1+1+2$-decomposition. Each case is then identified according to the
assumptions that characterise it ---the fields assumed to vanish are
indicated in parenthesis. We discuss in details the
\emph{instability} condition for the complete list of cases. Assuming
the radial acceleration and other kinematic variables to be zero, we
explore the implications of the assumption on the set up configuration
and its stability.
A first insight into the stability of the system can be inferred from the trace
 in terms of $\Sigma$ and $\Theta$ as
\be
{\rm{Tr}}\,\mathring{\mathbf{B}}=6 \mathring{\Sigma}-(7+v_s)\mathring{ \Theta}  -\sigma_J.
\ee
It then readily follows that ${\rm{Tr}}\,\mathring{\mathbf{B}}>0$ so
that the system is \emph{not  stable} if
\begin{equation}
\label{Eq:first_cond}
 \Sigma>\frac{(7+v_s) \Theta +\sigma_J}{6}.
\end{equation}
In the remainder of this article, we proceed to a more detailed
analysis in which we classify the results according to suitable
assumptions on the remaining system kinematic variables $(T, Q, \Phi,
\xi, \Omega)$. It should be pointed out, for ease of reference, that
the condition $T = 0$ is equivalent to the relation $\frac{2}{3}\Theta
=\Sigma$ between the expansion $\Theta$ of the 3-sheets and the radial
part of the shear of the 3-sheet $\Sigma$. Similarly, the condition
$Q=0$ is equivalent to ${\frac{1}{3}\Theta =-\Sigma}$, while the two
conditions $T=0$ and $Q=0$ imply $\Sigma=0$ and $\Theta=0$. We also
notice from the Maxwell equations that the evolution of the electric
(respectively, magnetic) field is coupled to the magnetic
(respectively, electric) field via the  twisting of the
2-sheet. Hence,  when $\xi= 0 $ the two evolution equations decouple
and evolve with the only common dependence on $T$.
\subsection{
{Preliminaries  on the symmetries of the system and classes of solutions}}
The configuration can expand  or contract along   the direction of symmetry,  according to $\Theta>0$ or $\Theta<0$ respectively,  but not accelerating as it is assumed the parallel acceleration $\mathcal{A}=0$.
The limit  situation of  null radial acceleration  has been adopted   to simplify the analysis of the unstable modes, in this scheme  we  are able  to provide a complete classification  of the  stable and unstable solutions for the Einstein-Maxwell-Euler system,   
   considering appropriate restrictions for the set of   the perturbed equations.
All the kinematic  quantities  and fields are  reduced to scalars by the  projections in the radial direction and on the orthogonal plane by the projector $N_{ab}$, the set of scalars  define  the vector variable $\overrightarrow{\mathbf{v}}$ introduced in Eq.\il(\ref{Eq:vu-vectorvdef}).
Any further restrictions on the system, as the vanishing of other dynamical variables:
\be
Q_i\in \overrightarrow{\nu}, \quad  \overrightarrow{\nu}\equiv \{T, Q, \Phi, \xi, \Omega\}\subset\overrightarrow{\mathbf{v}}
\ee
leads to   particular  different solutions of  the Einstein-Maxwell-Euler equations  with the  new symmetry conditions. The self-gravitating systems   will be especially constrained  by  the couple $(\mathcal{E},\mathcal{B})$, in the gravitating systems  the background geometry is assumed  to be in one of the classes of solutions including the  new symmetries. The vector $\overrightarrow{\nu}$ is  actually a restriction of the vector variable  $\vec{\mathbf{v}}$, where the fluid four-velocity $u_a$   defines the metric in its  $3 +1$  form and supplies the projected components of the Weyl tensor together with the direction of symmetry $(\xi,\Phi)$ related through  Eq.\il(\ref{Eq:trek}),   and the variation of the symmetry direction $n_a$, used  to construct the metric tensor in its  $2+1+1$ form.
The vanishing of the  $Q_i$ elements implies    further  restrictions  involving the annihilation of other quantities, or the constance of  $Q_j\neq Q_i$ during  the evolution along $u_a$. We can then refer to the general conditions $\mathfrak{C}$:
\bea\label{Eq:campi}
\mathfrak{C}\quad (Q_i=0):&&\quad(E^2, B^2,\mathcal{E}, \Omega)\mapsto \left(p+ \frac{\rho}{3}\right)
\\\label{Eq:vac-cin}
&&\qquad
(Q_j,\mathcal{B})=0
\\\label{Eq:further}
&&\qquad\dot{Q_j}=0.
\eea
Conditions in Eqs.\il(\ref{Eq:vac-cin}), specified in Table\il(\ref{Table:asterisco}),  define five principal classes of solutions according to the assumption of null $Q_i\in\overrightarrow{\nu}$.
Table\il(\ref{Table:Fasterisco})  shows the sub-classes  of solutions constructed by  the vanishing of a couple of  scalars $(Q_i, Q_j)$. Conditions (\ref{Eq:campi}) are listed in  Table\il(\ref{Tabel:Eich}), and state the relationship between the remaining field variables of the vector  $\overrightarrow{\mathbf{v}}$ i.e. the couple $(\mathcal{E},\mathcal{B})$, the electromagnetic fields  $(E, B)$,  and the matter density $\rho$.
  Finally, condition (\ref{Eq:further})  is made explicit in Table\il(\ref{Tab:Costanza}).
 Tables  (\ref{Table:asterisco}), (\ref{Table:Fasterisco}),(\ref{Tabel:Eich}) and (\ref{Tab:Costanza}) characterize entirely the system throughout  its sub-configurations ruled by the system  symmetries,
 providing  a complete classification of  the solutions.
 An analysis and a general  discussion of the solutions of the Einstein-Maxwell system in terms of the  magnetic and electric parts of the   Weyl  tensor   can be found in  \cite{KramerSte}.
The stability analysis will be performed on the systems with  $\mathcal{A}=0$, on the five classes  of configurations  and their sub-classes.
The
$\mathfrak{I}$-class with  $(\mathbf{\mathcal{A}T})$  and
$\mathfrak{II}$-class with  $(\mathbf{\mathcal{A}Q})$ are particularly significant.
Systems  $T=0$ are characterized by  $\Theta \Sigma>0$, i.e. the sign of the expansion is concordant with that of the radial shear, then positive for expanding systems, along the direction of symmetry, or negative for  contracting systems, see also Fig.\il(\ref{Fig:PrimaS}). On the other side
systems $Q=0$ correspond to the case $\Theta \Sigma<0$.
The relative sign of the scalars of the radial shear and expansion  is a significant element   affecting   the stability of the system  and the equilibrium configurations: even in the general case where the only assumption on the system  is  the  null radial acceleration $\mathcal{A}=0$,
the balance of the contributions given by the radial shear in the  systems in expansion ($\Theta>0$) or in contraction ($\Theta<0$), is relevant in determining the states  certainly unstable. For the contracting systems, or in expansion  but with  positive shear,  a threshold on the expansion rate $\Theta$ or equivalently the radial shear, appears for  the emergence of the  instability,  this is  a function of  the two model parameters  $(\sigma_J, v_s)$. The magnetic field, although it is not constant in the time along the direction of symmetry, has no specific role to establish  the stability of this  model. It is possible to show  that for the case $\mathbf{(\mathcal{A} T)}$  there is always  a threshold  for the emergence of unstable phases,  while in the case of  the expanding systems with  negative  shear, the difference in sign between the two scalar does not involve any instability threshold.

The conditions for the  equilibrium of these solutions involve quantities $Q_i\in\vec{\nu}$ exclusively related to the
 fluid dynamics
such as  $\mathcal{A} T\xi$ or $\mathcal{A}\xi \Omega$  and imply a serious constraint on the background, for  that instability is in some cases certainly verified assuming that $\mathcal{B}=0$.
Table\il(\ref{Tabel:Eich}) shows  conditions $\mathfrak{C}$ in  Eqs.\il(\ref{Eq:campi}):
only the  electric  part of the Weyl tensor  is determined by the matter fields  and the vorticity.
In general the classes  $\mathbf{(\mathcal{A}T)}$ and $\mathbf{(\mathcal{A}Q)} $ do not implies the vanishing of others variables by conditions $\mathfrak{C}$ in  Eqs.\il(\ref{Eq:vac-cin}):  that is, the systems do not require any additional symmetry as  the initial data $Q_j=0$ for the solutions (\ref{Eq:campi}).
Remarkably the $\mathbf{(\mathcal{A}Q T)}$ class corresponds to the case of null  shear and  null radial expansion.
Tables\il(\ref{Table:asterisco}) and (\ref{Table:Fasterisco}) show that the   couples   $(\xi,\Omega)$ and  $(\xi, \Phi)$ are related. The magnetic field must be constant along the fluid flow  for the solution  $(\mathcal{A},T, \xi)$.
and  conversely  $\mathcal{B}=0$  where  $\xi $ is zero.
The solutions $T=0$ have  radial vorticity
$\Omega$  constant during the evolution of the system as shown  in   Table (\ref{Tab:Costanza}). If  the radial vorticity    is initially zero, then   the expansion $ \Phi $  of the 2-sheet generated by $n_a$  is constant along $u_a$ and
the electric part of the Weyl tensor  is entirely determined by the matter field $\rho$.
If  initially $\Phi=0$ then $\xi=0$  and  therefore  $\mathcal{B}=0$,  providing finally a no vacuum solution  with null magnetic component of the Weyl tensor,  Table\il(\ref{Tabel:Eich}).

\begin{table}[hbpt]
\begin{tabular}{|l|l|r|c|c|}
\hline
& $\mathcal{A}$ &$\mathcal{A}T\quad\qquad$&$\mathcal{A}Q$&$\mathcal{A}\xi$
\\
\hline \hline
$\mathfrak{I}$-class&$\mathcal{A} T$&$\checkmark\quad\qquad$&$\checkmark$&$\checkmark$
\\\\
$\mathfrak{II}$-class&$\mathcal{A} Q$&$ \mathcal{A} T Q\quad\qquad$&$\checkmark$&$\checkmark$
\\\\
$\mathfrak{III}$-class&$\mathcal{A}\Phi=\mathcal{A}\Phi\Omega$&$\mathcal{A}T \Phi=T+\mathcal{A}\Phi\Omega$&$\mathcal{A}Q \Phi=Q+\mathcal{A}\Phi\Omega$&$\checkmark$\\
&\quad\qquad$\mathcal{A}\Phi\xi\mathcal{B}$&$\quad\qquad T+\mathcal{A}\Phi\xi\mathcal{B}$&\quad\qquad $Q+\mathcal{A}\Phi\xi\mathcal{B}$&
\\
\\
$\mathfrak{IV}$-class&$\mathcal{A} \xi$&$\mathcal{A}T \xi =T+\mathcal{A}\Phi\xi\mathcal{B}$&$\mathcal{A}Q\xi$&$\checkmark$
\\\\
$\mathfrak{V}$-class&$\mathcal{A} \Omega$&$\mathcal{A} T\Omega$&$\mathcal{A} Q\Omega$&$\mathcal{A}\xi \Omega=\mathcal{A}\xi\Omega\mathcal{B}$
\\
\hline
\end{tabular}
\centering
\caption{\footnotesize{Null radial  acceleration $\mathcal{A}$: the five principal  classes of solutions for the  linear stability the problem, identified by canceling the radial acceleration and a scalar quantity of the set $Q_i\in\vec{\nu}=\{T, Q, \Phi, \xi, \Omega\}$  and  $ \mathcal{B}$, the magnetic part of the Weyl tensor. The elements of the Table explicit conditions  $\mathfrak{C}$ in (\ref{Eq:vac-cin}).
Thus, for example the   $\mathfrak{III-}$class, conditions of   $\mathbf{(\mathcal{A} \Phi)}$ configurations  implies
 a null radial vorticity $\Omega$ or the null couple $\xi$ and $\mathcal{B}$. Then   we assumed that the system has a further constraint represented by a third  zero quantity $Q_i\in \vec{\nu}$ defining the  subclasses.
  The check marks indicate the cases already treated, the Table exhausts the five  classes  $\mathbf{(\mathcal{A},Q_i)}$ and the subclasses  ${(\mathcal{A},Q_i,Q_j)}$. Particularly the $\mathfrak{III-}$class $(\mathcal{A}\phi)$
shows the  symmetry between the solutions at zeros $\mathbf{(\mathcal{A}Q)}$, $\mathbf{(\mathcal{A}T)}$ and  $
\{\xi \Omega \mathcal{B}\}$. Subclasses,  ${(\mathcal{A},Q_i,Q_j,Q_k)}$ are  in Table.\il(\ref{Table:Fasterisco})}.}\label{Table:asterisco}
\end{table}
\begin{table}[hbpt]
\begin{tabular}{|l|r|c|c|l|}
\hline
 $\mathcal{A} T Q$ &$\mathcal{A}T\xi\quad\qquad$&$\mathcal{A}Q\Phi$&$\mathcal{A}\Phi\xi$
\\
\hline \hline
$\mathcal{A} T Q \Phi= Q+T+ \mathcal{A}\Phi\xi\mathcal{B}$&$\checkmark\quad\qquad$&$\checkmark$&$\checkmark$
\\
$\qquad\qquad Q+T+ \mathcal{A}\Phi\Omega$&$\checkmark\quad\qquad$&$\checkmark$&$\checkmark$
\\
$\mathcal{A} T Q \xi$&$ \mathcal{A} T Q\quad\qquad$&$\checkmark$&$\checkmark$
\\\\
$\mathcal{A}TQ\Omega$&$\mathcal{A}T \xi \Omega=T+\mathcal{A}\xi \Omega \mathcal{B}$&$\mathcal{A}Q\xi \Omega =Q+A\xi \Omega\mathcal{B}$&$\mathcal{A}\Phi \xi \Omega=\Phi+\xi\Omega\mathcal{B}$
\\
\hline
\\
$\mathcal{A}T Q \xi \Omega=T+Q+\mathcal{A}\xi \Omega \mathcal{B}$&&&
\\
\hline
\end{tabular}
\centering
\caption{\footnotesize{Subclasses  ${(\mathcal{A},Q_i,Q_j)}$ of  the five principal classes $\mathfrak{I}-\mathfrak{V}$  in Table\il(\ref{Table:asterisco}). The Table  highlights the role of the zeros couples   $(TQ)$ and
 and $\xi \Omega \mathcal{B}$.}}\label{Table:Fasterisco}
\end{table}
\begin{table}[hbpt]
\begin{tabular}{|l|c|}
\hline
  $\mathfrak{C}$&
  \\\hline
 $ \mathcal{C}(\la, T)$:&$\mathcal{E}+2 \Omega^2=p+\frac{\rho }{3}>0$
  \\
 $ \mathcal{C}(\la, Q)$:& $E^2+B^2+2\mathcal{E}=-(p+\rho/3)<0$
  \\
$  \mathcal{C}(\la,\xi)$:&$\mathcal{B}=\Phi \Omega$
  \\
$  \mathcal{C}(\la,T,\Omega)$:&$\mathcal{E}=p+\rho/3>0$
  \\
 $ \mathcal{C}(\la\, \xi,\Omega)$:&$\mathcal{B}=0$
 \\\hline
\end{tabular}
\centering
  \caption{Conditions $\mathfrak{C}$ as in Eq.\il(\ref{Eq:campi})  and Eq.\il(\ref{Eq:vac-cin}).}\label{Tabel:Eich}
\end{table}
\begin{table}[hbpt]
\begin{tabular}{|l|c|}
\hline
  \textbf{Class}&
  \\\hline
 $ (\la, T)$:&$\dot{\Omega}=0$
  \\
 $ (\la, T, \xi)$:& $\dot{\Omega}=\dot{\Phi}=\dot{\mathcal{B}}=\dot{B}=0$
  \\
$(\la,T,\Omega)$:&$\dot{\Phi}=0$
  \\\hline
\end{tabular}
\centering
  \caption{Condition $\mathfrak{C}$ in Eq.\il(\ref{Eq:further}). Classes of solutions and  null evolution  of the kinematic quantities.}\label{Tab:Costanza}
\end{table}
\subsection{{Analysis of the system stability, general conditions and results}}\label{Sec:submain}
The symmetries of the system  and the adapted $1+1+2$ formalism   highlight the emergence of states certainly unstable  for the linear (and non linear) perturbation. The  unstable  phases of the system  with $\mathcal{A}=0$ are  primarily     regulated by  the expansion (or the  contraction) and the shear in the radial direction. These scalars are   related  by the sound velocity   and the conductivity through a relation  $\mathfrak{F}$:
\bea\label{Eq:Rus-mil-eas}
\mathfrak{F}:&&\quad(\Sigma,\Theta)\mapsto (v_s, \sigma_J, \rho),
\\\label{Eq:density-on-afa}
\qquad&&\quad\rho\mapsto (v_s, \sigma_J).
\eea
Conditions $\mathcal{C}$  in Eqs.\il(\ref{Eq:campi},\ref{Eq:vac-cin},\ref{Eq:further}) describe  the system symmetries   but not its  stability, conversely the  relations $\mathfrak{F}$  in Eqs.\il(\ref{Eq:Rus-mil-eas}) relate  the only kinematic variables of the radial shear and radial expansion,  $\Sigma$ and  $\Theta$ respectively, to  the constants from the  state and constitutive equations: the velocity of sound $v_s $  and the conductivity  $\sigma_J$. We emphasize that   Eq.\il(\ref{Eq:Rus-mil-eas})  does not involves other dynamic variables such as  the Weyl scalar or the electromagnetic contribution.
Analogously  for  the five  classes of solutions and their subclasses    according to Eqs.\il(\ref{Eq:campi},\ref{Eq:vac-cin},\ref{Eq:further})
 the relationship in  (\ref{Eq:density-on-afa}) provides  an upper bound on the matter density as  a limiting value  determined by the couple $(v_s, \sigma_J)$.   The density, but not to its gradient, is  in fact a variable in the system and, since this is and iso-entropic  and barotropic  fluid,  one can use this to get information on the hydrostatic pressure the system is subjected to.
\subsubsection{Discussion on the stability of the system with zero acceleration}
\begin{figure}
\centering
\begin{tabular}{cc}
\includegraphics[scale=.5]{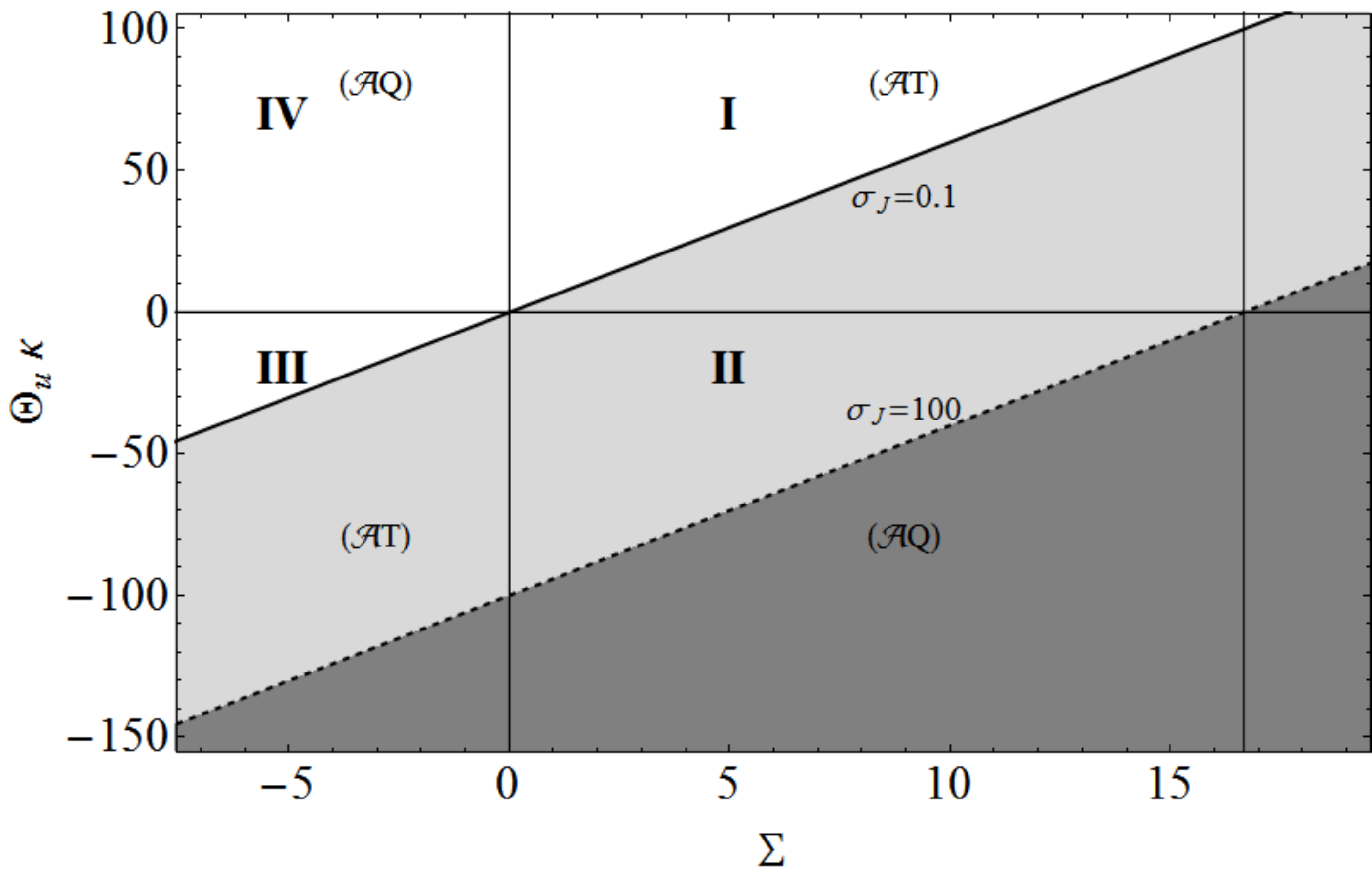}
\end{tabular}
\caption[font={footnotesize,it}]{\footnotesize{Plot of the limiting expansion  $\Theta_u \kappa$, defined in Eq.\il(\ref{Eq:thetau}) as function of the shear in the preferred radial direction where $\kappa\equiv v_s+7$. per $\sigma_J=0.1$ (black line ) e $\sigma_J=100$ (dashed black line )
Possibly an identical plot  for the expansion $\Theta_u$ in terms of  $\sigma_J/\kappa$
and $\Sigma\kappa$. The instability regions, for  $\Theta<\Theta_u$, are colored:
then for   $\sigma_J=0.1$ the gray  region, and light gray region determines the
instability for configurations with conductivity parameter  $\sigma_J=100$, gray region is then stable for this system. In the white region the system can be stable. Configurations $\mathbf{(\mathcal{A}T)}$ are in the  \textbf{I} and \textbf{III} quadrants ($\Sigma \Theta>0$) as in the $T=0$ class of solution , for $\Sigma=(2/3) \Theta$,  while the $\mathbf{(\mathcal{A}, Q)}$ class for $\Sigma=-(1/3)\Theta$  belongs to the \textbf{II} and \textbf{IV} quadrants ($\Sigma \Theta<0$).
The stability of a   system expanding along the direction of symmetry (but not accelerating, as it is  $\mathcal{A}=0$) is regulated as in the  $\mathbf{I}$ and $\mathbf{IV}$ quadrants. The contraction in  $\mathbf{II}$ and $\mathbf{III}$  quadrants, where the  radial shear  is positive and negative, respectively.
The classes are summarized in Table\il(\ref{Table:asterisco}). The contraction limit vanishes, and the stability limit is only on the radial expansion for  high enough shear according to $\Sigma= \sigma_J/6$.
}}\label{Fig:PrimaS}
\end{figure}

\begin{figure}
\centering
\begin{tabular}{cc}
\includegraphics[scale=.5]{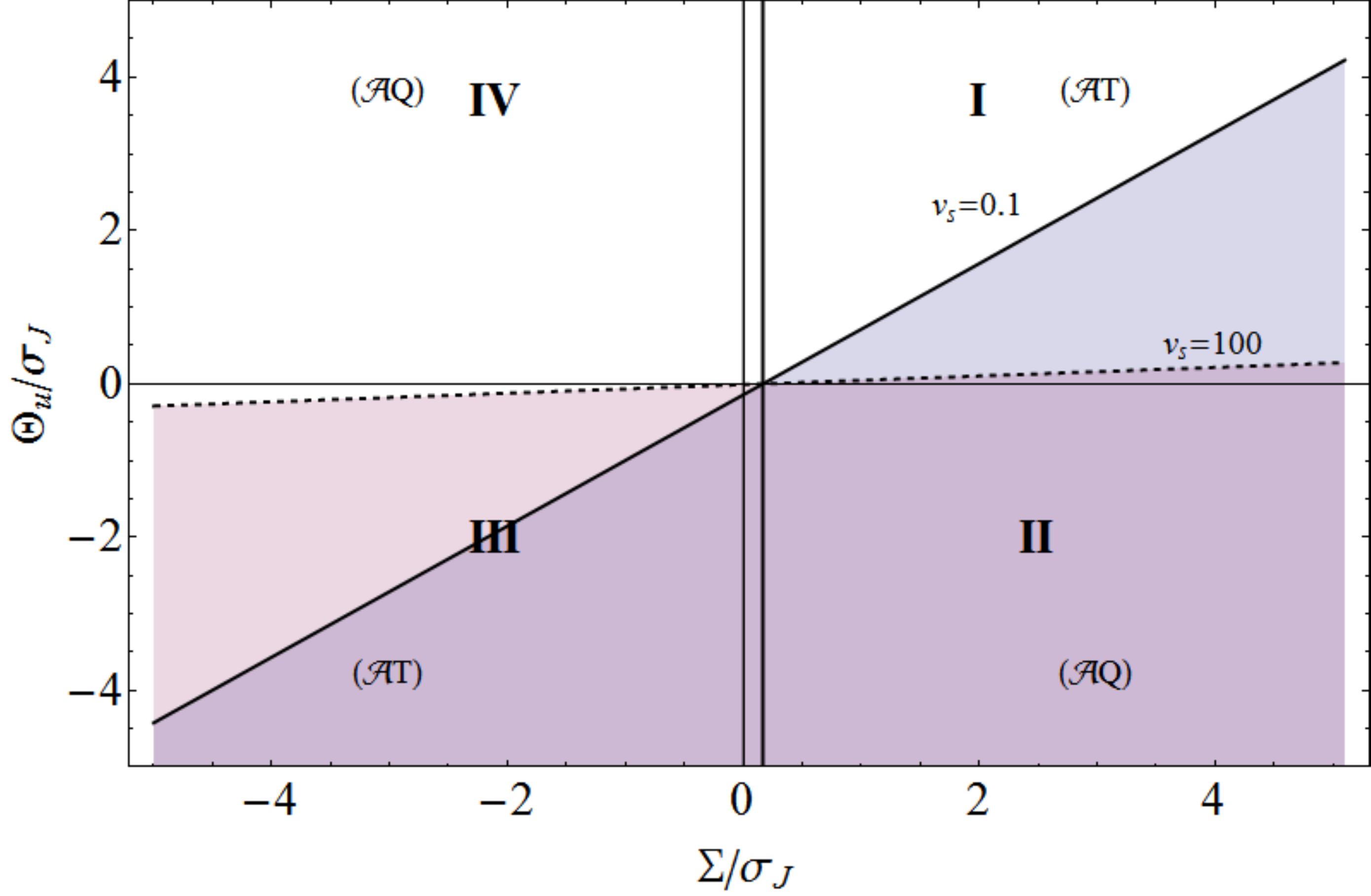}
\end{tabular}
\caption[font={footnotesize,it}]{\footnotesize{Plot of the limiting expansion  $\Theta_u /\sigma_J$,  in Eq.\il(\ref{Eq:thetau}), versus the radial shear $\Sigma /\sigma_J$  for $v_s=0.01$ (black line)  and  $v_s=100$ (dashed black line). The  Shaded regions marke the sections in the $\Theta-\Sigma$ plane where instability occurs as $\Theta<\Theta_u$. It is $\Theta_u=0$ for
$\Sigma/\sigma_J= 1/6$.
}}\label{Fig:P01min}
\end{figure}

A   relevant  constraint $\mathfrak{F}$, for the stability of the   system with  null radial  acceleration, is provided by condition  (\ref{Eq:first_cond}) on the trace ${\rm{Tr}}\,\mathring{\mathbf{B}}$,
and it  can be expressed  in terms of the expansion along the privileged direction as follows
\be\label{Eq:thetau}
\mathbf{(u)_0:}\quad
\Theta <\Theta_u\equiv \frac{6 \Sigma -\sigma_J }{\kappa },\quad\kappa\equiv 7+v_s>1,
\ee
this is  a first condition for the instability of the system, providing an upper bound, a threshold for the occurrence of the instability,  on  the expansion or contraction with respect to the shear projected along the direction of symmetry
i.e., the system is bound to be unstable if the expansion  is, algebraically, smaller  then  $\Theta_u$, function  of the radial shear,  the sound velocity and the conductivity parameter.
For  small or zero conductivity  the limit $\Theta_u$ is   a fraction of the positive shear, where  the system cannot contract or an  instability occurs, for negative  shear  the contraction is limited by the value $\Theta_u<0$, for contractions too fast i.e.  $\Theta\in]\Theta_u,0[$  the system is unstable according to Eq.\il(\ref{Eq:thetau}).
We can  summarize the analysis in the  as following points:
\begin{description}
\item[Systems in contraction  $\Theta<0$ and in expansion $\Theta>0$]
The \emph{expanding} configurations, described  the $\mathbf{I}$ and $\mathbf{\mathbf{IV}}$ quadrant of Fig.\il(\ref{Fig:PrimaS}), are  favored in their  stable phases  with respect to the systems in contraction:
the preference for the stable expansion is especially evident  for negative shear, as in  $\mathbf{IV}$ quadrant which describes also a part of the $(\mathcal{A},Q)$ systems. In the fourth quadrant the stability  conditions
are always guaranteed within the condition (\ref{Eq:thetau})
 as it is  $\inf{\sigma_J}>0$,
the condition of the trace  in Eq.\il(\ref{Eq:thetau}) is not sufficient  to establish the emergence of the instability for this system.
Contrasting verses of the scalars  $\Theta \Sigma<0$  appear to favor  the equilibrium,  this seems to be confirmed by the situation in the \textbf{I} quadrant, with positive shear, describing parts of the  $(\mathcal{A},T)$ systems.
 For slow    expansions $\dot{\Theta}\thickapprox0$  and small conductivity,  where the system is unstable,  a  threshold for the instability  appears  depending  on the conductivity and can be  easily seen by  Eq.\il(\ref{Eq:thetau}). Further restrictions on the symmetries of the system, in the subclasses of Table\il(\ref{Table:asterisco},\ref{Table:Fasterisco})  will turn in  a deformation of Eq.\il(\ref{Eq:thetau}) as shown by Eq.\il(\ref{Eq:gener-quel-e}).
An  increase of   the  shear corresponds in these cases to an increase of the instability regions in  the plane $\Theta-\Sigma$ in Fig.\il(\ref{Fig:PrimaS}). Considering the instability for  the contracting systems,  the \textbf{II} and \textbf{III} quadrants, we can see that
the threshold for  the emergence of the instability  $\Theta_u$  decreases in general  for positive shear  up to the limiting value $\Sigma_0\equiv\sigma_J v_s$, the contraction  is canceled and  for higher shear  the system begins an  expanding phase,
then the system will necessarily be unstable and the zone of stability will be limited to  positive shear in the range  $\Sigma\in]0,\Sigma_0[$ and  contraction  $\Theta\in]\Theta_0,0[$, where $\Theta_0\equiv-\sigma_J/\kappa<0$ is the threshold  for the stability  of the systems in contraction  with non-zero conductivity: for  faster contractions    $\Theta<\Theta_0$, the system with null shear is unstable. This region, however, increases with increasing conductivity, this means that the conductivity acts to stabilize  the contracting configurations   with  positive shear.
Systems $(\mathcal{A},Q)$ are examples of  configurations in  the    \textbf{III} quadrant: the  systems are stable   for high conductivity, negative shear and   sufficiently  low  contractions,  therefore the expanding  $\mathbf{(\mathcal{A} Q)}$ systems
are favored for the  stability.
In general for systems in contraction,  in the  \textbf{II} and \textbf{III} quadrants, a high conductivity is required
   for   stable systems
of    $\mathbf{(\mathcal{A} T)}$  and  $\mathbf{(\mathcal{A} Q)}$ classes respectively.
\item[The role of the radial shear]
In the    \textbf{I} and \textbf{II} quadrants the radial shear is  positive,  an increases of $\Sigma$ generally acts  to favor the instability of the configuration. An increase in magnitude of the  negative shear  in  the \textbf{III}  quadrant ($\Theta \Sigma>0$), for  contracting systems  tends to favor the system stability for fast contractions.
\item[The role of the conductivity]
 In general  an increase of the conductivity parameter particularly in  the \textbf{II} and \textbf{III} quadrants, has a stabilizing effect on the system according to the law  in Eq.\il(\ref{Eq:thetau}). In the \textbf{III} quadrant, for example  for systems of the $\mathfrak{I}-$class or  $\mathbf{(\mathcal{A} T)}$ configurations,
where the radial shear  sign is equal to  the contraction one, the stability region increases for low contractions and also for small conductivity as the shear is high enough in  magnitude; ultimately for high conductivity the system can remain stable even for high radial contractions.
 The expansion phase is stable for negative shear, whereas at low expansion the system is  unstable. The minimum threshold for the emergency of the  instability increases with the shear but decreases with the conductivity, which has then  a stabilizing effect on the system, for very high conductivity. The   stable regions  in the $\Theta-\Sigma$ plane increase by increasing the shear and therefore the stability is  advantaged for high conductivity, for the expansion phase and for shear negative zero or small conductive.
In particular the solutions of the $\mathfrak{II}$-class  $\mathbf{(\mathcal{A} Q)}$ are  a  wider class of  solutions admitting difference possible   symmetries  according to Table\il(\ref{Table:asterisco}),  stable for any conductivity at  negative shear,
while for expanding systems as in the \textbf{II} quadrants the instability  regions are maximum  for smaller shear and zero for greater shear,
 the measure of this region increases with the  conductivity. However the turning point of the expansion
 $\Theta$ is regulated by the  $\sigma_J /\kappa$, for null radial shear $\Sigma$ it is $\Theta_u=-\sigma_J/\kappa$,
the contraction is maximum  as provided by $\Sigma_x$  and  $\Theta_x$.
\item[The role of the velocity of sound]
Fig.\il(\ref{Fig:P01min}) emphasizes the role of the sound velocity in the determination of the unstable states of the system.
The   shear and expansion  scalars  have been normalized for the  finite conductivity $\sigma_J$.
For the systems included  in the \textbf{I}-\textbf{II} and \textbf{III} quadrants,  the velocity of sounds plays a significant role in the determination of the unstable phases, while there is no threshold for the emergence of the instability,  crossing  with  continuity the  \textbf{III} and \textbf{IV} quadrants of the $\Theta-\Sigma$ plane.
In the limit of very large $v_s$  the configuration stops the expansion  in the \textbf{I}  quarter or the contraction  phase in \textbf{III} one i.e.  the   threshold approaches
 $\Theta_u=0$.
For   approximately null velocity of sound  it is
$\Theta_u/\sigma_J\approx-{1}/{7}+(6/7) \Sigma$, then for  sufficiently small $\Sigma$  the scalar expansion changes sign, and the expansion stage cancels the threshold for the  instability, which instead appears for contracting systems. More specifically:  there is a set for positive but small radial  shear
where the  expansion threshold
 disappear, but for those shears the system is certainly unstable for sufficiently fast contractions. Generally the upper limit is small and the contracting system is always  unstable:  the expansion is null as the positive shear is  $\Sigma/\sigma_J=1/6$, then for this configuration, the contracting phases are unstable. The solutions $\Theta=0$  \emph{and} $\Sigma=0$  belongs to $\mathbf{(\mathcal{A}, T, Q)}$ class.
For  $\Sigma/\sigma_J>1/6$  the contracting  system  is always unstable for any couple $(\sigma_J, v_s)$.
Decreasing  the velocity of the sound, the region of $\Theta-\Sigma$ plane, for   expanding systems correspond to unstable  regions for an increase of the  positive shear.
As  the velocity of sound decreases for  positive shear  the instability region increases.
  The instability in the \textbf{I} quadrant is carried out for sufficiently high expansions and shear.
The increase of the speed of sound acts to stabilize the  expanding system for positive shear and destabilize the contracting systems  at $\Sigma/\sigma_J>1/6$,
in the window of  smaller but positive shear,
the increase of the speed of sound acts to destabilize the system while a decrease of this corresponds to an increase the stability provided that the contraction is sufficiently small in magnitude or  $\Theta/\sigma_J>
-1/(v_s+7)$.

We  now focus  on the effects the velocity of sound in the case of negative shear:
for the  expanding system in the \textbf{IV} quadrant, no threshold  exists.
For contracting systems  in the  \textbf{III} quadrant,   an increase  of the speed of sound increases the stability of the system  increasing in magnitude the threshold for very high contractions. A decrease of  the velocity  of sound,  for contracting systems  with negative shear,   induces the system unstable even for very small expansions.
The increase in magnitude of the radial shear increases the system stability.
Ultimately the system's equilibrium   depends on  the velocity of sound
in \textbf{I} and \textbf{III} quadrants, that is for concordant shear and expansion  or in the \textbf{IV} quadrant for systems in contraction with positive shear smaller than $1/6 \sigma_J$. In the limit of  very large velocity of sound $\Theta_u>0$  for  $\Sigma/\sigma_J>1/6$, for $\Sigma/\sigma_J\in]-\infty,1/6[$ it is $\Theta_u<0$.
\end{description}
\subsubsection{Stability of the subclasses  with restricted symmetries}
\begin{figure}
\centering
\begin{tabular}{cc}
\includegraphics[scale=.3]{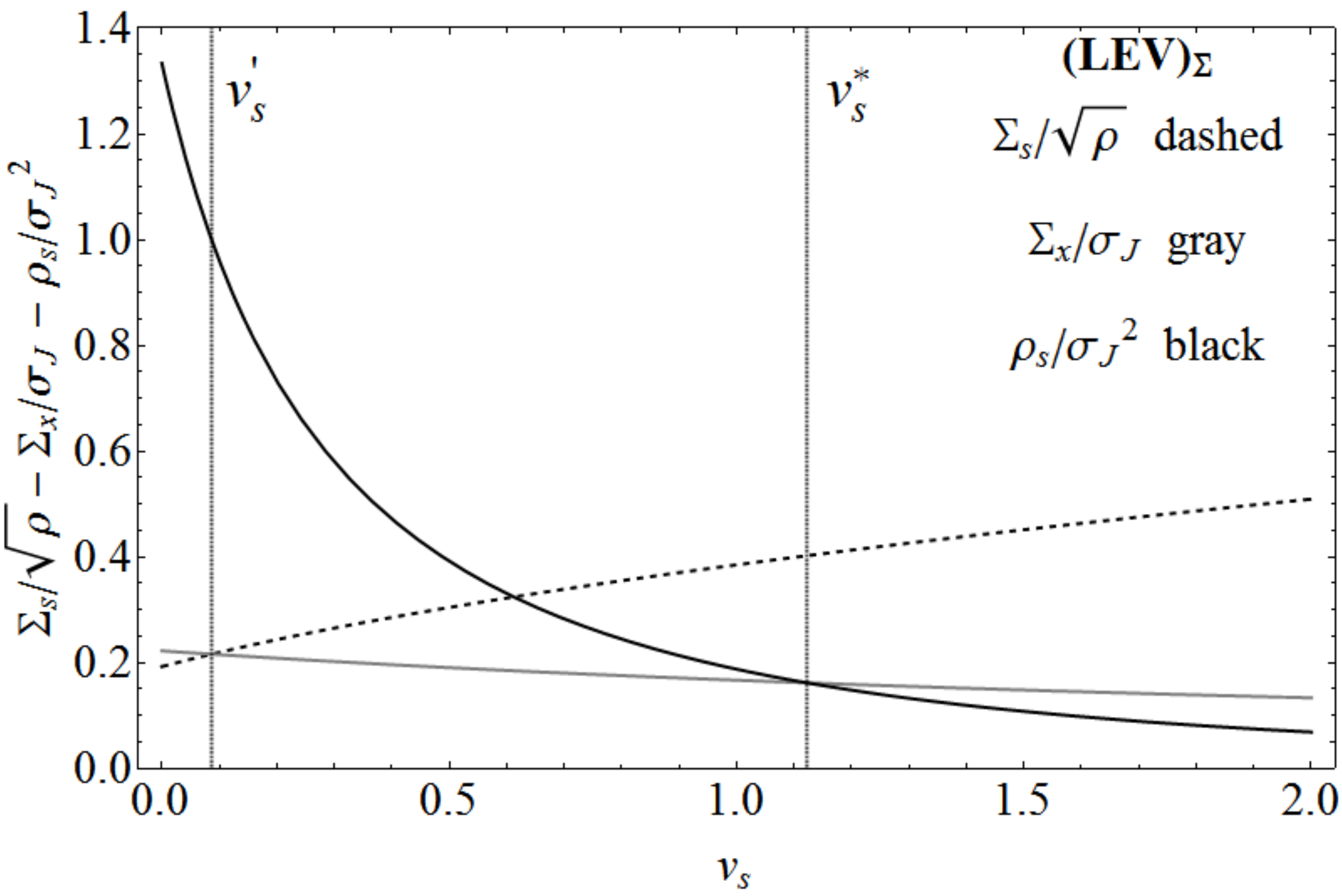}
\includegraphics[scale=.3]{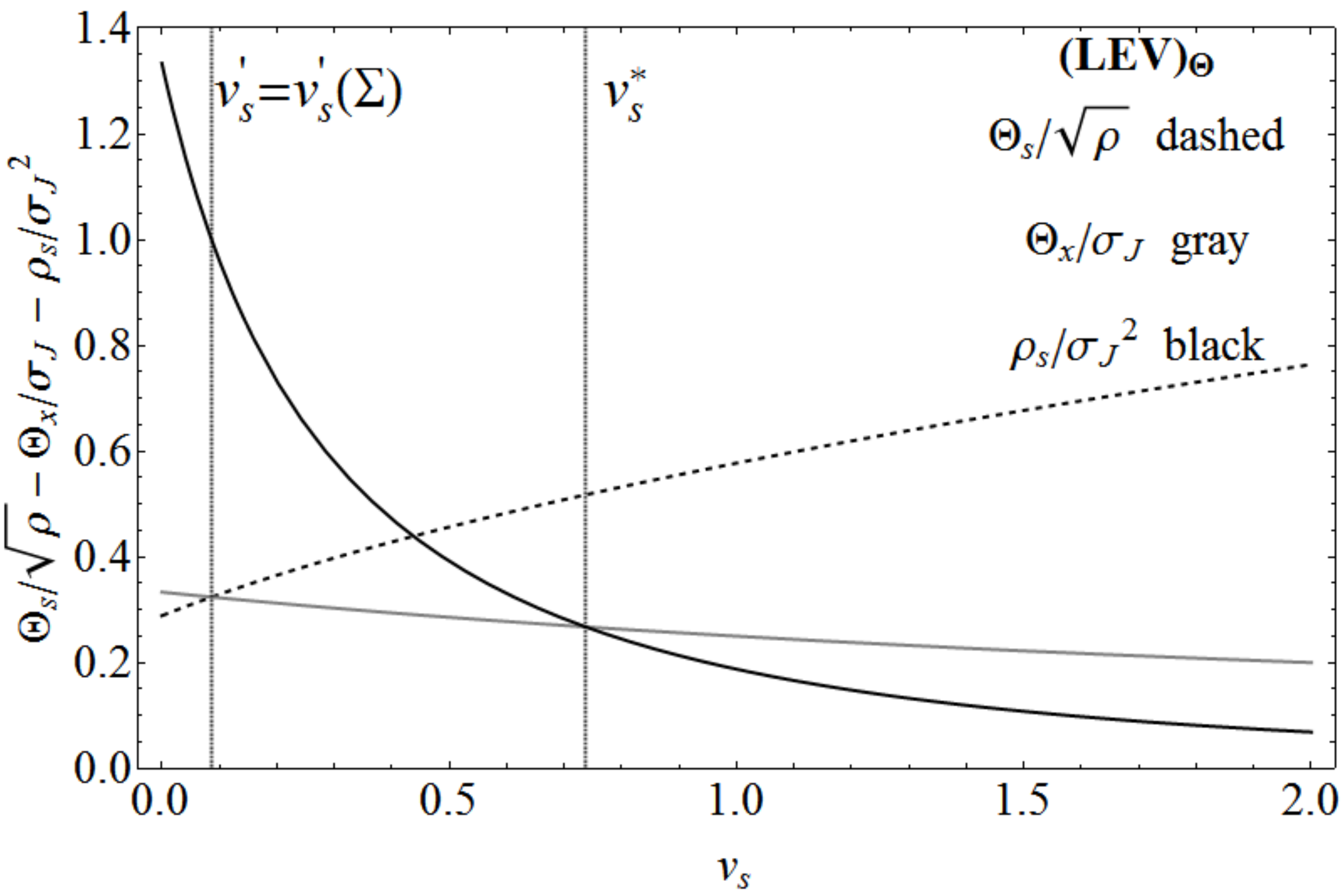}
\end{tabular}
\caption[font={footnotesize,it}]{\footnotesize{Ratios of the limiting extreme values \textbf{(LEV)}$_{\Sigma}$  (left panel) and  \textbf{(LEV)}$_{\Theta}$  introduced in Eq.\il(\ref{Eq:con3sSSinte}),  functions of the velocity of sound $v_s$. At at $v_s^*$ it is $\Sigma_x=\sigma_J^{-1} \rho_s=\frac{2 \sigma_J }{4+\sqrt{70}}\approx0.16 \sigma_J$ while
$\Theta_x=0.267592 \sigma_J$, in $v_s'$ it is
$\Sigma_s=\Sigma_x\sqrt{\rho}/\sigma_J\approx0.22 \sqrt{\rho}$ analogously  $\Theta_s=0.27 \sqrt{\rho}$, in the cross point in $]v_s',v_s^*[$ it  is
$\rho_s=0.32\sigma_J^2$ where   $\Sigma_s=0.32 \sqrt{\rho} $ in $\mathbf{(LEV)_{\Sigma}}$
while  the ratio is  $\rho_s/\sigma_J^2\approx 0.44$ in $\mathbf{(LEV)_{\Theta}}$.
}}\label{Fig:leex}
\end{figure}

Additional restrictions on the system with $\Sigma\neq0$ and $\Theta\neq0$, and therefore on the perturbations modify  the condition  (\ref{Eq:thetau}) on the reduced system as follows
\be\label{Eq:gener-quel-e}
\Sigma\geq \left.\Sigma_0\right|_{\mathcal{C}},\quad\left.\Sigma_0\right|_{\mathcal{C}}\equiv a [(b+c v_s)\Theta +c \sigma_J],\quad a<1\quad\mbox{or}\quad a\ll 1,\quad b\gg c,\quad
c\geq1,\qquad\{a,b,c\} \in\mathbb{N},
\ee
 the  quantities  $\{a, b, c\}$ change depending on the class of solutions.
In terms of the expansion  $\Theta $, analogously to Eq.\il(\ref{Eq:thetau}) one has
\be
\Theta \leq \left.\Theta_0\right|_{\mathcal{C}},\quad \left.\Theta_0\right|_{\mathcal{C}}\equiv \frac{\Sigma-a c \sigma_J}{a (b+ c v_s)}.
\ee
then the expansion limit is null $\left.\Theta_0\right|_{\mathcal{C}}=0$. For  $\Sigma_{max}\equiv a c \sigma_J$, for smaller shear  $\Sigma<\Sigma_{max}$ the expanding systems are always stables while a threshold for the instability of the contracting systems appears: for $\Theta<\left.\Theta_0\right|_{\mathcal{C}}$ the system is unstable.
It is noteworthy that the shear limit $\left.\Sigma_0\right|_{\mathcal{C}}=0$ is cancelled  for maximal contractions  equal to
$\Theta_{max}\equiv -c \sigma_J/(b + c v_s)$.
More generally, for each subclasses of Table\il(\ref{Table:asterisco}) and Table\il(\ref{Table:Fasterisco}) the unstable phase will be regulated by  some limiting extreme values \textbf{(LEV)} for the systems  ${\Theta}=0$ or  $\Sigma=0$  providing the  boundary values  for the regions  of the plane  $\Theta-\Sigma$ respectively  for the stability of the system:
\bea\label{Eq:con3sSSinte}
\mbox{\textbf{(LEV)}}_{\Sigma}:&& \quad \rho_s\equiv\frac{12 \sigma_J^2}{(3+v_s)^2 (1+3 v_s)},\qquad \Sigma_s\equiv\frac{1}{3} \sqrt{\left(\frac{1}{3}+v_s\right) \rho }, \qquad \Sigma_x\equiv\frac{2 \sigma_J}{3(3+ v_s)}.
\\
\mbox{\textbf{(LEV)}}_{\Theta}:&& \quad \rho_s=\left.\rho_s\right|_{\textbf{(LEV)}_{\Sigma}},
 \qquad\Theta_s\equiv\frac{3}{2}\Sigma_s, \qquad \Theta_x\equiv\frac{3}{2}\Sigma_x.
 \eea
 see Fig.\il(\ref{Fig:leex}).
A threshold exists for the \emph{stability} in the systems with the density in the two regimes   $\rho>\rho_s$ and $\rho<\rho_s$, according to  $\mathbf{(LEV)}_{\Sigma}$ or $\mathbf{(LEV)}_{\Theta}$, with  conditions on the radial shear or respectively the expansion or contraction.
The details of the analysis are considered  in Sec.\il(\ref{Sec:sottoclassi}).
For the stability in general one has the two following cases:
\bea
&&\rho<\rho_s\quad\Delta \in[-\Delta_x,-\Delta_s]\quad \mbox{or}\quad ]+\Delta_s, +\infty[
\\
&&\rho > \rho_s\quad ]-\Delta_s,\Delta_s[\quad \mbox{or}\quad  ]+\Delta_s, +\infty[,\quad \Delta=\Sigma \quad\mbox{for}\quad \mathbf{(LEV)}_{\Sigma}, \quad \Delta=\Theta \quad\mbox{for}\quad \mathbf{(LEV)}_{\Theta}
\eea
each of the two regions  are typically regulated  by additional conditions on $Q_i\in\vec{\nu}$ $\mathbf{(LEV)}_{\Sigma}$, for  example for the $\mathfrak{I}-$class,  $\mathbf{(\mathcal{A} T)}$ solution, are determined by a set of conditions on
 $\mathcal{B}$, $\xi$  and $E $.
We note that the limiting value on the density $\rho_s$,  increases with the conductivity but decreases with the velocity of sound: there are three ranges of $v_s$ to be considered as detailed in Fig.\il(\ref{Fig:leex}). However the boundary $ \Delta_s$ depends on the density $\rho$ and therefore  differently  regulated in  the three regions. For low density regimes
$\rho<\rho_s$, the density function   regulates  the system  stability in a reduced region of the $\rho-\Sigma$ plane through the conditions imposed on shear or the expansion especially for low values.
This situation is more evident for low velocity of sounds $v<v_s^*$. For larger values the  ranges  for low density variations are restricted to
 $\rho <0.2 \sigma_J$.
\subsection{Comments on the five principal classes of solutions}
\begin{description}
\item[
$\mathfrak{I}$-Class $ \mathbf{(\la\; T)}$:] this case is constrained by the condition
$\mathcal{C}(\la, T)$
obtained from Eq.\il(\ref{Eq:T_Ev}). This  relation  is a consequence
of the condition $T=0$, and it will occur in the other subcases
characterised by this assumption.  The evolution equation for the
parallel vorticity $\Omega$ is  $\dot{\Omega}=0$:  this
quantity remains constant along the fluid motion, for the perturbed
quantity as well as the unperturbed one. Furthermore,  the time evolution of the
matter density only  involves  the kinematic variable $Q$. We can draw
some conclusions on the stability of the system on the basis of the
 matrix trace. The  system is \emph{linearly unstable}
if with a  negative shear on
the radial direction which is bounded by $ \mathring{\Sigma} \leq
\Sigma_x$. It is worth noting that the limiting case is
defined only by the constitutive  equation, the conductivity
$\sigma_J$, and the equation of state by  $v_s$ ---fixed by the
reference solution. Furthermore, considering the coefficients $c_9=-1$
and $c_8={\rm{Tr}}\mathring{\mathbf{B}}<0$ we infer the condition for
the \emph{stability} $c_1<0$. Using the condition $\mathcal{C}(\la,
T)$, this condition can be rephrased as the following two
alternatives: (i) $\mathring{\rho}<\mathring{\rho}_s$ with
$\mathring{\Sigma}\in(-\mathring{\Sigma}_x,-\mathring{\Sigma}_s)\cup(\mathring{\Sigma}_s,+\infty)$
and (ii) $\mathring{\rho}\geq\mathring{\rho}_s$ with
$\mathring{\Sigma}\in(\mathring{\Sigma}_s,+\infty)$, where the $\mathbf{(LEV)_{\Sigma}}$ hold.
\\
\item[
$\mathfrak{II}$-\textbf{Class} $\mathbf{(\la\, Q)}$:] the condition ${\frac{1}{3}\Theta=-\Sigma}$
implies that the matter and field variables are related by the
constraint $\mathcal{C}(\la, Q)$
---see equation \eqref{Eq:Q_Ev}.  This relation is a consequence of
the assumption $Q=0$, and it will occur in the other subcases
characterised by this hypothesis. We observe that the assumption $Q=0$
implies in particular that the expansion of the 3-sheets and the
radial component of the shear of the 3-sheet must have opposite
sign. Studying the problem for the reduced system of nine variables we
find that
 the system is \emph{linearly unstable}
if ${\rm{Tr}}\oB>0$ ---that is if $\mathring{\Theta}$ is negative and
bounded above by $\mathring{\Theta} \leq\left.\Theta_0\right|_{\mathcal{C}}$ with $ (a=c=1; b=9)$. Again,
as in the case $T=0$, we obtain a constraint for the radial expansion
that depends on the unperturbed constants $(\sigma_J, v_s)$.
\\
\item[
$\mathfrak{III}-$\textbf{Class} $\mathbf{(\la\,\Phi)}$:] the assumptions $\la=0$ and $\Phi=0$ on
equations \eqref{Eq:dotphi} and \eqref{Eq:xi_Ev} give rise to two
possible subcases: (i) $\Omega=0$ or (ii) $\xi=0$ and
$\mathcal{B}=0$. We consider therefore these two subcases separately:
\begin{description}
\item[$\mathbf{(\la\,\Phi\, \Omega)}$:]
the trace
 implies the following condition of \emph{instability} $\mathring{\Sigma} \geq \left.\Sigma_0\right|_{\mathcal{C}}$ with $(a=2/9, b=6, c=1)$.
 We note that for the reference solution $\mathring{\Theta}$ and
$\mathring{\Sigma}$ are related by the square of the velocity of sound
and the conductivity through $\sigma_J$. However, the sign of the
$\Sigma \Theta$ is not constrained by this relation.
\item[
$\mathbf{(\la\,\Phi\, \xi\, \mathcal{B})}$:] in this case $T$ is the
only kinematical variable involved in the evolution of the
electromagnetic field. The criterion on the trace of the matrix
$\mathring{\mathbf{B}}$ implies the following condition of
\emph{instability}   $\mathring{\Sigma} \geq \left.\Sigma_0\right|_{\mathcal{C}}$ with $(a=2/21, b=16, c=3)$.
\end{description}
Thus, we finally note that the  subcases (i) and (ii) are
characterised by similar constraints on  $\mathring{\Sigma} $ and $\mathring{\Theta}$.
\item[
$\mathfrak{IV}-$\textbf{Class} $\mathbf{(\la\,\xi)}$: ]  equation \eqref{Eq:xi_Ev} leads to the
condition $\mathcal{C}(\la,\xi):$. On the other
hand the condition of \emph{instability} on the trace of the matrix
$\mathring{\mathbf{B}}$ leads to the inequality  $\mathring{\Sigma} \geq \left.\Sigma_0\right|_{\mathcal{C}}$ with $(a=1/33, b=40, c=6)$.
\item[$\mathfrak{V}-$\textbf{Class}  $\mathbf{(\la\,\Omega)}$:] in this case the system is \emph{unstable} if
 $\mathring{\Sigma} \geq \left.\Sigma_0\right|_{\mathcal{C}}$ with $(a=1/15, b=19, c=3)$.
\end{description}
We conclude this Section pointing out  that the  stability of the
configuration under consideration is constrained by similar relations
between the  unperturbed radial part of the shear of the 3-sheet
$\mathring{\Sigma}$ and the radial expansion
$\mathring{\Theta}$. These constraints only depend on the square of
the  sound velocity and the  conductivity. In Sec.\il(\ref{Sec:sottoclassi}) we
specialise this analysis to consider  the subclasses of Tables\il(\ref{Table:asterisco}) and (\ref{Table:Fasterisco}).
\section{Conclusions}\label{Sec:conclusions}
In this article we have explored the stability proprieties of an
ideal, LRS Einstein-Maxwell perfect fluid system. As
a first step we have formulated an hyperbolic initial value problem
providing a suitable (quasilinear) symmetric hyperbolic system by
means of which one can address the nonlinear stability of this system.
This first result allows us to provide   a suitable  formulation of an initial value
problem, that is a  necessary  issue for the construction of the  numerical solutions, ensuring  the  local
and global existence problems. Moreover the problem of the propagation of the constraints
can be   assumed  satisfied at
all time and then prove this result by fairly general arguments
as discussed in \cite{Reu98,Reu99}.
Our analysis is based on a $1+1+2$-tetrad formalism. In our
calculations certain choices of kinematic properties of  the
configuration  were motivated by the necessity to extract some
information from the rather complicated equations governing the
perturbations. We  studied five principal classes of solutions and  different subcases, considering systems with
particular kinematic configurations. The assumption of the radial
symmetry simplifies the problem and takes great advantage from the
$1+1+2$-decomposition \cite{Clarkson07}. Thus, in this work we have proceeded as follows:
we wrote the  $1+1+2$-equations for the LRS
system using the radial vector, pointing
along the axis of symmetry. Then, a discussion on the thermodynamical
quantities of the system was been provided. For the linear
perturbation analysis it was  useful to introduce a re-parametrised
set of evolution equations based of a suitable combination of  the
radially projected shear and  expansion. The resulting evolution system
was then used to analyse the stability  problem for small linear perturbations of the background.
We presented the   re-parameterized set of evolution equations
collected in a proper symmetric hyperbolic form and discussed  the
perturbation to the first order of the variables. We have presented
the main results concerning the linear stability of this system in a
classification of subcases that constitutes the principal result of
this paper.  In the Appendix to this article, we provide also an
alterative symmetric hyperbolic system for the fluid fields within the
$1+1+2$-decomposition followed by some general notes on the evolution
equations and hyperbolicity considerations. A suitable propagation
equation for the  fluid radial acceleration is there recovered
by the introduction of a new unknown field corresponding to the
derivative of the matter density projected along the radial
direction. suitable field and evolution equations can be obtained for
this quantity. The set of evolution equations is complemented by the
constraint and constitutive equations.  Restricting our attention
to isotropic fluids (entropy is a constant of both space and time) we
considered a one species particle fluid (simple fluid) and we
introduced a polytropic equation of state with a constant velocity of
sound. In order to close the system of evolution equations it is
necessary to specify the form of the conduction current. Accordingly,
we assumed the  Ohm's law so that a linear relation between the conduction current
and the electric field, involving a constant electrical conductivity
coefficient, holds. We have assumed that the pressure of the fluid $p$
and the charge density $\varrho_C$ to be functions of the matter
density $\rho$, and the charge current $j_c$ function of the electric
field $E$. Although viable, the perturbed equation for the radially
projected acceleration $\la$ turned out to be a very complicated
expression of the other variables and their derivatives.  Thus, in
order proceed in the stability analysis, we studied a simplified form
system by taking up some assumptions on the configuration. Assuming a
null radial acceleration for the reference solution, our analysis is
particularly focused on some specific cases defined by fixing the
expansions of the 3-sheets and 2-sheets, the radial part of the shear
of the 3-sheet, the twisting of the 2-sheet and the radial part of the
vorticity of the 3-sheet. In this way we are also able to provide
results concerning the structure of the associated LSS taking into
considerations all the different subcases. This analysis constitutes
the main result of the paper. In particular, \emph{we found that in many
cases the stability conditions can be strongly determined by the
constitutive equations by means of the square of the velocity of sound
and the electric conductivity}. In particular, this  is evident for the contracting and expanding  LRS configurations:   a threshold for the emergence of the instability appears in both cases.
The conditions bind mainly the expansion (viceversa contraction) along the preferred direction with respect to different regimes  of the radial shears. These results provide in a quite immediate manner information regarding  were certainly unstable system. The results for this type of configurations are illustrated in Sec.\il(\ref{Sec:submain}) and  schematically in Fig.\il(\ref{Fig:PrimaS}) and
Fig.\il(\ref{Fig:P01min}), in the four fundamental cases emphasizing the role of the couple of parameters  $(v_s,\sigma_J)$ moreover together with  these parameters the  relative sign of scalars  $\Sigma$ and $\Theta$ plays an essential role in the determination of the unstable phases of the systems. For expanding configurations with $\Sigma\Theta<0$, appears no threshold for the emergence of instability by means of  condition (\ref{Eq:first_cond}), this does not mean that expanding systems  with negative sher  are in any condition  certainly stable, but further  conditions can be provided  as we discussed in dealing with different classes of solutions. The interesting situation is in the remaining three cases, the role of the velocity of sound and the  conductivity  acts in a different way for the systems in the three regions of the Figs.\il(\ref{Fig:PrimaS},\ref{Fig:P01min}), according to our  analysis  to favor  or not  certainly unstable states of the system; in any case there is always is a specific threshold for the contraction or expansion, above which in the first case for contracting  systems  with a  fast contraction rate $|\Theta|>|\Theta_u(\sigma_J, v_s)|>0$ or  expansions at $0<\Theta<\Theta_u(\sigma_J, v_s)$ system is certainly unstable. The other cases and subcases show similar situations where the threshold provided by the density values,  and the couple $(\Sigma, \Theta)$ varies in  in different ranges depending on    $(\rho, v_s, \sigma_J)$. The magnetic field does not have a specific role in determining the stability of the system, and  the  Maxwell field and the geometrical  effects   enclosed by the magnetic and electric  part of the Weyl tensor.

Concerning the methods,  the core of the stability
analysis is  given by the study  of the non principal  part of the matrix $\mathring{\mathbf{B}} $ of the  system using some
relaxed stability eigenvalue conditions.  However, a full analysis of
the stability properties of the system turns extremely cumbersome
because of the form of the matrix. Thus, we  proceeded with the
analysis of the eigenvalues using an  indirect method aimed at
determining to know the sign of these.  Repeatedly  use
   the fact that a sufficient condition for the \emph{instability} of
   the system is the existence of at last one positive
   eigenvalues. This simple requirement is nevertheless able to
   provide a immediate way to show the conditions where the linear
   instability occurs. In several places we made use of  the
   Descartes criterion  to determine the \emph{maximum} number of
   positive and negative real roots of the  characteristic
   polynomial. In particularly simple cases one can make use of the
   so-called Routh-Hurwitz criterion to determine the number of roots
   with positive and negative real part of the polynomial by
   constructing the Routh associated matrix. It may
be possible to use some of these
criteria  for some specific cases and we expect future work  to
include a development of this paper in this direction.

\subsection*{Acknowledgments}

DP gratefully acknowledges support from the Blanceflor
Boncompagni-Ludovisi, n\'ee Bildt   and wishes to thank  the Angelo
Della Riccia Foundation and thank the institutional support
of   the Faculty of Philosophy and Science of the Silesian University of Opava.

\appendix
\section{A symmetric hyperbolic system for the fluid fields}\label{Sec:fluidfield}
In this Section we provide a brief discussion of a construction
leading to a suitable propagation equation for the radial acceleration
$\mathcal{A}$. In addition, we also provide an alternative set of the
propagation equations for the other kinematic and field variables. A
similar construction has been given in a slightly different context in \cite{LubbeKroon2011kz}.

For the sake of convenience let us define
$g_{ab}u^au^b=u_0u^0+u^iu_i=\epsilon$. In the present case
$\epsilon=-1$. Using the identity $u_a\nabla_bu^a=0$ we readily
conclude that
\begin{equation}
\label{E:(24b)}
\nabla_au^0=-\frac{u_i}{u_0}\nabla_au^i.
\end{equation}
Moreover, one has that
\(
\nabla_c \nabla_du^0+\nabla_c\nabla_d u^i=-\frac{1}{u_0}\nabla_cu_i \nabla_d u^i-\frac{u_iu^j}{(u_0)^2u^0}\nabla_du^i\nabla_cu_j.
\)
From the Bianchi identity and the inhomogeneous Maxwell equation we
have that
\begin{equation}
\label{E:convSiM}
\nabla^aF_{ac}=+\varepsilon J_c.
\end{equation}
In this last equation we introduce $\varepsilon=-1$, to match
equations \eqref{E:convSiM} and \eqref{E:maxz}. In addition, we have
the continuity equation for the matter density $\rho$
\begin{equation}
\label{E:(26a)}
u_a\nabla^a\rho+(p+\rho)\nabla^au_a-\varepsilon u^b F_b^{\phantom\ c}J_c=0,
\end{equation}
and the Euler equation
\begin{equation}
\label{E:(a)}
(p+\rho)u^a\nabla_a u^c-\epsilon h^{bc}\nabla_bp-\varepsilon\epsilon J_b F^{cb}=0.
\end{equation}
This last equation can be alternatively written as
\begin{equation}
\label{E:b}
(p+\rho)u^a\nabla_au^c-\epsilon\nabla^cp+u^c u^b \nabla_bp-\varepsilon\epsilon J_b F^{cb}=0.
\end{equation}
In what follows, we will consider a barotropic equation of state
$p=p(\rho)$ and  introduce here the quantity $v_s$ defined by the
relation
\(
\nabla_bp=\left(\frac{\partial p}{\partial \rho}\right)\nabla_b\rho=v_s\nabla_b\rho.
\)
Note that $v_s=\nu_s^2$, where $\nu_s$ is the speed of sound
introduced   in equation \eqref{E:vs}. Thus Eq.\il(\ref{E:b}) is now
\begin{equation}
\label{E:(26b)}
(p+\rho)u^a\nabla_au^c-\epsilon v_s\nabla^c\rho+v_su^c u^b \nabla_b\rho-\varepsilon\epsilon J_b F^{cb}=0.
\end{equation}
Using equations \eqref{E:(26a)} and \eqref{E:(26b)}, we obtain:
\begin{equation}
\label{E:(27)}
\nabla_c\rho=\epsilon \frac{(\rho+p)}{v_s}u^a\nabla_au_c-\epsilon (p+\rho)(\nabla_a u^a)u_c-\varepsilon\frac{J^bF_{cb}}{v_s}+\varepsilon\epsilon u^bF_{bd}J^d u_c.
\end{equation}
Using again the normalisation condition  we can write $u_0$ as:
\(
u_0=\epsilon u^0=\beta\sqrt{\epsilon(\epsilon-u_iu^i)},
\)
where $g^{00}=\eta^{00}=\epsilon$ and $\beta=\beta(\epsilon)$ is a
sign to be fixed  according to the metric signature conventions.

Now, using equations \eqref{E:(24b)} in equation \eqref{E:(26a)} we find
\begin{equation}
\label{E:(28)}
(u^0\nabla_0\rho+u^i\nabla_i\rho)+(\rho+p) \left(\nabla_iu^i-\frac{u_i}{u_0}\nabla_0u^i\right)-\varepsilon u^b F_{b}^{\phantom\ c}J_c=0,
\end{equation}
and from equation \eqref{E:(26b)}
\begin{eqnarray*}
&& \tilde{\mathcal{E}}^{(0)}\equiv u^a\nabla_a u^0-\frac{\epsilon v_s }{\rho+p}\left(\nabla_cg^{c0}\right)+\frac{u^0v_s}{\rho+p}u^b\nabla_b\rho-\varepsilon\epsilon J_bF^{0b}=0,
\\
&& \tilde{\mathcal{E}}^{(i)}\equiv u^a\nabla_a u^i-\frac{\epsilon v_s }{\rho+p}\left(\nabla_cg^{ci}\right)+\frac{u^iv_s}{\rho+p}u^b\nabla_b\rho-\varepsilon\epsilon J_bF^{ib}=0.
\end{eqnarray*}
Thus, introducing the zero-quantity:
\[
\varsigma^i\equiv\frac{u^i}{u^0}\tilde{\mathcal{E}}^{(0)}-\tilde{\mathcal{E}}^{(i)}=0,
\]
we obtain the equation
\begin{equation}
\label{E:(29)}
\varsigma^i= -\left(u^a\nabla_a u^i+\frac{u^iu_j}{u^0u_0}u^a\nabla_a u^j\right)+\frac{\epsilon v_s}{\rho+p}\nabla_c\rho\left(g^{ci}-g^{c0}\frac{u^i}{u^0}\right)+\epsilon\varepsilon J_b\left(F^{ib}-\frac{u^i}{u^0}F^{0b}\right)=0.
\end{equation}
Equations\eqref{E:(28)},\eqref{E:(29)}) constitute, after
multiplication by a suitable numerical factor, an hyperbolic system
for $\rho$ and $u^i$.

\medskip
In what follows, It turns to be necessary to introduce here the fields
\begin{equation}
\label{E:(31)}
\mu_a\equiv\nabla_a \rho,\quad U_{ab}\equiv \nabla_a u_b,
\end{equation}
where $\nabla_{[a}\mu_{b]}=0$. One readily can verify that
\(
U_{ab}u^b=0,\quad U_{a0}=-\frac{u^i}{u^0}U_{ai},
\)
and that
\begin{eqnarray}
\label{E:(32a)}
&& \nabla_aU_b^{\phantom\ 0}=-\frac{u_i}{u_0}\nabla_a U_b^{\phantom\ i}-\frac{U_a^{\phantom\ 0}}{u_0}U_{b0}-
\frac{U_a^{\phantom\ i}}{u_0}U_{bi},
\\
\label{E:(32b)}
&& u^b\nabla_aU_{cb}=-U_{cb}U_a^{\phantom\ b}.
\end{eqnarray}
Now, applying a covariant derivative to the equations of motion of the
fluid and commuting gives
\begin{equation}
Z_{cb}\equiv (1+v_s)u_b u^a\nabla_a\mu_c+(\rho+p)\big( u_b
  \nabla_aU_c^{\phantom\ a}+u^a\nabla_aU_{cb}\big)-\epsilon v_s
\nabla_c\mu_b+W_{cb}=0, \label{Zcb}
\end{equation}
where
\begin{eqnarray*}
&& W_{cb}\equiv (\rho+p)\big(R^{a}_{dca}u^d u_b+R_{cabd}u^d
  u^a\big)+U_{cb}u^a \mu_a(1+v_s)+u_bU_c^{\phantom\ a}\mu_a
(1+v_s)+\mu_c (1+v_s)\big( u_b U_a^a+u^aU_{ab}\big)
\\
&&\hspace{3cm}+ (\rho+p)
\big( U_{cb}U_{a}^a
+U_{c}^{\phantom\ a}U_{ab}\big)-\epsilon \nabla_cv_s\mu_b-\epsilon \nabla_c\big(J_a F_b^{\phantom\ a}\big)+u_b \mu_c u^a\nabla_av_s.
\end{eqnarray*}

Using equation \eqref{E:(32b)} we obtain
\begin{equation}
\label{E:(35a)}
u^bZ_{cb}=\epsilon u^a\nabla_a \mu_c+\epsilon (\rho+p)\nabla_aU_c^{\phantom\ a}+X_c,
\end{equation}
with
\(
X_c\equiv u^bW_{cb}-(\rho+p)u^aU_{cb}U_a^{\phantom\ b},
\)
where
\[
u^bW_{cb}=\epsilon\big((\rho+p)R^a_{dca}u^d+\mu_aU_c^{\phantom\ a}(1+v_s)+\mu_c (1+v_s)U_a^a-\mu_b
u^b \nabla_cv_s-u^b\nabla_c(J_a F_b^{\phantom\ a})+\mu_c
u^a\nabla_av_s \big).
\]
In addition, one has that
\(
h_d{}^bZ_{cb}=(\rho+p)u^a\nabla_a U_{cd}
-\epsilon v_s\nabla_c\mu_d+v_s u_d u^b \nabla_b \mu_c+Y_{cd}=0,
\)
with
\(
Y_{cd}=h_d{}^bW_{cb}+\epsilon (\rho+p)u_d U_{cb}U_a^{\phantom\ b} u^a.
\)

Equation \eqref{E:(35a)} can be written as
\begin{equation}
\label{E:(36)}
\epsilon\big[(u^0\nabla_0\mu_c+u^i\nabla_i \mu_c\big)+\epsilon(\rho+p)\big(\nabla_iU_c^{\phantom\ i}-\frac{u_i}{u_0}\nabla_0U_c^{\phantom\ i}\big)+\hat{X}_c=0,
\end{equation}
with
\be
\hat{X}_c\equiv X_c-\frac{1}{u_0}\big(U_0^{\phantom\ a}U_{ca}\big).
\ee
Now, the combination $(u^i/u_0)h^{0b}Z_{cb}-h^{ib}Z_{cb}$, leads to the equation
\begin{eqnarray}
&&-(\rho+p)\left(u^a\nabla_a U_c^{\phantom\ i }+\frac{u^i}{(u_0)^2}u_ju^a\nabla_aU_c^{\phantom\ j}\right)+
\epsilon v_s \left(g^{di}\nabla_d\mu_c-\frac{u^i}{u_0}g^{d0}\nabla_d\mu_c\right)+\left(
\frac{u^i}{u_0}g^{d0}-g^{id}\right)Y_{cd}\\
&&\hspace{2cm}-(\rho+p)\frac{u^i}{(u_0)^2}u^aU_a^{\phantom\ b}U_{cb}+v_s u^i
u^b\nabla_b\mu_c(\epsilon-1)=0. \label{E:(37)}
\end{eqnarray}
Viceversa, the combination $(u^i/u^0)h^{0b}Z_{cb}-h^{ib}Z_{cb}$
leads to the equation
\begin{equation}
(\rho+p)\left(u^a\nabla_a U_c^{\phantom\ i}+\frac{u^i u_j u^a \nabla_a U_c^{\phantom\ j}}{u_0u^0}\right)-
\epsilon v_s\left(g^{di}\nabla_d\mu_c-\frac{u^i}{u^0}(\nabla_d
  \mu_c)g^{d0} \right)+\hat{Y}_c^{\phantom\ i}=0 \label{E:(37b)}
\end{equation}
where
\[
\hat{Y}_c^{\phantom\ i}\equiv (\rho+p)\frac{u^a u^i}{u_0u^0}U_a^{\phantom\ b} U_{cb}-\left(
g^{d0}\frac{u^i}{u^0}-g^{di }\right)Y_{cd}.
\]
Equations \eqref{E:(36)} and \eqref{E:(37b)} constitute a symmetric hyperbolic system for
the fields  $\mu_a$ and $U_a^{\phantom\ i}$.

\section{$1+1+2$-decomposition}\label{App:2+1+1decom}
In this Section we study the $1+1+2$-decomposition of the term
$Z_{cb}$ in equation \eqref{Zcb}. In what follows we fix
$\epsilon=-1$. Recall that  up to now $(\rho, \mu_c, u^i,U_{ab} )$ have been used
as independent variables for our evolution system. We now consider
$(\rho, \mu_c, u^i)$  and the quantities $(\para \la, \ort \la^a,
\Theta, \Sigma, \Omega, \ort \Sigma_a,\ort \Omega^a, \ort \Sigma_a^b)$
as independent variables, the last being defined by the decomposition
of the $U_{ab}$.  This decomposition is obtained from considering $\para \mathcal{A}=u^b n^a U_{ba}$, and
\(
U_{ab}=- u_a n_b \para\la +\widetilde{U}_{ab}\il,
\)
where
\[
\widetilde{U}_{ab}\equiv+ n_an_b\left(
\tfrac{1}{2}\para\Theta +\para \Sigma \right) + N_{ab}\left(
\tfrac{1}{3}\para\Theta - \tfrac{1}{2}\para\Sigma\right)+\para \Omega\epsilon_{ab}+\ort \widetilde{U}_{ab},
\]
with
\[
\ort \widetilde{U}_{ab}\equiv-ua \ort \la_b+ n_a\left(\ort\Sigma_b + \epsilon_{bc}\ort\Omega^c\right)
 + \left(\ort \Sigma_a-\epsilon_{ac}\ort\Omega^c \right) n_b +\ort\Sigma_{ab}.
\]

Thus,  equation \eqref{Zcb}   is now
\begin{eqnarray*}
&&Z_{cb}= (1+v_s)u_b u^a\nabla_a\mu_c-\epsilon v_s \nabla_c\mu_b-(\rho+p)u_c\left[u_b n^a\nabla_a \para\la+n_bu^a\nabla_a\para \la\right]
\\
&&\hspace{1cm}+(\rho+p)\left[u_b n_c n^a\nabla_a \left(\tfrac{1}{3}\para\Theta +\para\Sigma\right)+u_b N_c^a\nabla_a \left(\tfrac{1}{3}\para\Theta- \tfrac{1}{2}\para\Sigma\right)+u_b \epsilon_c^{\phantom\ a}\nabla_a \para \Omega+n_cn_b u^a\nabla_a \left(\tfrac{1}{3}\para\Theta +\para\Sigma\right)\right.
\\
&& \hspace{1cm} \left.+N_{cb}u^a \nabla_a \left(\tfrac{1}{3}\para\Theta- \tfrac{1}{2}\para\Sigma\right)+\epsilon_{cb}u^a\nabla_a\para\Omega\right]+{\grave{Z}}_{cb}=0,
\end{eqnarray*}
with
\begin{eqnarray*}
{\grave{Z}}_{cb}\equiv W_{cb}+(\rho+p)\big(\left[u_b \nabla_a\ort U_c^{\phantom\ a}+u^a\nabla_a\ort U_{cb}\right]-\para\la \left[u_b\nabla_a(u_cn^a)+u^a\nabla_a(u_c n_b)\right]+\tfrac{1}{3}u_b \nabla_a \left(n^an_c+N_c^{a}\right)\para\Theta
\\
+\para\Sigma u^a\nabla_a
  \left(n_bn_c-\tfrac{1}{2}N_{bc}\right)+\tfrac{1}{3}u^a\nabla_a \left(n_cn_b
+N_{bc}\right)\para\Theta
+\para\Sigma u_b\nabla_a \left(n^a n_c-\tfrac{1}{2}N^a_{c}\right)+\para\Omega \left[u_b\nabla_a\epsilon_c^{\phantom\ a}+u^a\nabla_a\epsilon_{cb}\right]\big).
\end{eqnarray*}
Explicitly $W_{cb}$ is given by
\begin{eqnarray*}
&& W^{cb}=- R^{c}{}_{a} u^{a} u^{b} \bigl(p + \rho\bigr) + R^{b}{}_{a}{}^{c}{}_{d} u^{a} u^{d} \bigl(p + \rho\bigr) + u^{b} \bigl(1 + v_s\bigr) \mu^{c} D_{a}u^{a} + u^{a} \mu^{c} D_{a}u^{b} + u^{a} v_s \mu^{c} D_{a}u^{b} -  J^{a} u^{b} D^{c}E_{a} + J^{a} u_{a} D^{c}E^{b}
\\
&&+ E^{b} u^{a} D^{c}J_{a} -  E^{a} u^{b} D^{c}J_{a} + E^{b} J^{a} D^{c}u_{a} + u^{b} \mu^{a} D^{c}u_{a} + u^{b} v_s \mu^{a} D^{c}u_{a} + p D^{a}u^{b} D^{c}u_{a} + \rho D^{a}u^{b} D^{c}u_{a} -  E^{a} J_{a} D^{c}u^{b}
\\
 &&+ u^{a} \mu_{a} D^{c}u^{b} + u^{a} v_s \mu_{a} D^{c}u^{b} + p D_{a}u^{a} D^{c}u^{b} + \rho D_{a}u^{a} D^{c}u^{b} -  \epsilon^{b}{}_{ae}{}^{d} \bigl(J^{a} u^{e} D^{c}B_{d} + B^{a} (- u^{e} D^{c}J_{d} + J^{e} D^{c}u_{d})\bigr)
 \\
 &&+ u^{a} u^{b} \mu_{a} D^{c}v_s + \mu^{b} D^{c}v_s.
\end{eqnarray*}

Taking into account the above expressions, we obtain from the
evolution equations the following projections:

\medskip
$\mathbf{u^bu^cZ_{cb}}$:
\begin{equation}
\label{cancell}
u^bu^cZ_{cb}=-u^c u^a\nabla_a \mu_c-(\rho+p)n^a\nabla_a\para\la+u^bu^c{\grave{Z}}_{cb}=0.
\end{equation}

$\mathbf{n^bn^cZ_{cb}}$:
\begin{equation}\label{bmolle}
n^bn^cZ_{cb}=v_s n^c n^b  \nabla_b \mu_c +(\rho+p)u^a\nabla_a \left(\tfrac{1}{3}\para\Theta +\para\Sigma\right)+n^bn^c{\grave{Z}}_{cb}=0.
\end{equation}

$\mathbf{n^cu^bZ_{cb}}$:
\begin{equation}\label{tildeori}
n^cu^bZ_{cb}=-n^c u^a\nabla_a\mu_c-(\rho+p)n^a\nabla_a\left(\tfrac{1}{3}\para\Theta +\para\Sigma\right)+n^cu^b{\grave{Z}}_{cb}=0.
\end{equation}

$\mathbf{u^cn^bZ_{cb}}$:
\begin{equation}\label{tildevert}
u^cn^bZ_{cb}=v_s u^c n^b \nabla_b \mu_c+(\rho+p)u^a\nabla_a\para\la+u^cn^b{\grave{Z}}_{cb}=0.
\end{equation}

$\mathbf{g^{cb}Z_{cb}}$:
\begin{equation}\label{meno}
g^{cb}Z_{cb}=(1+v_s)u^b u^a\nabla_a\mu_b+(\rho+p)n^a\nabla_a\para\la +v_s \nabla^c\mu_c +(\rho+p)
u^a\nabla_a \para\Theta+g^{cb}{\grave{Z}}_{cb}=0.
\end{equation}

$\mathbf{h^{cb}Z_{cb}}$:
\begin{equation}\label{piu}
h^{cb}Z_{cb}=+v_s h^{cb} \nabla_c\mu_b +(\rho+p)
u^a\nabla_a \para\Theta+h^{cb}{\grave{Z}}_{cb}=0.
\end{equation}

$\mathbf{N^{cb}Z_{cb}}$:
\begin{equation}\label{uguale}
N^{cb}Z_{cb}=+v_s N^{cb} \nabla_c\mu_b +2(\rho+p)\
u^a\nabla_a \left(\tfrac{1}{3}\para\Theta-\tfrac{1}{2} \para\Sigma\right)+N^{cb}{\grave{Z}}_{cb}=0.
\end{equation}

$\mathbf{\epsilon^{cb}Z_{cb}}$:
\begin{equation}\label{xomega}
\epsilon^{cb}Z_{cb}=+2(\rho+p)u^a\nabla_a\para\Omega+\epsilon^{cb}{\grave{Z}}_{cb}=0.
\end{equation}
The field $\mu_b$ can be decomposed as
\(
\mu_b=\ort\mu_b+\para\mu n_b.
\)
This expression can be explicitly written as $\mu_a=(N^b_a \mu_b-u_au^b\dot{\mu_b})+n_an^b\mu_b$,
thus the previous decomposition can be further refined. Using   the
torsion-free condition $\nabla_a\mu_b=\nabla_b\mu_a$ we obtain  from  $n^c u^a(\nabla_a\mu_c-\nabla_c\mu_a)=0$,
\begin{equation}
\label{torsi}
u^a\nabla_a\para\mu=\ort\mu_c u^a\nabla_an^c+u^an^c\nabla_c\ort\mu_a-\para\mu n_an^c\nabla_cu^a\il,
\end{equation}
which can be read as a propagation equation for $\para\mu$.

\medskip
Using equation \eqref{torsi} we get from equation \eqref{cancell},
($u^bu^cZ_{cb}=0$):
\begin{equation}
\label{cancell1}
 (\rho+p)n^a\nabla_a\para\la=-u^c u^a\nabla_a \ort\mu_c+\para\mu\para\la-u^bu^c{\grave{Z}}_{cb}.
\end{equation}

From equation \eqref{bmolle}, ($n^bn^cZ_{cb}=0$):
\begin{equation}
\label{bmolle1}
(\rho+p)u^a\nabla_a \left(\tfrac{1}{3}\para\Theta +\para\Sigma\right)+v_sn^b\nabla_b\para\mu=v_s\ort\mu^cn^b\nabla_bn_c -n^bn^c{\grave{Z}}_{cb}.
\end{equation}

From equation \eqref{tildeori}, ($n^cu^bZ_{cb}=0$):
\begin{equation}
\label{tildeori2a}
u^a\nabla_a\para\mu+(\rho+p)n^a\nabla_a\left(\tfrac{1}{3}\para\Theta +\para\Sigma\right)=\ort\mu_cu^a\nabla_an^c+n^cu^b{\grave{Z}}_{cb}.
\end{equation}

Using equation \eqref{torsi} in equation \eqref{tildeori} one obtains
\begin{equation}
\label{tildeori1b}
(\rho+p)n^a\nabla_a\left(\tfrac{1}{3}\para\Theta +\para\Sigma\right)=- u^an^c \nabla_c\ort\mu_a+\para\mu n_an^c\nabla_cu^a+n^cu^b{\grave{Z}}_{cb}.
\end{equation}

From equation \eqref{tildevert} ($u^cn^bZ_{cb}=0$) one gets:
\begin{equation}
\label{tildevert1a}
(\rho+p)u^a\nabla_a\para\la=-v_s u^c n^b \nabla_b\ort \mu_c-v_s\para\mu u^cn^b\nabla_bn_c-u^cn^b{\grave{Z}}_{cb},
\end{equation}
and using equation \eqref{torsi} in equation \eqref{tildevert}
\begin{equation}
\label{tildevert1b}
(\rho+p)u^a\nabla_a\para\la+v_su^c\nabla_c\para\mu=v_s \ort \mu_b u^c \nabla_c n^b-u^cn^b{\grave{Z}}_{cb}.
\end{equation}

From equation \eqref{meno} ($g^{cb}Z_{cb}=0$) we get:
\begin{eqnarray}
\nonumber
&&  (\rho+p)
u^a\nabla_a \para\Theta+(\rho+p)n_c\nabla^c\para\la+v_s n_c\nabla^c\para\mu\\
&&\hspace{1cm}=-
(1+v_s)u^b u^a\nabla_a\ort\mu_b -v_s\nabla_c\ort\mu^c-\mu_bu^bu^a\nabla_av_s-(1+v_s)
\para\mu
u^bu^a\nabla_an_b-v_s \para\mu\nabla^cn_c-g^{cb}{\grave{Z}}_{cb}.\label{meno1}
\end{eqnarray}

From equation \eqref{piu}, ($h^{cb}Z_{cb}=0$):
\begin{equation}
\label{piu1}
(\rho+p)
u^a\nabla_a \para\Theta+v_s n^c \nabla_c\para\mu=-v_s h^{cb} \nabla_c\ort\mu_b -v_s\para\mu\nabla_cn^c -h^{cb}{\grave{Z}}_{cb},
\end{equation}

From equation \eqref{uguale}, ($N^{cb}Z_{cb}=0$):
\begin{equation}
\label{uguale1}
(\rho+p)
u^a\nabla_a \left(\tfrac{2}{3}\para\Theta- \para\Sigma\right)=-v_s N^{cb} \nabla_c\ort\mu_b-v_s N^{cb} \para\mu\nabla_cn_b -N^{cb}{\grave{Z}}_{cb}.
\end{equation}

From equation \eqref{bmolle1}, ($n^bn^cZ_{cb}=0$):
\begin{equation}
\label{bmolle2}
 (\rho+p)u^a\nabla_a (\tfrac{1}{3}\para\Theta +\para\Sigma)+v_sn^b\nabla_b\para\mu=v_s\ort\mu^cn^b\nabla_bn_c -n^bn^c{\grave{Z}}_{cb}.
\end{equation}

From  equation \eqref{piu} ($h^{cb}Z_{cb}=0$):
\begin{equation}
\label{piu1}
(\rho+p)
u^a\nabla_a \para\Theta+v_s n^c \nabla_c\para\mu=-v_s h^{cb} \nabla_c\ort\mu_b -v_s\para\mu\nabla_cn^c -h^{cb}{\grave{Z}}_{cb}.
\end{equation}

Finally,  we consider the equation
\(
\tfrac{3}{2}\frac{\rho+p}{v_s} (\mbox{\eqref{bmolle1}}-\tfrac{1}{3}\mbox{\eqref{piu1}}),
\)
which, explicitly, is given by
\begin{equation}
\label{evsigma}
\tfrac{3}{2}\frac{(\rho+p)^2}{v_s}u^a\nabla_a\para\Sigma+(\rho+p)n^c\nabla_c\para\mu=\tfrac{3}{2}\frac{(\rho+p)}{v_s}\left(v_s\ort\mu^cn^b\nabla_bn_c -n^bn^c{\grave{Z}}_{cb}+\tfrac{1}{3}(v_s h^{cb} \nabla_c\ort\mu_b +v_s\para\mu\nabla_cn^c +h^{cb}{\grave{Z}}_{cb})\right),
\end{equation}
and the equation
\(
\frac{\rho+p}{3v_s}\mbox{\eqref{piu1}}
\)
or
\begin{equation}
\label{piu2}
\frac{(\rho+p)^2}{3v_s}
u^a\nabla_a \para\Theta+\frac{\rho+p}{3} n^c
\nabla_c\para\mu=-\frac{\rho+p}{3v_s}\left(v_s h^{cb}
  \nabla_c\ort\mu_b +v_s\para\mu\nabla_cn^c
  +h^{cb}{\grave{Z}}_{cb}\right).
\end{equation}

\medskip
For a LSS the equations
Equations \eqref{tildevert1a}, \eqref{xomega}, \eqref{tildeori2a},
\eqref{piu2}, \eqref{evsigma} constitute a  symmetric hyperbolic system for the unknowns
$(\para\la,\para\Omega,\para\mu, \para \Theta, \para\Sigma)$. In
Section \eqref{Sec:Re-parametrized} we discussed an alternative set of
symmetric hyperbolic system where the evolution equation  for the
radial acceleration $\la$ is coupled with the evolution of equation
for the variable $Q \equiv \frac{2}{3}\Theta+2 \Sigma$. A  symmetric
hyperbolic system  has, therefore, been   given for the variables
$\mathbf{v}\equiv ({\rho
},{E},{B},{\mathcal{E}},{\mathcal{B}},{Q},{T},{\xi },{\Phi },{\Omega
},\la)$. Equation \eqref{Eq:A_Ev} for the variable $\la$ has been
recovered from the equation $\mbox{\eqref{tildevert1b}}-v_s
\mbox{\eqref{tildevert1a}}$. In a LSS
equation \eqref{tildevert1a} is explicitly given by
\[
 \la (1 +  v_s) \left( E j-  \la  \Theta (p+\rho)\right)+ j \varrho_C +
 E \varrho_C\left(\tfrac{2}{3} \Theta   - \Sigma\right)
 - 2 B \xi \varrho_C+ (p+\rho) u^{a} D_{a}\la  +\mu  u^{a} D_{a}v_s +v_s  u^{a} D_{a}\mu -  E u^{a} D_{a}\varrho_C=0,
\]
and 
\begin{eqnarray*}
&& \left(\tfrac{4}{3} \Theta +\Sigma\right)\left[ \la (p+\rho) + \mu(1+v_s)\right] -   \varrho_C\left[j
+ E \left( \Sigma+ \tfrac{1}{3} \Theta\right) \right]
  + E j (\Phi- \la)+ \tfrac{3}{2} (p+\rho) \Sigma \Phi  \\
&& \hspace{2cm} + \Omega \left[2 (p+\rho) \xi- 2 B j\right]  -  E n^{a} D_{a}j +(\rho+p)\left( \tfrac{1}{3} n^{a}  D_{a}\Theta + n^{a}  D_{a}\Sigma\right)+ u^{a} D_{a}\mu=0.
\end{eqnarray*}
It is possible to show,  using again the evolution equations and the
constraints that  the scalars $n^cn^b{\grave{Z}}_{cb}$,
$h^{cb}{\grave{Z}}_{cb}$, $u^c n^b{\grave{Z}}_{cb}$ and  $n^c
u^b{\grave{Z}}_{cb}$  do not contain derivatives of the variables. These terms are discussed with more detail in the following Section.
\section{Evolution equations and hyperbolicity considerations}\label{Sec:summary}
The system consisting of equations \eqref{E:HP_rho}, \eqref{E:fosev2},
\eqref{Eq:HP_omega}, \eqref{Eq:HP_xi}, \eqref{Eq:HP_phi},
\eqref{E:ma-re},  \eqref{Eq:HP_Weyl-B},
\eqref{Eq:HP_Weyl-E}, \eqref{Eq:HP_theta}, and \eqref{Eq:HP_sigma}
as discussed in Section \eqref{Sec:varandeq} and equations
\eqref{tildevert1a}, \eqref{tildeori2a}, \eqref{evsigma} and \ref{piu2} for the variables
\[
\mathbf{v}^{(b)}\equiv (\rho,s,
n, \para\Omega, \xi,   \Phi, \para E, \para
B, \para\mathcal{B}, \para\mathcal{E}, \para\la, \para\mu, \para\Sigma
,\para \Theta)
\]
can be written explicitly as
\begin{eqnarray}
&& \dot{\rho}=-(\rho+p)\Theta+\para E \para j, \label{permu7}
\\
&& \dot{s}=\frac{1}{nT}\para E\para j, \label{E:osev27}
\\
&& \dot{n}=-n \para \Theta, \label{E:nosev7}
\\
&& \dot{\para\Omega}=\para\mathcal{A}\xi
-\tfrac{2}{3}   \para\Theta\para \Omega+\para\Sigma \para\Omega, \label{qomega7}
\\
&& \dot{\xi}=\tfrac{1}{2}\para\mathcal{B}-\left(\tfrac{1}{3}\Theta-\tfrac{1}{2}\Sigma\right)\xi
+\left(\para\mathcal{A}-\tfrac{1}{2}\Phi\right)\para\Omega, \label{perxi7}
\\
&& \dot{\Phi}=\left(\tfrac{1}{3}\para\Theta+\tfrac{1}{2}\para\Sigma\right)\left(2 \para\mathcal{A}+\Phi
\right)-2\xi\para\Omega, \label{perfi7}
\\
&& \para\dot{{E}} =2\xi\para{B}
 -\left(\tfrac{2}{3}\Theta-\Sigma\right)\para{E}-\para{j}, \label{E:mawire7}
\\
&& \para\dot{{B}}=-2\xi\para{E}-\left(\tfrac{2}{3}\Theta-\Sigma\right)\para{B},\label{E:mawirb7}
\\
&& \para\dot{{\mathcal{B}}}=\left(\tfrac{3}{2}\Sigma
-\Theta\right)\para{{\mathcal{B}}}-3\xi{}\para {{\mathcal{E}}}-\xi
\left(\para E^2+\para B^2\right), \label{bdotv7}
\\
&& \para\dot{{\mathcal{E}}}=\left(\tfrac{3}{2}\Sigma
-\Theta\right)\para{{\mathcal{E}}}+3\xi\para{{\mathcal{B}}}-\tfrac{1}{2}
(\rho+p)\Sigma+\left(\tfrac{1}{2}\para\Sigma-\tfrac{1}{3}\para\Theta\right)(\para E^2+\para B^2)-\tfrac{2}{3}\para E \para j,
\label{E:We_R17}
\\
&& (\rho+p)u^a\nabla_a\para\la=-v_s\para\mu u^cn^b\nabla_bn_c-u^cn^b{\grave{Z}}_{cb},
 \\
&& u^a\nabla_a\para\mu+(\rho+p)n^a\nabla_a\left(\tfrac{1}{3}\para\Theta
   +\para\Sigma\right)=n^cu^b{\grave{Z}}_{cb}, \label{tildeori2a7}
\\
&& \tfrac{3}{2}\frac{(\rho+p)^2}{v_s}u^a\nabla_a\para\Sigma+(\rho+p)n^c\nabla_c\para\mu=\tfrac{3}{2}\frac{(\rho+p)}{v_s}
\left( \tfrac{1}{3}(v_s\para\mu\nabla_cn^c
    +h^{cb}{\grave{Z}}_{cb}) -n^bn^c{\grave{Z}}_{cb}\right), \label{evsigma7}
\\
&& \frac{(\rho+p)^2}{3v_s}
u^a\nabla_a \para\Theta+\frac{\rho+p}{3} n^c \nabla_c\para\mu=-\frac{\rho+p}{3v_s}\left(v_s\para\mu\nabla_cn^c +h^{cb}{\grave{Z}}_{cb}\right).
 \end{eqnarray}

In the above expressions notice that
\(
\dot{u}^a=\para\la n^a, \qquad
\dot{n}^a=\para\la u^a.
\)

Moreover, one has that
\begin{eqnarray*}
&& n^cn^b{\grave{Z}}_{cb}=-(\rho+p)(\para\la^2+2\para\Omega\para\xi)+(\rho+p)\left[\tfrac{1}{3}(3p-\rho)-\tfrac{1}{2}(p-\rho
-\para E^2-\para B^2)-\para\mathcal{E}\right]+\para\mu \para\la(1+v_s)
\\
&& \hspace{3cm} +(\rho+p)\left(\tfrac{1}{3}\Theta+\Sigma\right)\left(\tfrac{4}{3}\para\Theta+\para\Sigma\right)-\epsilon \para\mu \hat{v_s}+\para\hat{E}\varrho_C+\para E\left(\hat{\rho}_C+\para j \left(\tfrac{1}{3}\para\Theta+\para\Sigma\right)\right),
\end{eqnarray*}
\begin{eqnarray*}
  && h^{cb}{\grave{Z}}_{cb}=-\para \la (\rho+p)+\tfrac{1}{2}(\rho+p)(\rho+3p +\para E^2+\para B^2)+\para\mu\para\la(1+v_s)+(\rho+p)\left(\Theta^2+\left( \tfrac{1}{3}\Theta+\Sigma\right)^2+2(\tfrac{1}{3}\Theta-\tfrac{1}{2}\Sigma)^2\right)
\\
&&\hspace{3cm} -\epsilon \para\mu\hat{v}_s+\para\hat{E}\varrho_C+\para E (\hat{\rho}_C+\varrho_C\Phi+\para j \Theta),
\end{eqnarray*}
\begin{eqnarray*}
&& u^c n^b{\grave{Z}}_{cb}=-\para\la(\rho+p)\left(\para\Sigma +\tfrac{1}{3}\Theta\right)+\para\la(\rho+p)\left(
\tfrac{4}{3}\Theta+\Sigma\right)-\epsilon \para\mu \dot{v}_s+\para\dot{E}\varrho_C+\para E\dot{\rho}_C+\para E\para\la\para j,
\end{eqnarray*}

\begin{eqnarray*}
&& n^c u^b{\grave{Z}}_{cb}=-(\rho+p)\left(\tfrac{1}{2}\Phi\Sigma+\para\la\left(\tfrac{1}{3}\Theta+\Sigma\right)+
2\para\Omega\para\xi\right)-\para\mu(1+v_s)\left(\tfrac{4}{3}\Theta+\Sigma\right)+\dot{v}_s \para\mu-\para\hat{E}\para j-\para E \left(\para\hat{j}+\varrho_C \left(\para\Sigma +\tfrac{1}{3}\Theta\right)\right).
\\\label{E:-4-}
\end{eqnarray*}

To write down the above explicit expressions for
$(n^cn^b{\grave{Z}}_{cb})$,  $(n^c u^b{\grave{Z}}_{cb})$ and
$(h^{cb}{\grave{Z}}_{cb})$,   $(u^c n^b{\grave{Z}}_{cb})$ we  have
used, again, the evolution equation \eqref{E:mawire7}  and the
constraint  equations \eqref{E:contE} for $\para {E}$. With regards to
the term $\hat{v}_s$  in $(n^cn^b{\grave{Z}}_{cb})$ and
$(h^{cb}{\grave{Z}}_{cb})$ and $\dot{v}_s$  in $(u^c
n^b{\grave{Z}}_{cb})$  and $(n^c u^b{\grave{Z}}_{cb})$, involving the
derivatives of $v_s$ we note the following: the definition
$v_s=dp/d\rho$ has been used by setting an appropriate equation of
state $p=p(\rho)$. For example, in te case of a polytropic equation of
state one has that $v_s=$ and therefore set aside the terms $\hat{v}_s$ and
$\dot{v}_s$.  Otherwise, one can consider  a generic equation of  state
$\rho=\rho(p)$ such that $d{v}_s/d\rho=d^2p/d\rho^2\neq0$ and
$\nabla_av_s=\left(d^2p/d\rho^2 \right)\nabla_a \rho$ and assume
$\left(d^2p/d\rho^2 \right)=\mbox{constant}$. In the more general
case we refer to the discussion in\cite{Cho08} ---see Section
14.4. In fact, as discussed in Section \ref{Sec:on-the-thermod},  for
the more general case of a fluid which is not isentropic, i. e.
$s\neq$constant, the equation of state should be written in the form
$p=p(\rho, s)$ ---see \cite{PugKroon12} and  references therein. This
means that when considering derivatives of the pressure $p$, it should
have been taken into account that $\nabla_ap=v_s
\nabla_a\rho+(\partial p/\partial s)\nabla_as$, where the time
evolution of the entropy  is governed by  equation
\eqref{E:osev27}.

The evolution equations under consideration contains terms involving
$\nabla_a v_s$ and $\nabla_a s$. The evolution equation for the new
variable $s_a\equiv\nabla_a s$ can be obtained by covariant
differentiating equation \eqref{E:osev27}, commuting
covariant derivatives and using, again, the evolution equations ---see \cite{PugKroon12}.

Concerning the terms $\dot{\rho}_{C}$ in $(u^c n^b{\grave{Z}}_{cb})$
and $\hat{\rho}_{C}$ in $(n^cn^b{\grave{Z}}_{cb})$ and
$(h^{cb}{\grave{Z}}_{cb})$, one can assume, for example, that the
charge density $\varrho_C$ is a function of the matter density
$\varrho_C=\varrho_C(\rho)$. Here we assume it to be a constant
multiple of the fluid density, $\rho=\varrho \varrho_C$. In this case
one can use again equations \eqref{permu7} and \eqref{E:mapors} and
the definition of $v_s$. Note that the density $\varrho_C$ appears in
the constraint equations for the electric field and the pressure. The
discussion of Sections \eqref{Sec:varandeq} and
\eqref{Sec:Re-parametrized} shows that $\varrho_C$ is only involved in
the evolution equation for the radial acceleration.  Here $\varrho_C$
is inherited from the term $\para\mu$ containing information from the
propagation of the matter density and pressure. Finally, we can use
equation \eqref{E:contE} and \eqref{E:jpara} for $\para\hat{j}$ in
$(n^c u^b{\grave{Z}}_{cb})$, assuming $\para j=\para j(\para E)$. In
particular, we take $\para j=\para E \sigma_J$.

This system \eqref{permu7}-\eqref{evsigma7} can be written matricially as
\(
\mathbf{A}_{(a)(b)}^{a}\partial_a \mathbf{v}^{(b)}=\mathbf{B}_{(a)(b)} \mathbf{v}^{(b)}.
\)
It is convenient to write  $\mathbf{v}^{(b)}=(\mathbf{v}^{(i)}, \mathbf{v}^{(A)})$ where
\[
\mathbf{v}^{(i)}=(\para E, \para B, \para\mathcal{B}, \para\mathcal{E}, s,
n, \rho, \para\Omega, \xi, \Phi,\para\la),\qquad \mathbf{v}^{(A)}=(\para\mu, \para \Theta, \para\Sigma).
\]
Thus it follows that
\begin{eqnarray*}
&& A_{(i)(i)}^a=u^a,\quad A_{(i)(A)}^a=A_{(A)(i)}^a=0,
\\
&& A_{(i)(j)}^a=A_{(j)(i)}^a=0 \quad i\neq j,
\end{eqnarray*}
while for $(\para\mu, \para\Theta, \para\Sigma)$:
\be
\begin{array}{lcl}
 A_{(\mu)(\mu)}^a =u^a, &  A_{(\Theta)(\Theta)}^a =\frac{(\rho+p)^2}{3v_s}u^a, &A_{(\Sigma)(\Sigma)}^a =\frac{3(\rho+p)^2}{2v_s}u^a,
 \\
 \\
 \\
  A_{(\mu)(\Theta)}^a =\frac{(\rho+p)n^a}{3}, &  A_{(\Theta)(\mu)}^a =\frac{(\rho+p)n^a}{3}, &A_{(\Sigma)(\mu)}^a =(\rho+p)n^a,
  \\
  \\
  \\
   A_{(\mu)(\Sigma)}^a =(\rho+p)n^a, &  A_{(\Theta)(\Sigma)}^a =0,&A_{(\Sigma)(\Theta)}^a =0.
 \end{array}
 \ee
\section{Details on the subclasses of the   solutions}\label{Sec:sottoclassi}
\begin{description}
\item[{Subclasses of the $\mathfrak{I}$-class  $\mathbf{(\la,T)}$}]
In what follows we analyse in further detail the following subcases of
the configuration$\mathbf{(\la\; T)}$:
\begin{description}
\item[
$\mathbf{(\la\,T\, Q)}$: ] in this case the configuration is defined by
the conditions $\la=0$, $\Sigma=0$ and $\Theta=0$. The matter density
evolution only depends on the electric field and the current
density. The evolution of the electromagnetic fields is regulated by
the twist $\xi$  of the 2-sheet. Moreover, the radial vorticity  is
constant during the motion ---i.e. $\dot{\Omega}=0$. From the
evolution equations for $T$ and $Q$ we obtain, respectively,  the
conditions $\mathcal{C}(\la,T)$
and  $\mathcal{C}(\la,Q)$.
From these relations  we find  $\mathcal{E}=p+\tfrac{1}{3}\rho-2\Omega
^2$ and $4\Omega^2=B^2+E^2+3 p+\rho$. This case has at least one zero
eigenvalue associated with the vorticity. This eigenvalue has multiplicity 2.
 The sum of the eigenvalues of the  associated 7-rank matrix is
 ${\rm{Tr}}\mathring{\mathbf{B}}=-\sigma_J<0$\footnote{ The non-zero
 coefficients of the characteristic  polynomial are:
$ c_2=(6 \mathring{\xi }^2 \mathring{\Omega })^2>0$, $ c_3=\mathring{\xi }^2 (8 \sigma_J \mathring{E}^2+18
  \mathring{\mathcal{B}} \mathring{\xi }+9 \sigma_J \mathring{\Omega
  }^2)$, $
 c_4=\tfrac{1}{2} \mathring{\xi } \big(3 \sigma_J
  (3 \mathring{\mathcal{B}}-4 \mathring{B} \mathring{E})+2
  \mathring{\xi } \big(9 \mathring{\Omega }^2-4 (\mathring{B}^2+\mathring{E}^2-9 \mathring{\xi
      }^2)\big)\big)$, $
c_5=\tfrac{9}{2} \mathring{\xi }
(\mathring{\mathcal{B}}+2 \sigma_J \mathring{\xi })$, $
c_6=13 \mathring{\xi }^2>0 \qquad c_7=-{\rm{Tr}}\, \mathring{\mathbf{B}}>0. $}
\item[
$\mathbf{(\la\,T\, \Phi)}$:]  in this case one has $\dot{\Omega}=0$,
and $\mathcal{C}(\la,T)$.  The
conditions $\la=0$ and   $\Phi=0$   imply from equations
\eqref{Eq:dotphi}, \eqref{Eq:xi_Ev}) either (i) $\Omega=0$   or (ii)
$\xi=0$ and $\mathcal{B}=0$. We consider these two subcases
separately:

\begin{itemize}
\item[(i)] For $\mathbf{(\la\,T\, \Phi\,\Omega)}$ the system is
  characterised by the condition
  $\mathcal{C}(\la,T,\Omega)$. From the
  criterion  on the trace one can deduce that the system is
  \emph{unstable} if $\mathring{\Theta}\leq -(\sigma_J)/(3+v_s)$. Now, considering
  the determinant of the reduced $7\times7$-linearised
  matrix\footnote{
\(
\det\mathring{\mathbf{B}}=
(1+v_s) \mathring{\xi }^2 (2 \sigma_J \mathring{E}^2+3
  \mathring{\mathcal{B}} \mathring{\xi }) \big((1+3 v_s)
  \mathring{\rho }-3 \mathring{Q}^2\big).
\)} the necessary  condition for \emph{stability} is that
$\det\mathring{\mathbf{B}}<0$. Combining this condition with the
criterion of the trace  and using the fact that
$\mathring{p}=\mathring{\rho} v_s$ we obtain for \emph{stable}
configuration the following conditions:
\begin{itemize}
\item[(ia)] for $\mathring{\rho} <\mathring{\rho}_s$   then one has
  that  $\mathring{\Sigma}\in(-\mathring{\Sigma}_s,
  +\mathring{\Sigma}_s)$ and $\mathfrak{Q}(\mathring{\mathcal{B}},
  \mathring{\xi}, \mathring{E}^2)<0$, \emph{or}
  $\mathring{\Sigma}\in(-\mathring{\Sigma}_x,-\mathring{\Sigma}_s)
  \cup (\mathring{\Sigma}_s,\infty)$ and
  $\mathfrak{Q}(\mathring{\mathcal{B}}, \mathring{\xi},
  \mathring{E}^2)>0$;

\item[(ib)] For $\mathring{\rho} \geq{\rho}_s$   then one has
  $\mathring{\Sigma}\in(-\mathring{\Sigma}_x, +\mathring{\Sigma}_s)$
  and $\mathfrak{Q}(\mathring{\mathcal{B}}, \mathring{\xi},
  \mathring{E}^2)<0$, \emph{or}
  $\mathring{\Sigma}\in(+\mathring{\Sigma}_s, \infty)$ and
  $\mathfrak{Q}(\mathring{\mathcal{B}},\mathring{\xi},
  \mathring{E}^2)>0$. With $\mathring{\xi}\neq0$, $\mathfrak{Q}({\mathcal{B}},{\xi}, {E}^2)\equiv3 \mathcal{B} \xi +2 E^2 \sigma_J$ and considering the $\mathbf{(LEV)_{\Sigma}}$
\end{itemize}
We note that the  reference  matter density and shear  $\rho_s$ and
$\Sigma_x$, depend on the constants $(v_s,\sigma_J)$ and the limit
$\Sigma_s$   depends only on the matter density and the square of the
sound velocity.

\item[(ii)] $\mathbf{(\la\,T\, \Phi,\xi,\mathcal{B})}$: in this case
  the Maxwell equations simply became $\dot{B}=0$ and $\dot{E}=-j$.
  Moreover $\dot{\Omega}=0$ and the condition  $\mathcal{C}(\la, T)$
  holds. The corresponding reduced system has  six unknowns. We notice
  here the eigenvalues: $\lambda_0=0$ with a subspace of dimension 3
  and
\[
\lambda_1= -\sigma_J, \qquad \lambda_{\pm}=\tfrac{1}{2} \big(-(3+v_s)\mathring{\Theta} \pm\sqrt{2 \mathring{\rho } (v_s+1)^2+(v_s-1)^2(\mathring{\Theta })^2}\big).
\]
 The conditions on the  sign of $\lambda_{\pm}$  imposes severe
 restrictions on the radial expansion. These depend on the sound
 velocity and the background density matter: the condition
 $\lambda_{\pm}<0$  (for the system \emph{stability}) is satisfied if
 $\mathring{\Theta}>\sqrt{(1+v_s) \mathring{\rho}/2 }$.

\end{itemize}
\item[
$\mathbf{(\la\,T\, \xi)}$:] this case implies the equations
$\dot{B}=\dot{\mathcal{B}}=\dot{\Omega}=\dot{\Phi}=0$ while the
electric field satisfies $\dot{E}=-j$.  Moreover, one has the
condition $\mathcal{C}(\la, \xi)$ with
$\mathcal{C}(\la, T)$. We study the
associated $8\times 8$ matrix: the eigenvalues of this matrix are
$\lambda_{\pm}$ and $\lambda_i$ as defined for the case
$\mathbf{(\la\,T\, \Phi,\xi,\mathcal{B})}$ and a zero eigenvalue with
multiplicity $5$ due to the variables $(B, \mathcal{B}, \Omega, \Phi)$.
\item[
$\mathbf{(\la\,T\, \Omega)}$:] in this case one has  $\dot{\Phi}=0$.
From equation \eqref{Eq:T_Ev}  one finds
$\mathcal{C}(\la,T,\Omega)$. The time evolution
of the electric and magnetic fields  are regulated by the  twisting
$\xi$ of the 2-sheet ---see equations
\eqref{Eq:dotE}-\eqref{Eq:dotB}).   The temporal evolution of $\xi$ is
fixed explicitly   by the magnetic part of the Weyl tensor ---see
equation \eqref{Eq:xi_Ev}. On the other hand,  from the trace of the
associated rank 7  matrix  we infer  that the system is
\emph{unstable} if $\mathring{\Sigma }< -\Sigma_x$. In
order to obtain necessary conditions for \emph{stability} one requires
$
c_7=-{\rm{Tr}}\mathring{\mathbf{B}}>0$, and $c_1>0$.
 These inequalities lead to the following cases with $\mathbf{(LEV)_{\Theta}}$:
\begin{itemize}
\item[(i)] if $\mathring{\rho} <\mathring{\rho}_s$   then
  $\mathring{\Theta}\in(-\mathring{\Theta}_x,
  -\mathring{\Theta}_s)\cup (+\mathring{\Theta}_s,\infty)$  and
  $\mathring{\xi}\neq0$, \emph{or}
  $\mathring{\Theta}\in(-\mathring{\Theta}_x,+\mathring{\Theta}_s)$
  and $\mathring{\xi}\mathring{\mathcal{B}}<0$;

\item[(ii)] if $\mathring{\rho} \geq{\rho}_s$   then
  $\mathring{\Theta}\in(-\mathring{\Theta}_s, +\mathring{\Theta}_s)$
  and  $\mathring{\xi}\mathring{\mathcal{B}}<0$, \emph{or}
  $\mathring{\Theta}\in(+\mathring{\Theta}_s, \infty)$ and
  $\mathring{\xi}\neq0$
\end{itemize}
\end{description}
\item[
{Subclasses of the $\mathfrak{II}$-class  $\mathbf{(\la,Q)}$}]
In this subsection we focus on the configurations with
$\frac{1}{3}\Theta =-\Sigma$. Taking into account the results on the
system  $\mathbf{(\la\, Q)}$, we consider the following subcases.
\begin{description}
\item[
$\mathbf{(\la\, Q\, \Phi)}$: ] the condition $\mathcal{C}(\la, Q)$ holds. Moreover, the conditions
$\la=0$ and $\Phi=0$, imply from equation \eqref{Eq:dotphi}    two
subcases,  (i)  $\Omega=0$ and (ii) $(\xi=0,\mathcal{B}=0)$,
respectively.

\begin{itemize}
\item[(i)] $\mathbf{(\la\, Q\, \Phi\, \Omega)}$: from the trace of the
  reduced rank 7 matrix 
the configuration is bound to be
  \emph{unstable} if $\mathring{\Sigma} \geq
  \Sigma_s\equiv{2\sigma_J}/{3(15+2 v_s)}$.  On the other hand,
  studying the sign of the characteristic polynomial coefficient
  $c_5<0$ we infer that  to obtain  \emph{stability}: for
  $\mathring{\mathcal{E}}<\mathring{\mathcal{E}}_p$ and (a)  $\mathring{T}>-{2 \sigma_J}/({15+2 v_s})$ with
  $\mathring{\xi}<-\mathring{\xi}_p$ or $\mathring{\xi}
  >\mathring{\xi} _p$, or (b)
  $\mathring{\mathcal{E}}<\mathring{\mathcal{E}}_p$ and
  $\mathring{T}>\mathring{T}_p$. Otherwise, for $\mathring{\mathcal{E}}\geq \mathring{\mathcal{E}}_p$ with  $\mathring{T}>-{2 \sigma_J}/(15+2 v_s)$,
with the following definitions:
\begin{eqnarray*}
&& \mathring{\mathcal{E}}_p\equiv \big(\tfrac{5}{6}+\tfrac{5}{2}v_s+v_s^2\big)\mathring{\rho} +\frac{2 \big(50+v_s (15+2 v_s)\big) \sigma_J^2}{(15+2 v_s)^2},
\\
&& \mathring{T}_p \equiv \frac{-3 (13+2 v_s) \sigma_J+\sqrt{6 (95+26 v_s) \bigg(\big(5+3 v_s (5+2 v_s)\big) \mathring{\rho }-6 \mathring{\mathcal{E}}\bigg)+9 (13+2 v_s)^2 \sigma_J^2}}{285+78 v_s},
\\
&& \mathring{\xi}_p= \frac{\sqrt{39} \sqrt{2 \big(5+3 v_s (5+2 v_s)\big) \mathring{\rho}-12\mathring{\mathcal{E}}-3 \mathring{T}^2 (95+26 v_s) -6 \mathring{T}(13+2 v_s) \sigma_J}}{78},
\end{eqnarray*}
with $T=-3\Sigma$. Once again,  the limiting conditions on the  radial
part of the shear of the 3-sheet, $\Sigma$, and the electric part of
the Weyl tensor is completely regulated by the reference density and
the constants $(\sigma_J, v_s)$. This \emph{stability} conditions can
be alternatively expressed as follows:
\[
\rho>0,  \qquad \mathring{\Sigma}>\Sigma_s, \qquad  \mbox{and}  \qquad
\mathring{\xi}^2> \mathring{\xi}^2_s\equiv\tfrac{1}{52} \bigg(2
(\mathring{B}^2+\mathring{E}^2)+4 \big(1+{v}_s (3+{v}_s)\big)
\mathring{\rho} +3 \mathring {\Sigma}  \big(2\sigma_J(13+2 v_s)-3
(95+26 v_s) \mathring{\Sigma} \big) \bigg).
\]

\item[(ii)] $\mathbf{(\la\, Q\, \Phi\, \xi\, \mathcal{B})}$:  with the
  condition $\mathcal{C}(\la, Q)$, while the time evolution of the
  radial projected vorticity is entirely regulated  by the radial part
  of the shear ---see equation \eqref{eq:Omega}. The system, with a
  rank 6 matrix, is \emph{unstable} if  $\mathring{\Sigma}
  \geq\Sigma_s\equiv2 \sigma_J/3(13+2 v_s)$. Otherwise, it is
  \emph{stable} if (a)
  $\mathring{\mathcal{E}}<\mathring{\mathcal{{E}}}_p$  with
  $\mathring{T}>-(2 \sigma_J)/(13+2 v_s)$  and {$ \mathring{\Omega}
  >\mathring{\Omega} _p \cup  \mathring{\Omega} <-\mathring{\Omega}
  _p$}  \emph{or} $-\mathring{\Omega} _p\leq \mathring{\Omega} \leq \mathring{\Omega} _p$  with $ \mathring{T}>\mathring{T}_p $; (b)  $\mathring{\mathcal{E}}\geq\mathring{\mathcal{{E}}}_p$ and $\mathring{T}>-(2 \sigma_J)/(13+2 v_s)$,
where
\begin{eqnarray*}
&& \mathring{\Omega} _p \equiv \frac{1}{2 \sqrt{6}}\sqrt{-3 \mathring{\mathcal{E}}+2 \big(2+3 v_s (2+v_s)\big) \mathring{\rho} +\frac{6 \left(73+26 v_s+4 v_s^2\right) \sigma_J^2}{(13+2 v_s)^2}},
\\\nonumber
&& \mathring{T}_p \equiv \frac{-3 (11+2 v_s) \sigma_J}{210+66 v_s}+\\
&&
+\frac{\sqrt{3} \sqrt{8 (35+11 v_s) \big(2+3 v_s (2+v_s)\big) \mathring{\rho} +3 (11+2 v_s)^2 \sigma_J^2-12 \mathring{\mathcal{E}} (35+11 v_s)-96 (35+11 v_s) \mathring{\Omega} ^2}}{210+66 v_s},
\\
&& \mathring{\mathcal{E}}_p \equiv \tfrac{2}{3} \big(2+3 v_s
(2+v_s)\big) \mathring{\rho} +\frac{2 \left(73+26 v_s+4 v_s^2\right)
  \sigma_J^2}{(13+2 v_s)^2}.
\end{eqnarray*}

Alternatively,   these conditions can be reexpressed as
\[
\mathring{\Sigma}<\Sigma_s \qquad \mbox{and} \qquad \mathring{\Omega}^2> \bigg(\mathring{B}^2 (3
\mathring{E}^2-2)-2\mathring{E}^2+\big(3+v_s (9+4 v_s)\big)  \mathring{\rho} +6
\mathring{\Sigma}\big((11+2 v_s) \sigma_J-3 (35+11 v_s)  \mathring{\Sigma} \big)\bigg)/16.
\]
\end{itemize}
\item[
$\mathbf{(\la\,Q\, \xi)}$:] in this case  one has $\Sigma=-\Theta/3$
with $\mathcal{C}(\la,\xi)$ from equation
\eqref{Eq:xi_Ev} and $\mathcal{C}(\la\,Q)$. The radial shear is the only kinematical variable that explicitly
regulates the time evolution of the variables  $(E, B, \rho,\Phi,
\mathcal{E},\mathcal{B})$. The radial vorticity and the shear are
related by the two evolution equations \eqref{eq:Omega} and
\eqref{Eq:T_Ev}),  respectively. This is a rank 8 matrix problem. The
system is \emph{unstable} if  $\mathring{\Sigma} \geq (2
\sigma_J)/(51+6 v_s)$.  The system is \emph{stable} if either (a)
$\mathring{\mathcal{E}}<\mathring{\mathcal{E}}_p$  and
$\mathring{T}>\mathring{T}_p$  \emph{or} $-{2 \sigma_J}/({17+2
  v_s})<\mathring{T}<\mathring {T}_p$   and $\mathring{\Omega}
<-\mathring{\Omega} _p\cup \mathring{\Omega} >\mathring{\Omega} _p$;
or (b) $\mathring{\mathcal{E}}\geq\mathring{\mathcal{E}}_p$  and
$\mathring{T}>-{2 \sigma_J}/({17+2 v_s})$. In the above conditions we used the following definitions:
\begin{eqnarray*}
&& \mathring{\Omega} _p \equiv \frac{\sqrt{3} \sqrt{4 \big(2+3 v_s (2+v_s)\big) \mathring{\rho}-6 \mathring{\mathcal{E}}-15 \mathring{T}^2 (25+6 v_s) -6 \mathring{T} (15+2 v_s) \sigma_J}}{12},
\\
&& \mathring{\mathcal{E}}_p \equiv \tfrac{2}{3} \big(2+3 v_s (2+v_s)\big) \mathring{\rho} +\frac{4 \big(65+v_s (17+2 v_s)\big) \sigma_J^2}{(17+2 v_s)^2},
\\
&& \mathring{T}_p \equiv \frac{\sqrt{30 (25+6 v_s) \bigg(2 \big(2+3
    v_s (2+v_s)\big) \mathring{\rho}-3 (15+2 v_s) \sigma_J-3
    \mathring{\mathcal{E}} \bigg)+9 (15+2 v_s)^2 \sigma_J^2}}{375+90
  v_s}.
\end{eqnarray*}
Alternatively, these conditions can be written as
\[
\mathring{\Sigma}<{2 \sigma_J}/({51+6 v_s}), \qquad  \mbox{and} \qquad
\mathring{\Omega}^2>\tfrac{1}{16}
\bigg(\mathring{B}^2+\mathring{E}^2+\big(3+v_s (9+4 v_s)\big) \mathring{\rho}
+3 \mathring{\Sigma}  \big((30 +4 v_s) \sigma_J-15 (25+6 v_s)
\mathring{\Sigma} \big)\bigg).
\]
\item[
$\mathbf{(\la\, Q\, \Omega)}$:]  in this case one has that $T=-3\Sigma$
and $\Sigma=-\Theta/3$ and  the condition $\mathcal{C}(\la, Q):
B^2+E^2+2 \mathcal{E}=-(p+\rho/3)$ holds. The system is
\emph{unstable} if $\mathring{\Sigma }\geq\Sigma_s\equiv\sigma_J/3(8+
v_s)$. A \emph{stable} system   meets the following conditions: (a)
$\mathring{\mathcal{E}}<\mathring{\mathcal{E}}_p$, with
$\mathring{T}>\mathring{T}_p$  \emph{or} $-3\Sigma_s<\mathring{T}\leq
\mathring{T}_p$and   $(\mathring{\xi }<-\mathring{\xi}
_p,\;\mathring{\xi }>\mathring{\xi}_p)$; or (b)
$\mathring{\mathcal{E}}\geq \mathring{\mathcal{E}}_p$ and $
\mathring{T}>-3\Sigma_s$. In the above expressions we have used the
definitions
\begin{eqnarray*}
&& \mathring{T}_p \equiv \frac{-3 (7+v_s) \sigma_J+\sqrt{3 (55+14 v_s) \big((5+3 v_s (5+2 v_s)) \mathring{\rho}-6 \mathring{\mathcal{E}} \big)+9 (7+v_s)^2 \sigma_J^2}}{165+42 v_s},
\\
&& \mathring{\xi} _p \equiv \frac{\sqrt{(5+3 v_s (5+2 v_s)) \mathring{\rho}-6 \mathring{\mathcal{E}}-3 \mathring{T}^2 (55+14 v_s) -6 \mathring{T} (7+v_s) \sigma_J}}{\sqrt{78}},
\\
&& \mathring{\mathcal{E}}_p \equiv \frac{(8+v_s)^2 (5+3 v_s (5+2 v_s)) \mathring{\rho} +3 (57+2 v_s (8+v_s)) \sigma_J^2}{6 (8+v_s)^2}.
\end{eqnarray*}
The conditions can be expressed, alternatively, as
\[
\mathring{\Sigma}<\Sigma_s \qquad  \mbox{and} \qquad 16 \mathring{\xi}^2>
\big(\mathring{ B}^2+\mathring{ E}^2+2 (1+v_s (3+v_s))\mathring{
    \rho} -9 (55+14 v_s)\mathring{\Sigma}^2+6 (7+v_s) \mathring{\Sigma
  } \sigma_J\big).
\]
\end{description}
\item[{Subclasses of the $\mathfrak{IV}$-class  $\mathbf{(\la,\xi)}$}]
$\mathbf{(\la\, \xi\, \Omega)}$: the assumptions $\la=0$, $\xi=0$ and
$\Omega=0$, lead from equation \eqref{Eq:xi_Ev} to the condition
$\mathcal{C}(\la\, \xi,\Omega):\mathcal{B}=0$. Thus, we analyse the
case: $\mathbf{(\la\, \xi\, \Omega\,\mathcal{B})}$, with a rank 7
matrix. The system is \emph{unstable} if
\(
\mathring{\Theta}\leq -3 (2
\sigma_J+(v_s-5)\mathring{ \Sigma})/4 (7+v_s).
\)
\item[{The case $\la=0$, $T=0$, and $Q=0$}]

$\mathbf{(\la\,T\, Q\, \Phi)}$: in this special case we assume the
radial acceleration to be zero. In addition,  the expansion of the
2-sheet $\Phi$ vanishes and also $\Sigma=\Theta=0$. It is worth noting
that the assumption $\la=\Sigma=\Theta=\Phi=0$ implies from equation
\eqref{Eq:dotphi} that $\xi \Omega=0$ ---that is, the twist of the
2-sheet $\xi$ together with the magnetic part of the Weyl tensor
vanish. This system can only accelerate in the radial direction or
otherwise radially projected vorticity will vanish. The basic
assumptions in this case directly lead to $\dot{\Omega}=0$,
$\dot{\xi}=\mathcal{B}/2$ and the conditions  $\mathcal{C}(\la,T)$ and $\mathcal{C}(\la,Q)$  hold.  Thus, two subcases occur:
\begin{itemize}
\item[(i)] $\mathbf{(\la\,T\, Q\, \Phi\,\xi\, \mathcal{B})}$; in this
  case one further has $\dot{B}=\dot{\Omega}=0$, $\dot{E}=-j$. The
  only non zero eigenvalue is $\lambda=-\sigma_J$.

\item[(ii)] $\mathbf{(\la\,T\, Q\, \Phi\,\Omega)}$; in this case, it
  can be shown that the trace of the reduced matrix  $
  {\rm{Tr}}\mathring{\mathbf{B}} = - \sigma_J< 0$  and  this system
  has a zero eigenvalue. Nevertheless, in this case  the conditions
  $\mathcal{C}(\la,T,\Omega)$ and
  $\mathcal{C}(\la,Q)$ is $B^2+E^2+2 \mathcal{E}=-\left(p+\frac{\rho
    }{3}\right)<0$ holds. These conditions are inconsistent with the
  type of  equation of state considered in this work. We notice that
  this situation   always  occurs when the two conditions
  $\mathcal{C}(\la,T,\Omega)$ and $\mathcal{C}(\la,Q)$ must be
  satisfied simultaneously.
\end{itemize}
\begin{description}
\item[
$\mathbf{(\la\,T\, Q\,  \xi)}$: ]  in this case the problem is
simplifies considerably and we find the relations
$\mathcal{C}(\la,\xi)$, $\mathcal{C}(\la,T)$  and $\mathcal{C}(\la,Q)$. Moreover, one has that $\dot{E}=-j$. The
evolution equations ${\dot{{B}}}=0$, $\dot{{{\mathcal{B}}}}=0$, ${\dot{{\Omega}}}=0$,
and $\dot{{{\Phi}}}=0$ give rise to repeated zero eigenvalues. In
addition one has the eigenvalue $\lambda=-\sigma_J<0$.
\item[
$\mathbf{(\la\,T\, Q\,  \Omega)}$:]  in this case one readily has that
$\dot{\Phi}=0$. The associated rank 5 matrix has a zero eigenvalue
with multiplicity 2.  In addition, one has that  $c_6=
{\rm{Tr}}\mathring{\mathbf{B}} = -\sigma_J< 0$. Thus, imposing the
condition $c_6 c_7>0$  leads to
$\mathring{\xi}^2>-(\mathring{B}^2+\mathring{E}^2+3\mathring{\mathcal{E}})/{26}$.
However, in this case the relations $\mathcal{C}(\la, T,\Omega)$ and $\mathcal{C}(\la,Q)$ hold. These cannot be satisfied for
ordinary matter density (i.e. such that $\rho>0$) and the given equation of state.
\end{description}
\item[{The case $\la=0$, $T=0$, $\xi=0$}]
$\mathbf{(\la\, T\, \xi\,  \Omega)}$:  the assumptions $\la=0$,
$\xi=0$ and $\Omega=0$ lead using  equation \eqref{Eq:xi_Ev}  to
$\mathcal{B}=0$.  From equation \eqref{Eq:T_Ev} one obtains the
conditions $\mathcal{C}(\la, T,\Omega)$, and
$\dot{\Phi}=\dot{B}=0$.  Thus,  we analyse the system $\mathbf{(\la\,
  T\, \xi\,  \Omega\, \mathcal{B})}$. There is a zero eigenvalue with
multiplicity 3. In addition, one has  $\lambda_1=-\sigma_J<0$ and
$\lambda_{\pm}=\frac{1}{4} (-(3+v_s) \mathring{3 \Sigma }\pm\sqrt{8
  (1+v_s)^2 \mathring{\rho }+(v_s-1)^2 (\mathring{3 \Sigma })^2})$,
the condition $\lambda_{\pm}<0$ is satisfied for
\[
\mathring{\Sigma} >2 (1+v_s) \sqrt{{ \mathring{\rho}
  }/({3(v_s+5+2\sqrt{3})(v_s+5-2\sqrt{3})})}.
\]

\item[{The case $\la=0$, $Q=0$, $\xi=0$}]
$\mathbf{(\la\,Q\, \xi \, \Omega)}$: this case reduces to
$\mathbf{(\la\,Q\, \xi \, \Omega\, \mathcal{B})}$ with
$\mathcal{C}(\la,Q)$. From the
associated rank 6 matrix trace  we infer that the system is certainly
\emph{unstable} if $\mathring{\Sigma}\geq \Sigma_s\equiv\sigma_J/3(6+
v_s)$. On the other hand, if the system is \emph{stable}, then the
following conditions must be verified: (a)
$\mathring{\mathcal{E}}>\mathring{\mathcal{E}}_p$ with $\mathring{T}>-3\Sigma_s$, or (b) $\mathring{\mathcal{E}}\leq \mathring{\mathcal{E}}_p$ and $\mathring{T}>\mathring{T}_p$ where we introduced the following notation
\begin{eqnarray*}
&& \mathring{\mathcal{E}}_p \equiv \frac{4 (6+v_s)^2 (2+3 v_s (2+v_s))\mathring{\rho} +3 (61+4 v_s (6+v_s))\sigma_J^2}{6 (6+v_s)^2},
\\
&& \mathring{T}_p \equiv \frac{-6 (5+v_s) \sigma_J+\sqrt{6}\sqrt{ (59+20 v_s) \big(2 (2+3 v_s (2+v_s)) \mathring{\rho}-3 \mathring{\mathcal{E}} \big)+6 (5+v_s)^2 \sigma_J^2}}{177+60 v_s}.
\end{eqnarray*}
This condition correspond to the requirements  $c_4>0$ and
${\rm{Tr}}\mathring{\mathbf{B}}<0$ arising, in turn, from
$c_5=-{\rm{Tr}}\mathring{\mathbf{B}}>0$ and  $c_4 c_5>0$. It  can be also written as: $\0o{\Sigma}<\Sigma_s$ and  $(\0o{E})^2+(\0o{B})^2<-[3+v_s (9+4 v_s)] \0o{\rho} +3 \0o{\Sigma}  [3 (59+20 v_s) \0o{\Sigma }-4 (5+v_s) \sigma_J]$.
\item[{The case $\la=0$, $\Phi=0$ and $\xi=0$}]
The case $\mathbf{(\la\,\Phi\,  \xi\,  \Omega)}$  reduces to
$\mathbf{(\la\,\Phi \, \xi \, \Omega\, \mathcal{B})}$. We can say that
the rank 6 system is \emph{unstable} when $\mathring{\Theta }\leq 3(5
\mathring{\Sigma} -2 \sigma_J)/2(14+3 v_s)$.
\item[{The case $\la=0$, $T=0$, $Q=0$ and $\xi=0$}]

In this case one has $\mathbf{(\la\,T\, Q\,  \xi\, \Omega)}$ which reduces to
$\mathbf{(\la\,T\, Q\,  \xi\, \Omega\, \mathcal{B})}$. This system is
characterised by $\dot{{\Phi}}=0$, $\dot{{B}}=0$,
$\dot{\mathcal{E}}=-j$. There  are the eigenvalues $\lambda=0$ with
multiplicity 4 and $\lambda_1=-\sigma_J<0$. Nevertheless, one has that
$\mathcal{C}(\la,T,\Omega)$ and
$\mathcal{C}(\la,Q$. These conditions cannot be satisfied  by the assumptions
$\rho>0$ and the given equation of state.
\end{description}


%

\begin{thebibliography}{90}
%



\bibitem{EtiLiuSha10}
Z.~B. Etienne, Y.~T. Liu, \& S.~L. Shapiro,
\newblock {\em Relativistic magnetohydrodynamics in dynamical spacetimes: a new
  AMR implementation},
\newblock Phys. Rev. D {\bf 82}, 084031 (2010).

\bibitem{Fon03}
J.~A. Font,
\newblock {\em Numerical hydrodynamics in general relativity},
\newblock Living Rev. Rel. {\bf 6} (2003).

 \bibitem{Font2008}
   J.~A.~Font,
    \newblock {\em Numerical hydrodynamics and magnetohydrodynamics in general relativity},
   \newblock Living Rev.\ Rel.\  {\bf 11}, 7 (2007) .


   \bibitem{Alc08}
M.~Alcubierre,
\newblock {\em Introduction to $3+1$ numerical Relativity},
\newblock Oxford University Press, 2008.

\bibitem{GiaRez07}
B.~Giacommazo \& L.~Rezzolla,
\newblock {\em WhiskeyMHD: a new numerical code for general relativistic MHD},
\newblock Class. Quantum Grav. {\bf 24}, S235 (2007).

 \bibitem{Lichnerowicz67}
  A. Lichnerowicz,
  \newblock {\em  Relativistic hydrodynamics and magnetohydrodynamics},
  \newblock Benjamin, New York, 1967.



 \bibitem{Anton2006} L. Anton, O.Zanotti, J.A.Miralles, J. M. Marti, J. M. Ibanez, J. A. Font and J. A. Pons,
 \newblock {\em Numerical 3+1 general relativistic magnetohydrodynamics: A Local characteristic approach},
 \newblock  Astrophys.\ J.\  {\bf 637}, 296 (2006).

\bibitem{Radice:2013xpa}
  D.Radice, L.Rezzolla and F.Galeazzi,
\emph{High-Order Fully General-Relativistic Hydrodynamics: new Approaches and Tests,
}  Class.\ Quant.\ Grav.\  {\bf 31},075012 (2014).


\bibitem{Witek:2013ora}
  H.Witek,
  \emph{Numerical Relativity in higher-dimensional space-times,
 } Int.\ J.\ Mod.\ Phys.\ A {\bf 28}, 1340017 (2013).


\bibitem{PugKroon12}
  D.~Pugliese and J.A. Valiente Kroon,
 \newblock{\em On the evolution equations for ideal magnetohydrodynamics in curved spacetime},
  Gen.\ Rel.\ Grav.\  {\bf 44}, 2785 (2012).
  
\bibitem{Cho08}
Y.~Choquet-Bruhat,
\newblock {\em General Relativity and the Einstein equations},
\newblock Oxford University Press, 2008.




  \bibitem{ShiSek05}
M.~Shibata \& Y.~Sekiguchi,
\newblock {\em Magnetohydrodynamics in full general relativity: formulations
  and tests},
\newblock Phys. Rev. D {\bf 72}, 044014 (2005).


\bibitem{BauSha03}
T.~W. Baumgarte \& S.~L. Shapiro,
\newblock {\em General relativistic magnetohydrodynamics for the numerical
  construction of dynamical spacetimes},
\newblock Astrophys. J. {\bf 585}, 921 (2003).


\bibitem{PalGarLehLie10}
C.~Palenzuela, D.~Garrett, L.~Lehner, \& S.~Liebling,
\newblock {\em Magnetospheres of black hole systems in force-free plasma},
\newblock Phys. Rev. D {\bf 82}, 044045 (2010).

\bibitem{Anile89}
 A.M. Anile
\newblock {\em Relativistic fluids and magneto-fluids: With applications in astrophysics and
plasma physics},
  \newblock  Cambridge UniversityPress, Cambridge, U.K.; New York, U.S.A., 1989.


\bibitem{Disconzi(2014)} M. M. Disconzi, 
\emph{On the well-posedness of relativistic viscous fluids,
}Nonlinearity \textbf{27}, 1915 (2014).

\bibitem{Ho60} E. Horst, \emph{Symmetric Plasmas and Their Decay},
Commun. Math. Phys. \textbf{126}, 613-633 (1990).

\bibitem{Mult-12}
S. C. Hsu, T. J. Awe, S. Brockington, A. Case, J.
T. Cassibry, G. Kagan, S. J. Messer,
M. Stanic, X. Tang, D. R. Welch, and F. D.
Witherspoon,
\emph{Spherically Imploding Plasma Liners as a Standoff
Driver for Magnetoinertial Fusion}, Ieee transactions on plasma science \textbf{40},
5  (2012).

\bibitem{Tan-B86}
Tai-Ho Tan
and Joseph E. Borovsky,
\emph{Spherically symmetric high-velocity plasma
expansions into background gases},
Journal of Plasma Physics  \textbf{35},  02 239
(1986).


\bibitem{Las-Lun07}
P. D. Laskyand A. W. C. Lun
\emph{Gravitational collapse of spherically symmetric plasmas in Einstein-Maxwell spacetimes
}.
Phys. Rev.  D {\bf 75}, 104010 (2007).
\bibitem{VCxL97} R. L. Viana, R. A. Clemente and S. R. Lopes,
\emph{Spherically symmetric stationary MHD equilibria
with azimuthal rotation},
 Plasma Phys. Control. Fusion \textbf{39}, 197 (1997).
\bibitem{Gu-Sha1999}
Yan Guo and A. Shadi Tahvildar-Zadeh,
\emph{Formation of singularities in relativistic
fluid dynamics
and in spherically symmetric plasma
dynamics},
Contem. Math.,
\textbf{238}, 151-161 (1999).


\bibitem{Reu98}
O.~Reula,
\newblock {\em Hyperbolic methods for Einstein's equations},
\newblock Living Rev. Rel. {\bf 3}, 1 (1998).

\bibitem{Reu99}
O.~Reula,
\newblock {\em Exponential decay for small nonlinear perturbations of expanding
  flat homogeneous cosmologies},
\newblock Phys. Rev. D {\bf 60}, 083507 (1999).

\bibitem{Clarkson07}
  C.~Clarkson,
  \emph{A Covariant approach for perturbations of rotationally symmetric spacetimes},
  Phys.\ Rev.\ D {\bf 76}, 104034 (2007).


 \bibitem{EllEls98}
G.~F.~R. Ellis \& H.~van Elst,
\newblock {\em Cosmological models: Cargese lectures 1998},
\newblock NATO Adv. Study Inst. Ser. C. Math. Phys. Sci. {\bf 541}, 1 (1998).



\bibitem{Clar-Mark-Bet-Dusns04} C. A. Clarkson, M. Marklund, G. Betschart, and P. K. S.
Dunsby,
\emph{The Electromagnetic Signature of Black Hole Ring-Down
}, Astrophys. J. {\bf 613}, 492 (2004).


\bibitem{Marklund:2004qz}
  M.~Marklund and C.~Clarkson,
\emph{The General relativistic MHD dynamo},
  Mon.\ Not.\ Roy.\ Astron.\ Soc.\  {\bf 358} (2005) 892.

\bibitem{Bets-Clark04}
G. Betschart,
and C. A. Clarkson,
\emph{Scalar field and electromagnetic perturbations on
locally rotationally symmetric spacetimes
},
Class. Quantum Grav. {\bf 21}, 5587 (2004).




 \bibitem{Bur-cqg-08}R. B. Burston,
\emph{1+1+2 gravitational perturbations on LRS class II
spacetimes: decoupling gravito-electromagnetic tensor
harmonic amplitudes
}, Class. Quantum Grav. {\bf 25} 075004 (2008).


\bibitem{Clark-Bar03}
C. A. Clarkson
and R. K. Barrett,
\emph{Covariant perturbations of Schwarzschild black holes
}, Class. Quantum Grav. {\bf 20} 3855 (2003).



\bibitem{Els-Ellis96}
H. van Elstyxand, G. F. R. Ellis,
\emph{The covariant approach to LRS perfect fluid spacetime
geometries
},Class. Quantum Grav. {\bf13}, 1099  (1996).


\bibitem{Burston:2007wt}
  R.~B.~Burston and A.~W.~C.~Lun,
  \emph{1+1+2 Electromagnetic perturbations on general LRS space-times: Regge-Wheeler and Bardeen-Press equations},
  Class.\ Quant.\ Grav.\  {\bf 25} (2008) 075003.


\bibitem{Burston:2007ws}
  R.~B.~Burston,
  \emph{1+1+2 Electromagnetic perturbations on non-vacuum LRS class II space-times: Decoupling scalar and 2-vector harmonic amplitudes},
  Class.\ Quant.\ Grav.\  {\bf 25}  075002 (2008).


\bibitem{Komiss}
 S. S. Komissarov,  Mon. Not. R. Astron. Soc.  \textbf{368},  993-1000 (2006).


\bibitem{Zanotti:2014haa}
  O.~Zanotti and D.~Pugliese,
  \emph{Von Zeipel's theorem for a magnetized circular flow around a compact object},
Gen.\ Rel.\ Grav.\  {\bf 47} (2015) 4,  44.


 \bibitem{10}
  Bekenstein, J.D., Oron, E.: New conservation laws in general-relativistic magnetohydrodynamics.
Phys. Rev. D 18, 1809"1¤71819 (1978). DOI 10.1103/PhysRevD.18.1809
\bibitem{11}
Bekenstein, J.D., Oron, E.: Interior magnetohydrodynamic structure of a rotating relativistic
star. Phys. Rev. D 19, 2827"1¤72837 (1979). DOI 10.1103/PhysRevD.19.2827

\bibitem{KramerSte}H. Stephani, D. Kramer, M. MacCallum, C. Hoenselaers, E. Herlt  ,\emph{Exact Solutions of Einstein's Field Equations}, Cambridge Monographs on Mathematical Physics, Paperback 2009.

\bibitem{LubbeKroon2011kz}
  C.~Lubbe and J.~A.~Valiente~Kroon,
 \emph{ A conformal approach for the analysis of the non-linear stability of pure radiation cosmologies},
  Annals Phys.\  {\bf 328}, 1 (2013).


\bibitem{proc}
 J.M. Stewart, and M. Walker, Proc. R. Soc. London \textbf{A} \textbf{431},  49 (1974).


\bibitem{AlhMenVal10}
A.~Alho, F.~C. Mena, \& J.~A. {Valiente Kroon},
\newblock {\em The Einstein-Friedrich-nonlinear scalar field system and the
  stability of scalar field Cosmologies},
\newblock In {\tt arXiv:1006.3778}, (2010).

\bibitem{AnalPropPol2002}
Q. I. Rahman and G. Schmeisser,
\newblock {\em Analytic Theory of Polynomials: Critical Points, Zeros and Extremal Properties}.
\newblock {London Mathematical Society Monographs}, Clarendon Press, 2002.

\bibitem{Barnett1971}
S. Barnett,
\emph{A New Formulation of the Theorems of
Hurwitz, Routh and Sturm}
J. Inst. Maths Applies {\bf 7} 240 (1971).

\bibitem{Acta98}
H. O. Kreiss and J. Lorenz,
\newblock {\em Stability for time-dependent differential equations}.
\newblock {Acta Numerica} {\bf 7}, 203 (1998).


\bibitem{1997funct.an..3003K}
H. O. Kreiss, O. E. Ortiz and O. A. Reula,
\newblock {\em Stability of quasi-linear hyperbolic dissipative systems}.
\newblock { Journal of Differential Equations} {\bf 142}, 78 (1998).

\bibitem{JMP1}
H. O. Kreiss, G. B. Nagy, O. E. Ortiz and O. A. Reula,
\newblock {\em Global existence and exponential decay for hyperbolic dissipative relativistic fluid theories}.
\newblock {Journal of Mathematical Physics} {\bf 38}, 5272 (1997).

\bibitem{JMP2}
O. E. Ortiz,
\newblock {\em Stability of nonconservative hyperbolic systems and relativistic dissipative fluids}.
\newblock {Journal of Mathematical Physics} {\bf 42}, 1426 (2001).

















%
%
%
%
%
%
%
%
%
%
%
%
%
%
%
%
%
%
%
%
%
%
%
%
%
%
%
%
%
%
%
%
%
%
%
%
%
%
%
%

























%


































\end{thebibliography}
\end{document}